\documentclass[twocolumn,aps,superscriptaddress,showpacs,floatfix,prc,noshowpacs]{revtex4-2}
\usepackage{mathrsfs}
\usepackage{amssymb}
\usepackage{amsmath}
\usepackage{graphicx}
\usepackage[normalem]{ulem}
\usepackage[dvips]{color}
\usepackage{bm}
\usepackage{longtable}\usepackage{slashed}

\usepackage{enumitem}

\usepackage{times}
\usepackage{fouriernc}

\usepackage[bb=ams, cal=cm, scr=boondox, frak=euler]{mathalpha}

\usepackage[breaklinks=true]{hyperref}
\hypersetup{
  colorlinks=true,
  citecolor=cyan,
  linkcolor=black,
  urlcolor=cyan,
}

\renewcommand\sout{\bgroup \color{red} \ULdepth=-.5ex \ULset}

\renewcommand{\rm}[1]{\textrm{#1}}
\renewcommand{\d}{\mathrm{d}}

\begin{document}

\title{Central Speed of Sound, Trace Anomaly and Observables of Neutron Stars from Perturbative Analyses of Scaled TOV Equations}

\author{Bao-Jun Cai\footnote{bjcai87@gmail.com}}
\affiliation{Quantum Machine Learning Laboratory, Shadow Creator Inc., Shanghai 201208, China} 
\author{Bao-An Li\footnote{Bao-An.Li$@$tamuc.edu}}
\affiliation{Department of Physics and Astronomy, Texas A$\&$M
University-Commerce, Commerce, TX 75429-3011, USA}
\author{Zhen Zhang\footnote{zhangzh275@mail.sysu.edu.cn}}
\affiliation{Sino-French Institute of Nuclear Engineering and Technology, Sun Yat-Sen
University, Zhuhai 519082, China}

\date{\today}

\begin{abstract}
The central speed of sound (SS) measures the stiffness of the Equation of State (EOS) of superdense neutron star (NS) matter. Its variations with density and radial coordinate in NSs in conventional analyses often suffer from uncertainties of the specific nuclear EOSs used. Using the central SS and NS mass/radius scaling obtained from solving perturbatively the scaled Tolman-Oppenheimer-Volkoff (TOV) equations, we study the variations of SS, trace anomaly and several closely related properties of NSs in an EOS-model independent manner. 
We find that the SS increases with the reduced central pressure $\widehat{P}_{\rm{c}}\equiv P_{\rm{c}}/\varepsilon_{\rm{c}}$ (scaled by the central energy density $\varepsilon_{\rm{c}}$), and the conformal bound for SS tends to break down for NSs with masses higher than about 1.9$M_{\odot}$. The ratio $P/\varepsilon$ is upper bounded as $P/\varepsilon\lesssim0.374$ around the centers of stable NSs. We demonstrate that it is an intrinsic property of strong-field gravity and is more relevant than the perturbative QCD bound on it. While a sharp phase transition at high densities characterized by a sudden vanishing of SS in cores of massive NSs are basically excluded, the probability for a continuous crossover signaled by a peaked radial profile of SS is found to be enhanced as $\widehat{P}_{\rm{c}}$ decreases, implying it likely happens near the centers of massive NSs. Moreover, a new and more stringent causality boundary as $R_{\max}/\rm{km}\gtrsim 4.73M_{\rm{NS}}^{\max}/M_{\odot}+1.14$ for NS M-R curve is found to be excellently consistent with observational data on NS masses and radii. Furthermore, new constraints on the ultimate energy density and pressure allowed in NSs before collapsing into black holes are obtained and compared with earlier predictions in the literature.
\end{abstract}

\maketitle


\setcounter{equation}{0}
\section{Introduction and Conclusions}
The speed of sound squared (SSS) is defined as $s^2=\d P/\d\varepsilon$\,\cite{Lan87} where $P$ and $\varepsilon$ are respectively the pressure and energy density of the matter under consideration.
The conditions of stability and causality together require that $0\leq s^2\leq 1$ (adopting $c=1$).
Due to the superdense nature of neutron star (NS) matter\,\cite{Walecka1974,Chin1976,Baym1976,Freedman1977,Baluni1978,
Migdal1978,Morley1979,Bailin1984,Shuryak1980,Wiringa1988,Akmal1998,
LP01,LP04,LP07,Steiner2007,Alford2008,LCK08,Wat16,Oertel2017,Isa18,Dri21,Bur21,Lovato22,Soren2023} especially near its core\,\cite{Pand76,Glen85,Glen92, Heis1993,Glen98,Alford1998,Sch98,Bie02,Kur14,Fra14,Ann20N},  the speed of sound (SS) can become very close to 1\,\cite{Zel61} or even exceed it in certain models of dense matter. 
As a measure of the stiffness of nuclear EOS, the behavior of SS is closely related to possible phase transitions (PTs) and/or a continuous crossover in NSs.
In particular, a sharp vanishing of SSS (i.e., $s^2=0$ or a constant $P$ in a finite range of $\varepsilon$) signals the occurrence of a first-order PT while a smooth reduction of $s^2$ indicates a continuous crossover. Several possible mechanisms for the PTs and/or crossover in NSs have been proposed in the literature, see, e.g.,  Refs.\,\cite{Baym18,Bai19,Ors2019,BALI19,Dexh21,Lat21,Kojo21,Armen} for recent reviews. While possible imprint of the SS on NS observables has been studied extensively,  the SS in NSs is not a quantity directly measurable and there are still many interesting issues about it to be addressed. 

Much progress has been made in understanding the internal structures and observational properties of NSs during the last decade thanks especially to the new opportunities provided by the era of multimessenger astronomy since the discovery of GW170817\,\cite{Abbott2017,Abbott2018,Bau17,Radice2018,Most2018,Fat2018,Bose2018,Ann18,ZhouY2019,Bau20,Diet20,ZLi,Tan20,Landry2020,Bombaci2021,Raa20,Raa21,Huth2022,Ann22,Cao22,Essick2020,Breschi2022,Pang21}. For instance, the NICER (Neutron Star Interior Composition Explorer) and XMM-Newton simultaneously measured the mass-radius for PSR J0740+6620\,\cite{Fon21,Riley21,Miller21,Salm22} and that for PSR J0030+0451\,\cite{Riley19,Miller19}. However, the relevant phases of matter in NS cores and therefore the EOSs there still remain ambiguous\,\cite{Rait16,Rait17,Rait18,Rait21,De2018,Ruiz2018,Nat21,Lim2018,Lim2019,Dris20,Han21,Huang2022,Biswas2022,
Weih2020,Xie20,Most2019,Bau19,Montana2019,Malfatti2019,Greif19,Baym2019,Zhao2020,Fuku2020,Kap21,Dri21-a,Dri22,Kojo22,Fuji22-a,Leg21,Mam2021,Perego2022,Han23,Som23,Brandes23,Fuku20,Fuji23,ZLi22,ZLi23,Rai23,Chiba23}. This is partially because of the lack of direct accesses to the core EOS without relying on the still uncertain structures and EOSs in other parts of NSs especially the NS crust. Thus, many interesting and critical questions regarding properties of superdense NS matter remain to be addressed. To outline our motivations and set the context of this work, we list below a few examples:
\begin{enumerate}[label=(\alph*),leftmargin=*]
\item What is the EOS of densest visible matter existing in the Universe? Can it be accessed/constrained directly using certain astrophysical data such as observed NS radii and/or masses without using any EOS model?

\item There exists an ultimate limit for the pressure/energy density in NSs above which the system becomes unstable. What is this ultimate limit? Can the observational data on NSs tell this quantity?

\item Is the limit on the ratio $P/\varepsilon\leq1$ (from the principle of causality) relevant/sufficient in NS cores? If it is insufficient, how and to which extend can this limit be improved?
In fact, $P\leq\varepsilon$ is equivalent to $s^2\leq1$ only for a linear EOS $P\propto\varepsilon$, while the EOS in NSs (especially in their cores) can be significantly nonlinear. As a result, the upper bound for $P/\varepsilon$ in NSs is expected to be smaller than 1.
What are the implications of this refined upper bound for $P/\varepsilon$ on certain NS properties which are determined by the balancing acts of gravitational attraction and nuclear repulsion? 
Is this refined limit a fundamental bound from General Relativity (GR) theory of gravity or intrinsically related to microscopic theories of dense matter EOS such as perturbative QCD (pQCD)?

\item How can one understand the possible violation of the conformal bound (CB, i.e., SSS is less than 1/3 based on the pQCD prediction) for SSS in massive NSs?
Can one calculate/estimate certain characteristic  quantities related to the conformality via the general-relativistic NS structural (TOV) equations alone without using any EOS model for NS matter?

\item Is the available causality boundary in the literature for NSs on the mass-radius (M-R) curve a tight one?
Can one refine or improve it? (Such boundary should be consistent with NS observational data besides microscopic physics constraints).
Similarly,  how large can the compactness mass/radius of a NS be?

\item What is the radial profile of SS in NSs? Can one locate its peak if it exists?

\end{enumerate}

\begin{figure}[h!]
\centering
\includegraphics[width=6.cm]{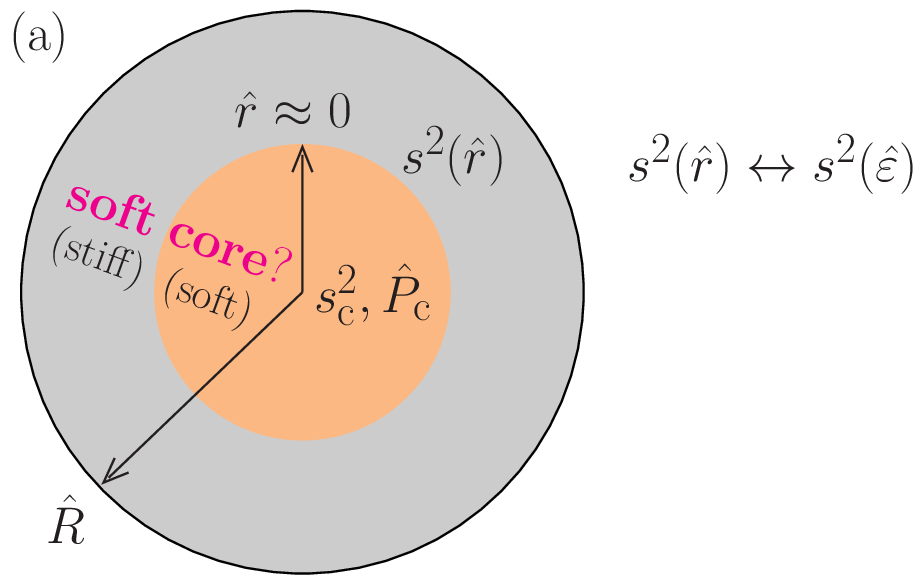}\\[0.2cm]
\includegraphics[width=3.5cm]{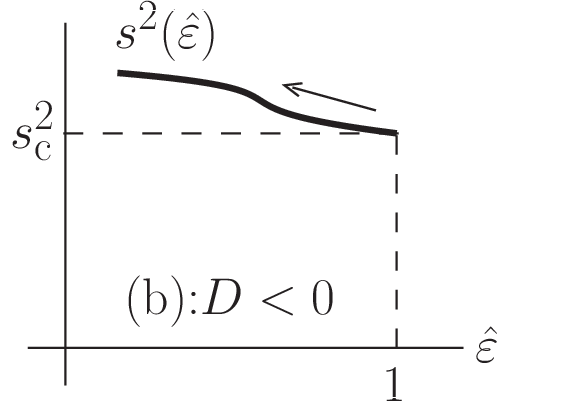}\quad
\includegraphics[width=3.5cm]{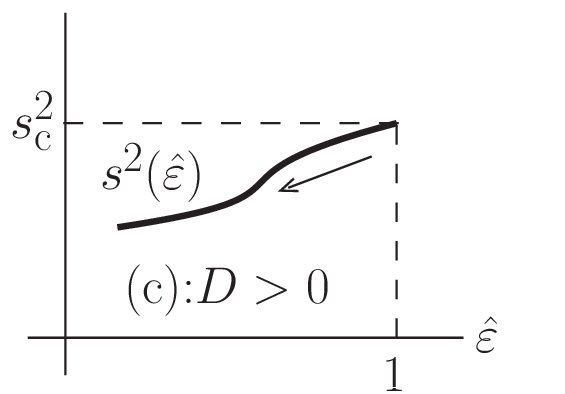}
\caption{Sketch of the SSS near NS centers (panel (a)). 
The sign of coefficient $D$ (see Eq.\,(\ref{ref-D})) characterizes the possibility of $s_{\rm{c}}^2<s^2(\widehat{\varepsilon})$ (indication of a crossover/soft core, panel (b)) or  $s^2_{\rm{c}}>s^2(\widehat{\varepsilon})$ (the center EOS is stiffer than its surroundings, panel (c)). }\label{fig_sr2_sk}
\end{figure}

Recently, we studied the question (a) above using a novel approach\,\cite{CLZ23} which facilitates extracting the central EOS of NS matter in the maximum-mass configuration (where $\d M_{\rm{NS}}/\d\varepsilon_{\rm{c}}=0$ with $\varepsilon_{\rm{c}}$ being the central energy density on the mass-radius curve) directly from the observational data without relying on any specific EOS model. The method was developed from analyzing perturbatively the dimensionless Tolman-Oppenheimer-Volkoff (TOV) equations for NS internal variables scaled by $\varepsilon_{\rm{c}}$.
In particular, a formula for SSS in the center of these NSs was obtained as,
\begin{equation}\label{sc2}
s_{\rm{c}}^2=\widehat{P}_{\rm{c}}\left(1+\frac{1}{3}\frac{1+3\widehat{P}_{\rm{c}}^2+4\widehat{P}_{\rm{c}}}{1-3\widehat{P}_{\rm{c}}^2}\right),
\end{equation}
here $\widehat{P}_{\rm{c}}\equiv P_{\rm{c}}/\varepsilon_{\rm{c}}$ is the ratio of the central pressure $P_{\rm{c}}$ over the central energy density $\varepsilon_{\rm{c}}$.
From Eq.\,(\ref{sc2}), one immediately finds that $s_{\rm{c}}^2>0$, therefore excluding a sharp PT (at high densities) occurring in cores of maximum-mass configuration NSs.
Whether it allows a continuous crossover in the cores depends on the behavior of the SSS away from the center as sketched in panel (a) of FIG.\,\ref{fig_sr2_sk}.
A continuous crossover indicates $s^2(\widehat{\varepsilon})>s_{\rm{c}}^2$ (panel (b) where the SSS at the center is smaller than that at a finite distance away) and a possible peak in $s^2$ somewhere away from the center as the SSS should become zero at the core-crust transition density, here $\widehat{r}$ is the dimensionless distance from the center (Section \ref{SEC2}), $\widehat{\varepsilon}=\varepsilon/\varepsilon_{\rm{c}}$ and $D$ is a coefficient characterizing the variation of $s^2$ with $\widehat{r}$ (Section \ref{SEC3}). It is seen from Eq.\,(\ref{sc2}) that the chance for $s^2(\widehat{\varepsilon})>s_{\rm{c}}^2$ is reduced as $\widehat{P}_{\rm{c}}$ increases since $s_{\rm{c}}^2$ is upper bounded. Therefore, for NSs with their $\widehat{P}_{\rm{c}}$'s close to its upper bound of about 0.374 from causality \cite{CLZ23}, the crossover near the center is unfavored with high probabilities. In the limiting case of $s_{\rm{c}}^2\to1$, the probability naturally approaches zero and it is probably in this situation $s_{\rm{c}}^2>s^2(\widehat{\varepsilon})$ (panel (c) of FIG.\,\ref{fig_sr2_sk}).
Moreover,  we can deduce from Eq.\,(\ref{sc2}) that $s_{\rm{c}}^2\to1/3$ occurs earlier than $\widehat{P}_{\rm{c}}\to1/3$. This also explains why the CB for SSS is highly likely to break down in NS cores.

In this paper, we report results of our investigations into the remaining issues listed above ((b)$\sim$(f)) in essentially the same framework as in our earlier work  \cite{CLZ23}. When possible, we also compare our results with existing ones by others in the literature. Our main contributions/findings include:
\begin{enumerate}[label=(\arabic*),leftmargin=*]
\item The NS radius $R$ for a given dense matter EOS is shown to scale with $\nu_{\rm{c}}\equiv\varepsilon_{\rm{c}}^{-1/2}[\widehat{P}_{\rm{c}}/(1+3\widehat{P}_{\rm{c}}^2+4\widehat{P}_{\rm{c}})]^{1/2}$ along its M-R curve, extending the results of Ref.\,\cite{CLZ23}.
This scaling could be used to reduce uncertainties of the core EOS for a certain NS considering its mass/radius measurements. See FIG.\,\ref{fig_fk1-fk2} for illustrations of the R scaling.

\item The SSS is shown to increase with $M_{\rm{NS}}^{\max}$ under the assumption that NSs with masses about $M_{\rm{NS}}^{\max}/M_{\odot}=1.3\mbox{$\sim$}2.3$ ($M_{\odot}=$ solar mass) have similar radii about 12\,km.
As a result, the CB for SSS\,\cite{Hoh09,Cher09} is found to break down if $M_{\rm{NS}}^{\max}\gtrsim1.9M_{\odot}$. See FIG.\,\ref{fig_sckk}.

\item Using the upper bound for the reduced central pressure $\widehat{P}_{\rm{c}}\lesssim0.374$ to satisfy the causality requirement and the universal correlations given in Ref.\,\cite{CLZ23}, a new causality boundary for the NS M-R curve as $R_{\max}/\rm{km}\gtrsim 4.73M_{\rm{NS}}^{\max}/M_{\odot}+1.14$ is obtained. It is shown to be consistent with observational data of several NSs albeit in certain tension with predictions using a few empirical EOSs for NS matter in the literature. As a direct corollary,  the radius of the PSR J0952-0607\,\cite{Romani22} with a mass about $2.35M_{\odot}$ is found to be $\gtrsim$12.25\,km. See FIG.\,\ref{fig_MR-C} and FIG.\,\ref{fig_Compt}.

\item  The ultimate energy density and pressure allowed in NSs are estimated,  e.g., the existence of a $2.08M_{\odot}$ NS leads to $\varepsilon_{\rm{ult}}\lesssim1.32\,\rm{GeV}/\rm{fm}^3$ and $P_{\rm{ult}}\lesssim494\,\rm{MeV}/\rm{fm}^3$, which are found to be consistent with several recent microscopic model analyses/calculations. See FIG.\,\ref{fig_ULTeps}.

\item By adopting the empirical criteria $\Theta=[(1/3-P/\varepsilon)^2+(s^2-P/\varepsilon)^2]^{1/2}\lesssim0.2$ and $\gamma=\d\ln P/\d\ln\varepsilon\lesssim1.75$ for identifying conformality, we find it is unlikely to be realized in cores of massive NSs.
See TAB.\,\ref{sstab}, FIG.\,\ref{fig_Th-g} and Eqs.\,(\ref{gr-1})$\sim$(\ref{TGG}).

\item The $\widehat{P}_{\rm{c}}$ is shown to be a key quantity determining the onset of crossover at high densities (near the center). Specifically, the probability of $s_{\rm{c}}^2<s^2(\widehat{\varepsilon})$ for $\widehat{\varepsilon}\lesssim1$ (near the center) for the maximum-mass configurations is found to be larger than about 50\% for $\widehat{P}_{\rm{c}}\lesssim0.3$.
See FIG.\,\ref{fig_NEG-K3}, FIG.\,\ref{fig_NEG}, FIG.\,\ref{fig_AD} and FIG.\,\ref{fig_s2peak}.

\item Continued with the last point, the existence of a peak in the derivative part of $s^2$ as well as the non-existence of such peak in the non-derivative part of $s^2$ are generally demonstrated, as introduced and discussed in Ref.\,\cite{Fuji22}.
See Eqs.\,(\ref{tpk}) and (\ref{tpk1}).

\item By extending the upper bound $\widehat{P}_{\rm{c}}\lesssim0.374$ holding at NS centers for the maximum-mass configurations,  we demonstrated that the reduced pressure $P/\varepsilon=\widehat{P}/\widehat{\varepsilon}$ is also bounded from above to about 0.374 at finite distances near centers of stable NSs.
Such bound for $P/\varepsilon$ is a unique consequence of strong-field gravity since it is much smaller in conventional (low-density) nuclear-physics problems.
As a direct corollary, the trace anomaly (TA) $\Delta\equiv 1/3-P/\varepsilon$ is bounded from below to about $\Delta\gtrsim-0.041$ near the centers of massive NSs.
See FIG.\,\ref{fig_TAc}, Eqs.\,(\ref{pk-3}) and (\ref{pk-3a}).

\end{enumerate}
Here, our findings (1)$\sim$(5) concern properties at NS centers, while the points (6)$\sim$(8) are extended to positions away from the centers (see FIG.\,\ref{fig_Pec} for a notation guide).
A few empirical relations in the literature are also explained using our method (e.g., FIG.\,\ref{fig_Rrhoc} and  FIG.\,\ref{fig_RSca}).
Since our results are obtained without using any specific input EOS for NS matter, they provide generally EOS-model independent insights on several interesting properties of NSs especially near their cores where direct observational data is lacking.

The rest of this paper is organized as follows: in Section \ref{SEC2} we review briefly the methods of Ref.\,\cite{CLZ23} on directly extracting the NS central EOS from observational data and then in Section \ref{SEC_xx} we estimate the central SS as a function of $M_{\rm{NS}}^{\max}$ using Eq.\,(\ref{sc2}); 
we study the causality boundary for the NS M-R curve in Section \ref{SEC4} and the closely related 
compactness parameter in Section \ref{SEC_Compt}; and Section \ref{SEC5} gives the estimate on the ultimate (maximum) energy density as well as the pressure allowed in NS cores. Section \ref{SEC3} is devoted to the analysis on the possible occurrence of continuous crossover in NS cores and in Section \ref{SEC_BRP} and Section \ref{SEC_CSD} we generalize the limit $\widehat{P}_{\rm{c}}\lesssim0.374$ which holds at NS centers for the maximum-mass configurations to stable NSs along the M-R curve either at centers or finite distances away.
We then summarize this work in Section \ref{SEC6}. In the appendices, we give relevant mathematical details of expressions in the main text and estimates on certain quantities involved.

\section{A Brief Review: Neutron Star Mass and Radius Scalings from Perturbative Solutions of the Scaled TOV Equations}\label{SEC2}
For completeness and ease of our following discussions, we recall here the main points of solving perturbatively the scaled TOV equations and the resulting mass and radius scalings for NSs at the maximum-mass configuration. More details of our approach can be found in Ref.\,\cite{CLZ23}. We also discuss here intuitively some general features of NSs from the scaled TOV equations that are independent of the EOS-model and present the general radius scaling on the NS mass-radius curve. 

The TOV equations\,\cite{TOV39-1,TOV39-2,Misner1973} describe the evolution of the pressure $P$ and mass $M$ as functions of the distance $r$ from the NS centers (adopting $c=1$),
\begin{align}
 \frac{\d}{\d r}P=&
-\frac{G(\varepsilon+P)(M+4\pi r^3P)}{r^2(1-2GM/r)} ,~~
\frac{\d}{\d
r}M=4\pi r^2\varepsilon.\end{align}
They are conventionally integrated from the center to surface (given a EOS for NS matter) to give the NS mass $M_{\rm{NS}}\equiv M$ and radius $R$ defined as the vanishing point of pressure (i.e., $P(R)=0$).
We can rewrite the TOV equations by scaling the NS mass by $W\equiv G^{-1}(4\pi G\varepsilon_{\rm{c}})^{-1/2}$ and the radius by $Q\equiv (4\pi G \varepsilon_{\rm{c}})^{-1/2}$\,\cite{CLZ23},
\begin{equation}\label{TOV-ds}
\frac{\d}{\d\widehat{r}}\widehat{P}=-\frac{(\widehat{P}+\widehat{\varepsilon})(\widehat{r}^3\widehat{P}+\widehat{M})}{\widehat{r}^2-2\widehat{M}\widehat{r}},~~\frac{\d}{\d\widehat{r}}\widehat{M}=\widehat{r}^2\widehat{\varepsilon},
\end{equation}
here $\widehat{P}\equiv P/\varepsilon_{\rm{c}}$ and $\varepsilon_{\rm{c}}$ is the central energy density.
The above dimensionless TOV equations (\ref{TOV-ds}) can be solved by expanding the (reduced) energy density,  pressure and mass as $\widehat{\varepsilon}=1+\sum_{k=1}^Ka_k\widehat{r}^k$, $\widehat{P}=\widehat{P}_{\rm{c}}+\sum_{k=1}^Kb_k\widehat{r}^k$ and $\widehat{M}=\sum_{k=1}^Kc_k\widehat{r}^k$ with $\widehat{r}=r/Q$, here $K$ is the effective truncation order of the polynomial expansions. As shown in detail in Ref.\,\cite{CLZ23}, $c_1=c_2=0$ and $c_k=a_{k-3}/k$ for $k\geq3$, 
and the leading nonzero coefficient is $b_2=-6^{-1}(1+3\widehat{P}_{\rm{c}}^2+4\widehat{P}_{\rm{c}})$.
Truncating the pressure to order $b_2$, namely $\widehat{P}_{\rm{c}}+b_2\widehat{R}^2=0$, gives the reduced radius $\widehat{R}\sim[\widehat{P}_{\rm{c}}/(1+3\widehat{P}_{\rm{c}}^2+4\widehat{P}_{\rm{c}})]^{1/2}$ and the physical radius $R\sim\widehat{R}Q$ is found to scale according to\,\cite{CLZ23}
\begin{equation}\label{gk-1}
R\sim\nu_{\rm{c}}\equiv\frac{\widehat{P}_{\rm{c}}^{1/2}}{\sqrt{\varepsilon_{\rm{c}}}}\left(\frac{1}{1+3\widehat{P}_{\rm{c}}^2+4\widehat{P}_{\rm{c}}}\right)^{1/2}.
\end{equation}
The NS mass $M_{\rm{NS}}\sim \widehat{R}^3/\sqrt{\varepsilon_{\rm{c}}}$ is found to scale as\,\cite{CLZ23}
\begin{equation}\label{gk-2}
M_{\rm{NS}}\sim\Gamma_{\rm{c}}\equiv\frac{\widehat{P}_{\rm{c}}^{3/2}}{\sqrt{\varepsilon_{\rm{c}}}}\left(\frac{1}{1+3\widehat{P}_{\rm{c}}^2+4\widehat{P}_{\rm{c}}}\right)^{3/2}.
\end{equation}
In Ref.\,\cite{CLZ23},  the correlations $M_{\rm{NS}}^{\max}$-$\Gamma_{\rm{c}}$ and $R_{\max}$-$\nu_{\rm{c}}$ are then verified using 87 widely used phenomenological and 17 microscopic EOSs, 
here $R_{\max}$ is the corresponding radius at $M_{\rm{NS}}^{\max}$.
Specifically\,\cite{CLZ23},
\begin{align}
R_{\max}/\rm{km}&\approx1.05\times 10^3\nu_{\rm{c}}+0.64,\label{Rmax-n}\\
M_{\rm{NS}}^{\max}/M_{\odot}&\approx0.173\times 10^4\Gamma_{\rm{c}}-0.106.\label{Mmax-G}
\end{align}
The SSS of Eq.\,(\ref{sc2}) is obtained via $\d M_{\rm{NS}}/\d\varepsilon_{\rm{c}}=0$ at the maximum-mass configuration on the M-R curve.

\begin{figure}[h!]
\centering
\includegraphics[width=6.4cm]{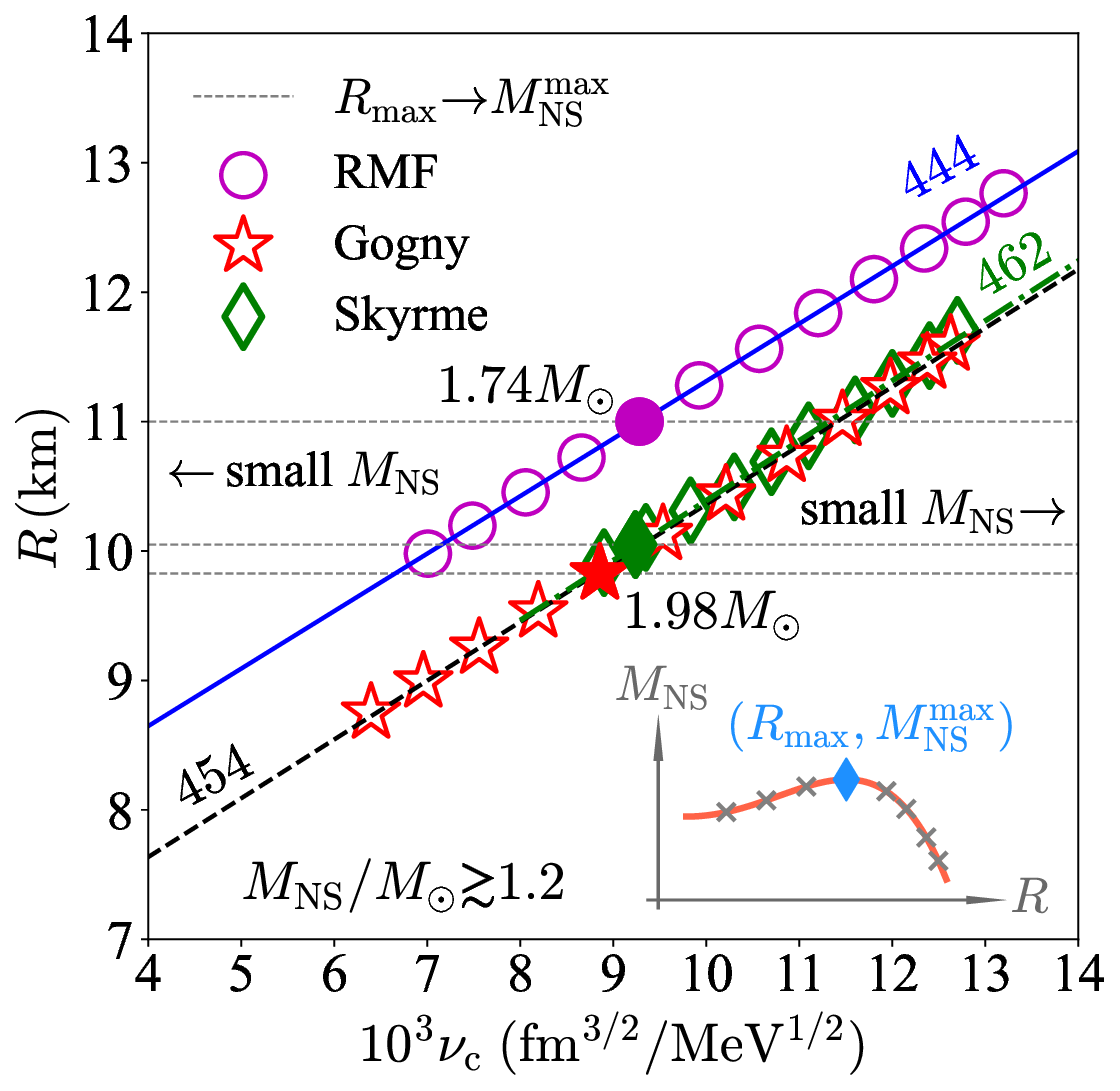}
\caption{Correlation between the NS radii $R$ and the factor $\nu_{\rm{c}}$ obtained from the non-relativistic EDFs (of the Gogny-like and Skyrme types) and the nonlinear RMF model EOSs.
The solid symbol is for the maximum-mass configuration of each model (corresponding to the solid symbol shown in the inset).
}\label{fig_fk1-fk2}
\end{figure}

The scalings of (\ref{gk-1}) and (\ref{gk-2}) could be understood intuitively by combining the quantum-degeneracy of nuclear pressure and the self-gravitating nature of NSs.
Since NSs are self-gravitating systems\,\cite{Lan32,Rho74,Hartle77,Hartle78,Lind84, Lind92,KB96,LP05,LP11,Shapiro1983}, one expects that a larger energy density $\varepsilon$ corresponds to a smaller radius $R$.
By temporally neglecting the general relativistic effects\,\cite{Cha10}, one has $\d P/\d r=-GM\varepsilon/r^2$ and thus $P\sim-GM\varepsilon/r\sim Gr^2\varepsilon^2$ where $\d M/\d r=4\pi r^2\varepsilon$ or $M\sim r^3\varepsilon$ is used.
Thus $r\sim1/\sqrt{G\varepsilon}$ because $P$ and $\varepsilon$ have the same dimension.
On the other hand, NSs are supported mainly by the neutron degenerate pressure, a larger pressure $P$ naturally leads to a larger radius $R$, i.e., $R\sim P^{\phi}$ where $\phi>0$.
In order to infer the value of $\phi$, one notices that $P\sim\int M\varepsilon/r^2\d r$ as a function of $r$ is even, since the $\varepsilon$ is even of $r$ and thus the mass $M$ is odd of $r$ because $M\sim\int r^2\varepsilon\d r$. Consequently, $P\sim\rm{const}.+Br^2+\cdots$, from which one obtains $\phi=1/2$ and $R\sim P^{1/2}+\cdots$.
The absence of linear term ($\propto r$) in the expansion of $P$ over $r$ could also be understood by the boundary condition $\d P/\d r=0$ at $r=0$ (pressure cannot have a cusp-like singularity).
By combining the self-gravitating and quantum-degenerate nature of NSs, we have $R\sim (P/\varepsilon)^{1/2}/\sqrt{G\varepsilon}+\cdots$, since $P/\varepsilon$ is the relevant dimensionless quantity from combining $P$ and $\varepsilon$. The NS M-R relation is obtained by integrating the TOV equations from a given central energy density $\varepsilon_{\rm{c}}$,  thus $R\sim \widehat{P}_{\rm{c}}^{1/2}/\sqrt{G\varepsilon_{\rm{c}}}+\cdots\sim\widehat{P}_{\rm{c}}^{1/2}/\sqrt{\varepsilon_{\rm{c}}}+\cdots$ where $\widehat{P}_{\rm{c}}=P_{\rm{c}}/\varepsilon_{\rm{c}}$.
Considering general relativistic effects, the scaling for $R$ will be modified to $R\sim\widehat{P}_{\rm{c}}^{1/2}/\sqrt{\varepsilon_{\rm{c}}}\cdot \vartheta(\widehat{P}_{\rm{c}})$ with the correction $\vartheta(\widehat{P}_{\rm{c}})$ to be revealed by analyzing structures of the TOV equations\,\cite{CLZ23}.
In particular,  Eq.\,(\ref{gk-1}) tells that $\vartheta(\widehat{P}_{\rm{c}})=(1+3\widehat{P}_{\rm{c}}^2+4\widehat{P}_{\rm{c}})^{-1/2}<1$, i.e., the strong gravity in GR than Newtonian's reduces the NS radius.
Similar arguments give for the mass as $M_{\rm{NS}}\sim R^3\varepsilon_{\rm{c}}\sim\widehat{P}^{3/2}/G^{3/2}\sqrt{\varepsilon_{\rm{c}}}+\cdots\sim\widehat{P}_{\rm{c}}^{3/2}/\sqrt{\varepsilon_{\rm{c}}}+\cdots$.

At this point, it is necessary to point out the potential degrees of uncertainty inherent in our results, due to the pertubative nature of our approach.
While some of them are general and robust, e.g., the correlations between $M_{\rm{NS}}^{\max}$ and $\Gamma_{\rm{c}}$ and that between $R_{\max}$ and $\nu_{\rm{c}}$\,\cite{CLZ23} and the consequent prediction on the causality boundary of FIG.\,\ref{fig_MR-C}, the compactness constraint of FIG.\,\ref{fig_Compt} and the ultimate energy density and pressure allowed in NSs of FIG.\,\ref{fig_ULTeps}, others may contain uncertainties.
For example, when estimating the possible peak of $s^2$ as a function of $\widehat{r}$ (see Eq.\,(\ref{ef-2})), we expand the $s^2(\widehat{r})$ to order $\widehat{r}^4$ which neglects the higher-order contributions from $\widehat{r}^6$ and $\widehat{r}^8$, etc.,  and these higher-order terms may induce quantitative effects on the estimate for the peak. However, when investigating whether there exists a peak in $s^2(\widehat{r})$, the estimate on the sign of the coefficient of $\widehat{r}^2$ in the expansion of $s^2(\widehat{r})$ is enough, and in this sense the result avoids suffering from contributions from the higher-order terms. Similarly, when bounding the ratio $P/\varepsilon$ in Section \ref{SEC_BRP}, the conclusion is effective only near the NS cores and whether it is still effective at places far from the center needs further analysis by including higher-order terms in $\widehat{r}$.
When necessary, we may point out the relevant uncertainties and/or the limitations of our results, see, e.g.,  the last second paragraph of Section \ref{SEC3}.

Besides the correlation between $M_{\rm{NS}}^{\max}$ and $\Gamma_{\rm{c}}$ discussed in Ref.\,\cite{CLZ23},  the NS radius for a given dense matter EOS is also found to be strongly correlated with $\nu_{\rm{c}}$ along the M-R curve.
We show
in FIG.\,\ref{fig_fk1-fk2} the dependence of radius $R$ on $\nu_{\rm{c}}$ using three kinds of EOSs: the covariant relativistic mean-field (RMF) models\,\cite{Serot1986} and the non-relativistic energy density functionals (EDFs) of the Gogny-like type\,\cite{CaiLi2022KK} as well as the conventional Skyrme-type\,\cite{Vau1972,Sky,ZC16}. 
Very good correlations between $R$ and $\nu_{\rm{c}}$ are found in all cases as indicated by the very similar slopes (444, 454 and 462 in unit of $[R]/[10^3\nu_{\rm{c}}]$ shown by the captions near the lines), while the intercept has certain model-dependence (considering the two classes).
A cutoff of $M_{\rm{NS}}\gtrsim1.2M_{\odot}$ is necessary to mitigate potential influences of uncertainties in modeling the crust EOS\,\cite{BPS71,Iida1997,XuJ} for low-mass NSs.
The radii $R_{\max}$ are also shown using solid symbols ($M_{\rm{NS}}^{\max}/M_{\odot}\approx2.05$ in the Skyrme model using green diamond). Points on the left side of $R_{\max}$ correspond to unstable NS configurations (i.e., for such NSs one has $\d M_{\rm{NS}}/\d\varepsilon_{\rm{c}}<0$).
The $R$-$\nu_{\rm{c}}$ correlation could be used to reduce uncertainties of the central EOS for a NS with certain mass via its radius constraints/measurements within a specific EOS model.

\begin{figure}[h!]
\centering
\includegraphics[width=6.4cm]{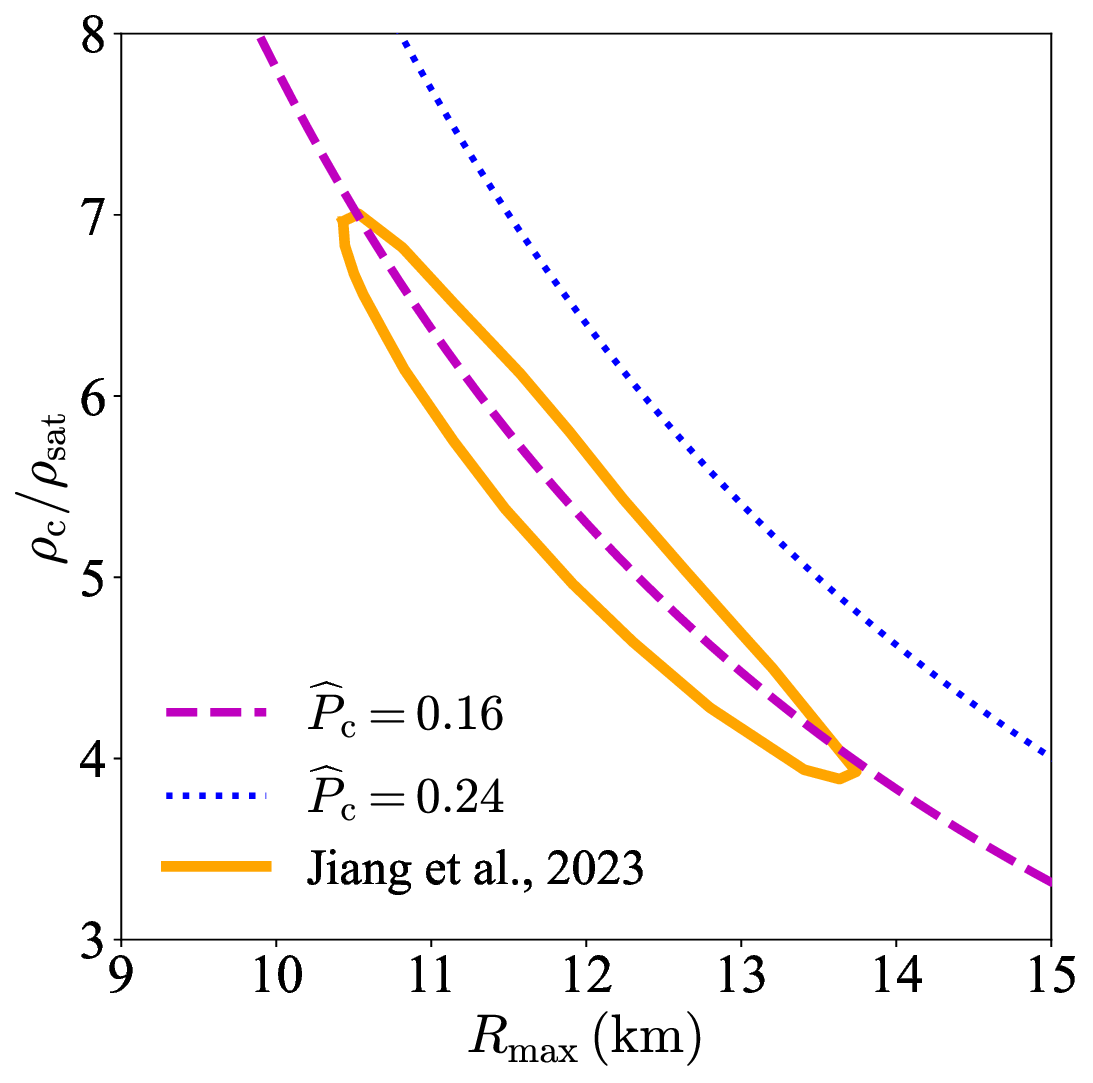}
\caption{Correlation between $\rho_{\rm{c}}/\rho_{\rm{sat}}$ and radius $R_{\max}$ adopting two reference values for $\widehat{P}_{\rm{c}}$ (0.16 and 0.24), here $\rho_{\rm{sat}}\approx0.16\,\rm{fm}^{-3}$ is used.
An empirical prediction on $\rho_{\rm{c}}/\rho_{\rm{sat}}$-$R_{\max}$ from Ref.\,\cite{Jiang23} (orange solid contour) is shown for a comparison.
}\label{fig_Rrhoc}
\end{figure}

By using the scaling in Eq.\,(\ref{Rmax-n}) we can derive a relation between NS central baryon density $\rho_{\rm{c}}$ and the radius $R_{\max}$. The result is,
\begin{equation}\label{Rrhoc}
\frac{\rho_{\rm{c}}}{\rho_{\rm{sat}}}
\approx\frac{7.35\times10^3\widehat{P}_{\rm{c}}}{1+3\widehat{P}_{\rm{c}}^2+4\widehat{P}_{\rm{c}}}\left(\frac{R_{\max}}{\rm{km}}-0.64\right)^{-2},
\end{equation}
where $\rho_{\rm{sat}}\approx0.16\,\rm{fm}^{-3}$\,\cite{Coester1970} is the nuclear saturation density and the approximation $\rho_{\rm{c}}\approx\varepsilon_{\rm{c}}/M_{\rm{N}}$ is adopted with $M_{\rm{N}}\approx939\,\rm{MeV}$ the static nucleon mass.
Shown in FIG.\,\ref{fig_Rrhoc} is the relation between  $\rho_{\rm{c}}/\rho_{\rm{sat}}$ and the radius $R_{\max}$ using two values of $\widehat{P}_{\rm{c}}$ (0.16 and 0.24). For a comparison,  an algorithmic/empirical prediction from Ref.\,\cite{Jiang23} is also shown (orange solid contour). It is seen that the overall correlation between $\rho_{\rm{c}}/\rho_{\rm{sat}}$ and $R_{\max}$ from our analyses here is consistent with that from Ref.\,\cite{Jiang23}. Moreover, one can see from Eq.\,(\ref{Rrhoc}) that roughly $\rho_{\rm{c}}/\rho_{\rm{sat}}\sim R_{\max}^{-2}\cdot[1+\mbox{corrections of }R_{\max}^{-1}]$
once a $\widehat{P}_{\rm{c}}$ is specified.
The correlation shown in FIG.\,\ref{fig_Rrhoc} may help us to estimate the maximum central (baryon) density when future radius measurements of very massive NSs are available\,\cite{Tang23}.
Moreover, the above correlation is obtained under the approximation $\varepsilon_{\rm{c}}\approx M_{\rm{N}}\rho_{\rm{c}}$ and principally one needs to use $E(\rho_{\rm{c}})+M_{\rm{N}}$ to replace $M_{\rm{N}}$ where $E(\rho_{\rm{c}})$ defined as the energy per nucleon at $\rho_{\rm{c}}$ has its own uncertainties.
Doing so will shift down the lines in FIG.\,\ref{fig_Rrhoc} as $E(\rho_{\rm{c}})>0$. 

\begin{figure}[h!]
\centering
\includegraphics[width=6.4cm]{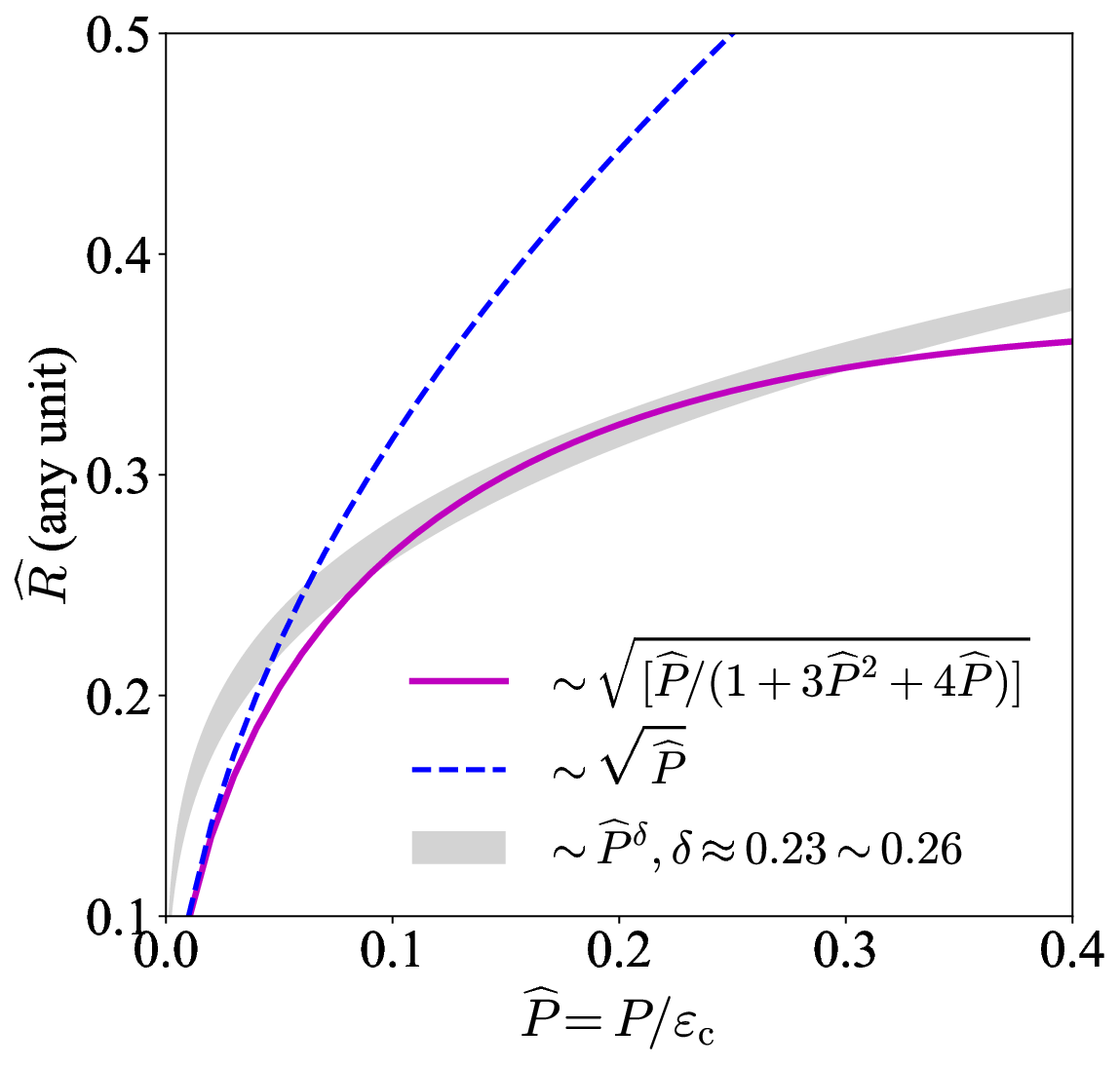}
\caption{A comparison of the radius scalings ($R$ vs reduced pressure $\widehat{P}$): the empirical power law $R\sim P^{\delta}$ with $\delta\approx0.23\mbox{$\sim$}0.26$ from Ref.\,\cite{LP01}, 
the Newtonian scaling $\widehat{R}\sim \widehat{P}^{1/2}$ and our full scaling $\widehat{R}\sim[\widehat{P}/(1+3\widehat{P}^2+4\widehat{P})]^{1/2}$.}\label{fig_RSca}
\end{figure}

It is appropriate at this point to compare our radius-pressure scaling with existing ones in the literature in FIG.\,\ref{fig_RSca}. The grey band shows the empirical power law that the NS radii depend on the pressure at 1-2 times $\rho_{\rm{sat}}$, i.e., $R\sim P^{\delta}$ with $\delta\approx0.23\mbox{$\sim$}0.26$\,\cite{LP01}.
One finds that the Newtonian prediction $\widehat{R}\sim \widehat{P}^{1/2}$ (by neglecting the general-relativistic correction $3\widehat{P}^2+4\widehat{P}$) indicated by the blue dashed line deviates significantly from the empirical power law $\widehat{P}^\delta$. On the other hand, our full scaling $\widehat{R}\sim[\widehat{P}/(1+3\widehat{P}^2+4\widehat{P})]^{1/2}$ (magenta line) is rather consistent with the empirical scaling, especially for $0.1\lesssim\widehat{P}\lesssim0.3$ (which is a reasonable region for $P/\varepsilon$ in NS cores).

\section{Speed of Sound and Violation of the Conformal Bound in Massive Neutron Stars}\label{SEC_xx}

\begin{figure}[h!]
\centering
\includegraphics[width=6.4cm]{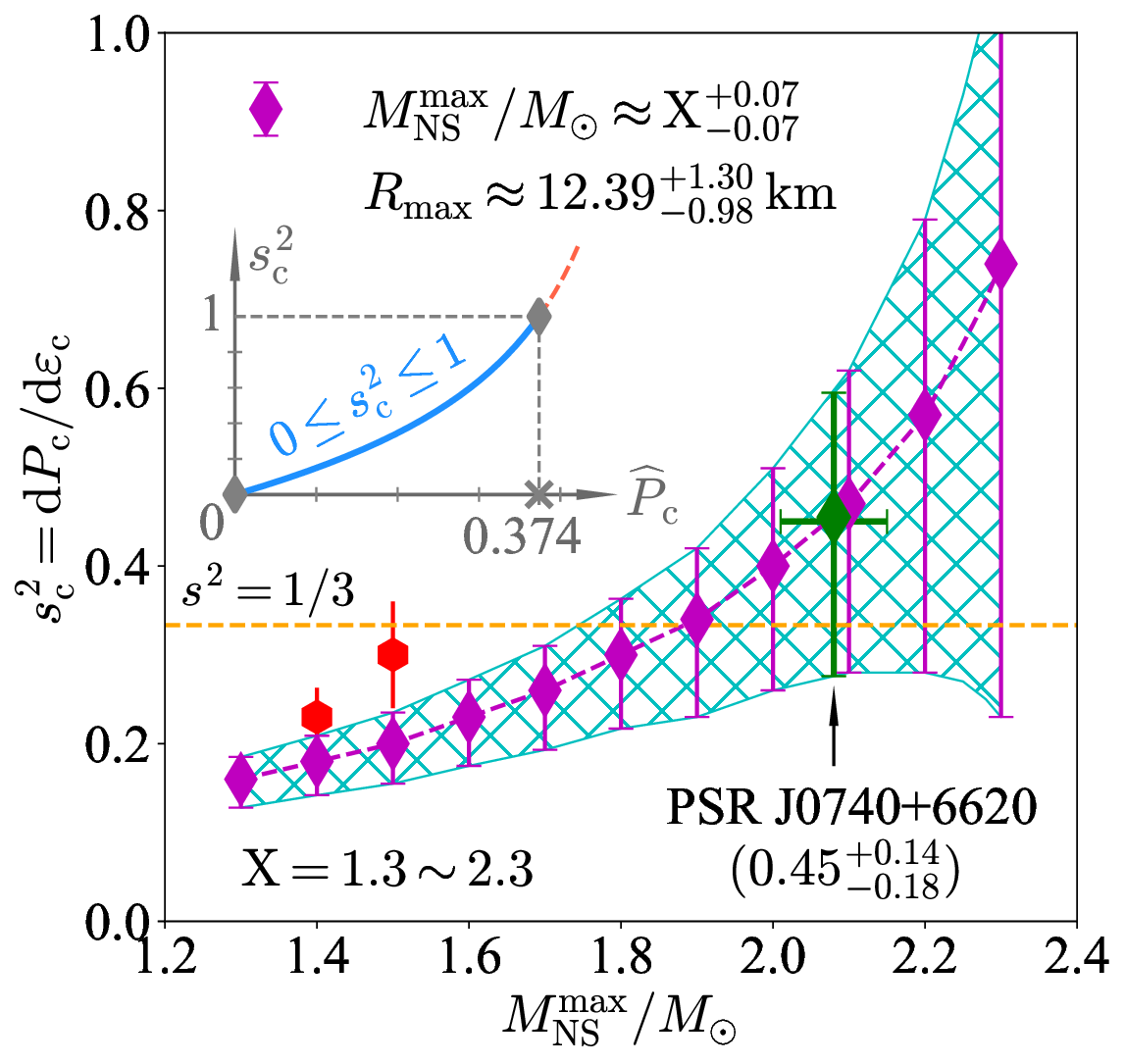}
\caption{Dependence of $s_{\rm{c}}^2$ on the maximum mass $M_{\rm{NS}}^{\max}/M_{\odot}$ (of a given EOS for NS matter), the CB on SSS is indicated by the dashed orange line.
The inset plots the $s_{\rm{c}}^2$ as a function of $\widehat{P}_{\rm{c}}$, where the lightblue curve is allowed by conditions of stability and causality ($0\leq s_{\rm{c}}^2\leq1$).
}
\label{fig_sckk}
\end{figure}

The dependence of $s_{\rm{c}}^2$ on $\widehat{P}_{\rm{c}}$ could be straightforwardly transformed into its dependence on $M_{\rm{NS}}^{\max}/M_{\odot}$.
In the following, we assume that NSs with masses $\rm{X}_{-0.07}^{+0.07}M_{\odot}$ where $\rm{X}=1.3\mbox{$\sim$}2.3$ have similar radii as indicated by the NICER observations to extract the $\nu_{\rm{c}}$ from Eq.\,(\ref{Rmax-n}). Specifically, NICER found that the radius of PSR J0740+6620 (mass $\approx2.08_{-0.07}^{+0.07}M_{\odot}$) is about $12.39_{-0.98}^{+1.30}\,\rm{km}$\,\cite{Riley21} and that of the canonical PSR J0030+0451 (mass $\approx1.34_{-0.16}^{+0.15}M_{\odot}$) is about $12.71_{-1.19}^{+1.14}\,\rm{km}$\,\cite{Riley19} (see also Ref.\,\cite{Miller19}). They are very similar, indicating that the NS radius is less sensitive to the central EOS\,\cite{CLZ23}.
Therefore, we obtain $10^3\nu_{\rm{c}}\approx11.2\,\rm{fm}^{3/2}/\rm{MeV}^{1/2}$ by using $R_{\max}\approx12.39_{-0.98}^{+1.30}\,\rm{km}$\,\cite{Riley21}.
Perturbatively, we have for the SSS at NS center from expanding Eq.\,(\ref{sc2}) as,
\begin{align}
s_{\rm{c}}^2&\approx\frac{4}{3}\widehat{P}_{\rm{c}}\left(1+\widehat{P}_{\rm{c}}+\frac{3}{2}\widehat{P}_{\rm{c}}^2+3\widehat{P}_{\rm{c}}^3\right)+\mathcal{O}\left(\widehat{P}_{\rm{c}}^5\right)\label{cdk-1}\\
&\approx\frac{4}{3}\beta\left(1+5\beta+\frac{57}{2}\beta^2+175\beta^3\right)+\mathcal{O}\left(\beta^5\right),\label{cdk-2}
\end{align}
where $\beta\approx0.052(M_{\rm{NS}}^{\max}/M_{\odot}+0.106)\ll1$.
Keeping only the leading-order term of (\ref{cdk-1}) gives the Newtonian prediction $s_{\rm{c}}^2=4\widehat{P}_{\rm{c}}/3$ from which one infers $\widehat{P}_{\rm{c}}\lesssim3/4$.
The general-relativistic contributions in the TOV equations have an effect about 100\% on the upper limit for $\widehat{P}_{\rm{c}}$.
Essentially,  the $s_{\rm{c}}^2$ increases as $M_{\rm{NS}}^{\max}$ increases as shown in FIG.\,\ref{fig_sckk}.
Results of FIG.\,\ref{fig_sckk} provide us a straightforward way to infer the central sound speed $s_{\rm{c}}^2$ once the radii/masses are known (measured). It also implies that more compact NSs have larger $s_{\rm{c}}^2$, e.g., NSs of masses $1.5M_{\odot}$ and $1.4M_{\odot}$ with radii 9.9$\sim$11.2\,km\,\cite{Oze16} and $11.0_{-0.6}^{+0.9}\,\rm{km}$\,\cite{Capano20} have $s_{\rm{c}}^2\approx0.31_{-0.07}^{+0.07}$ and $s_{\rm{c}}^2\approx0.23_{-0.05}^{+0.03}$,  respectively, shown as red hexagons in FIG.\,\ref{fig_sckk}.
While for PSR J0740+6620 as indicated with the green diamond, we have $s_{\rm{c}}^2\approx0.45_{-0.18}^{+0.14}$ using $\widehat{P}_{\rm{c}}\approx0.24_{-0.07}^{+0.05}$\,\cite{CLZ23}.

In FIG.\,\ref{fig_sckk}, the CB for SSS\,\cite{Hoh09,Cher09}, namely $s^2=1/3$ is also indicated (by the dashed orange line).
Our result on $s_{\rm{c}}^2$ implies that the CB for SSS tends to break down for $M_{\rm{NS}}^{\max}/M_{\odot}\gtrsim1.89$.
The latter is very close to the predicted most probable critical maximum-mass about $M_{\rm{NS}}^{\max}/M_{\odot}\approx1.87$ given in Ref.\,\cite{Bed15}, above which the CB is likely to break down (see their FIG.\,1 and FIG.\,2). See also Ref.\,\cite{Alt22}, which predicted that the largest mass of NSs being consistent with the CB should be $\lesssim1.99M_{\odot}$.
Equivalently, the CB $s_{\rm{c}}^2\leq1/3$ is broken for $\widehat{P}_{\rm{c}}\gtrsim0.195$ from Eq.\,(\ref{sc2}).
The finding here is consistent with previous studies on the same issue\,\cite{McL19,Tan22,Tan22-a,Alt22,Ecker22,Ecker23,Alsing2018,Tews18,Leo20,Miao2021},  indicating that Eq.\,(\ref{sc2}) grasps the main features of the SS. However, we emphasize that our method adopts no specific model for NS matter EOS. We also make no assumptions about the composition of NSs (nucleons, hyperons and/or quarks), and therefore provides a EOS-model independent way for investigating the SS in NSs.
Since the NS EOSs are often softened considering the exotic components such as hyperons, our formula (\ref{sc2}) indicates that the CB in these NSs containing non-nucleonic particles is relatively easier to be obeyed, see, e.g., Refs.\,\cite{Stone21,Otto20,Mott21} on related issues.

To avoid causing confusion, we emphasize here that although NSs with lower maximum-mass have $s_{\rm{c}}^2\lesssim1/3$, it does not imply that these NSs exhibit conformal symmetry (in their cores) since the latter emerges at extremely high energies where the strong force coupling constant diminishes, leading to the ``asymptotic freedom" of quarks and gluons. This symmetry is broken as the energy scale lowers or the theory is probed at larger scales. At the energy scales relevant to NS cores, the presence of massive particles and strong gravitational fields disrupt conformal symmetry. 
The latter being not exact even in quark-gluon plasma is generally expected to break down as the system transitions to a hadronic phase. Thus, the observation of $s_{\rm{c}}^2\lesssim1/3$ in low-mass NSs should not be interpreted to imply that the conformal symmetry is reached there \cite{referee}.

\section{Causality Boundary for Neutron Star Mass-radius Curve}\label{SEC4}

\begin{figure}[h!]
\centering
\includegraphics[width=6.4cm]{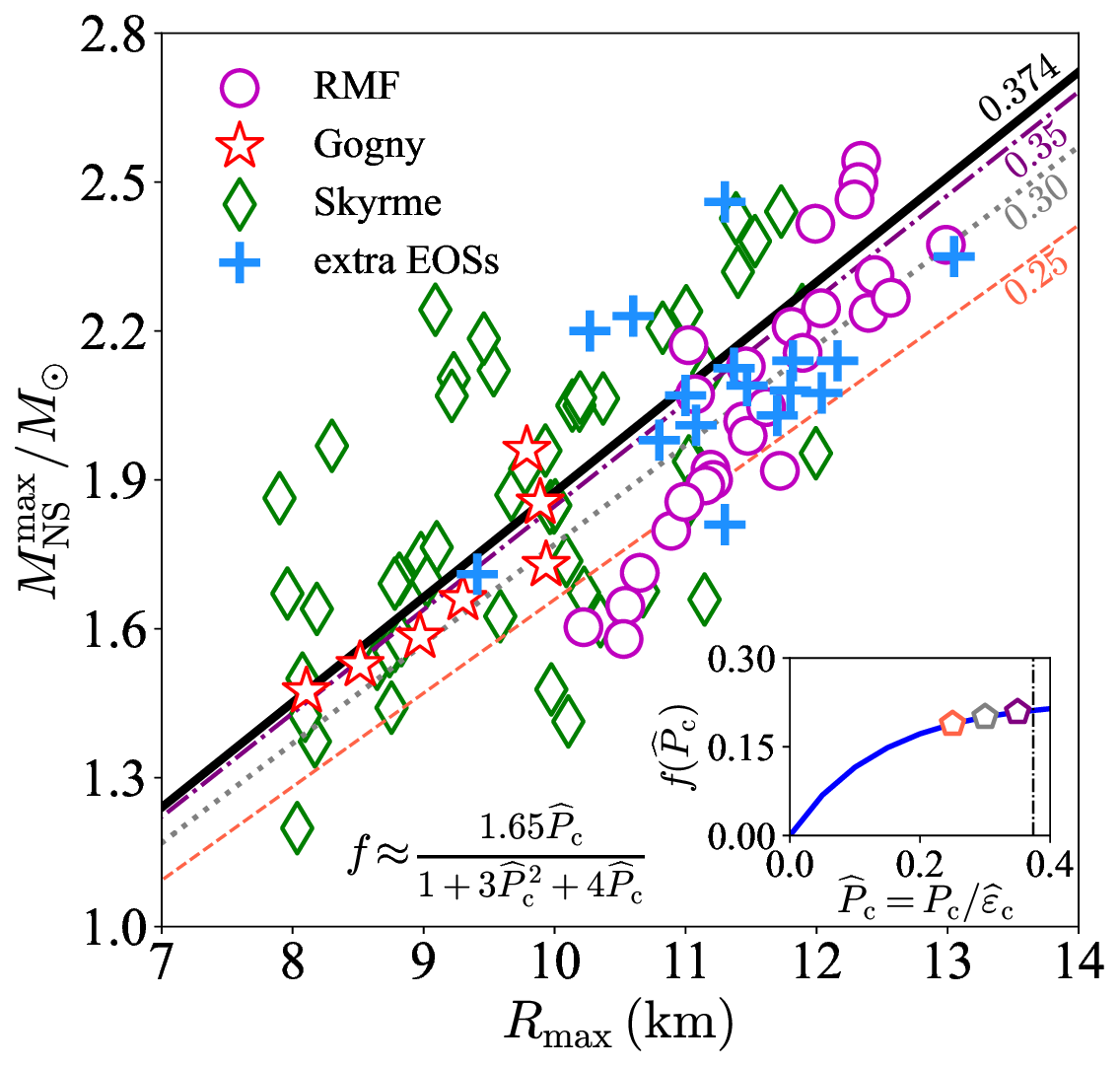}
\caption{Scatters of $M_{\rm{NS}}^{\max}$ and $R_{\max}$ with the covariant RMF approaches, the non-relativistic EDFs of Gogny-like and Skyrme types and several extra microscopic EOSs\,\cite{CLZ23}.
The captions near each lines are the fiducial values for $\widehat{P}_{\rm{c}}$.
}\label{fig_MRmm}
\end{figure}

\begin{figure}[h!]
\centering
\includegraphics[width=8.5cm]{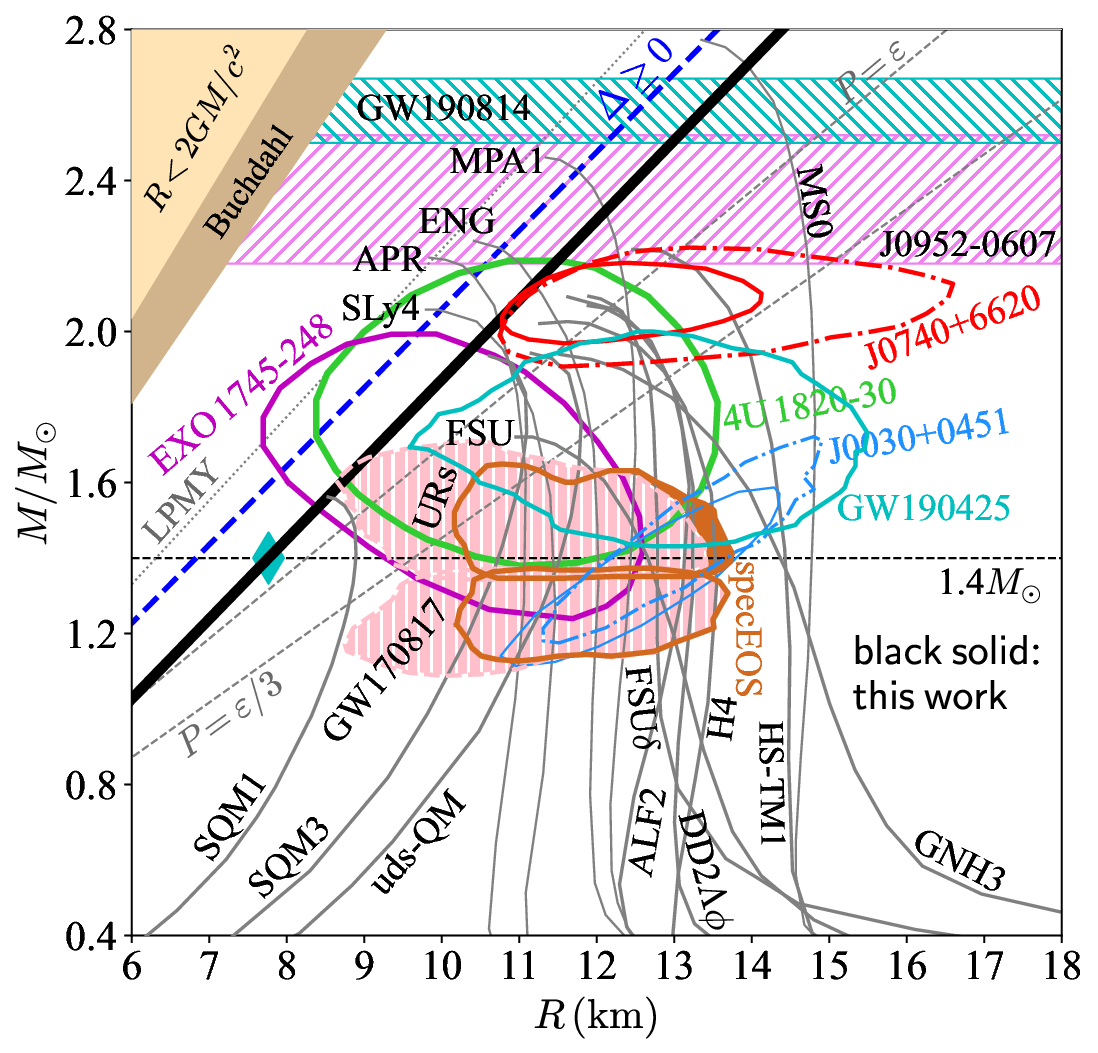}
\caption{Causality boundaries for NS M-R curve, here the black solid line is based on inequality (\ref{rel-2}) (see the corresponding black line in FIG.\,\ref{fig_MRmm}).  The prediction using the trace anomaly $\Delta\equiv1/3-P/\varepsilon\geq0$ (blue dashed line) as well as the previous constraint $R_{\max}/\rm{km}\gtrsim  4.51M_{\rm{NS}}^{\max} /M_{\odot}$ (grey dotted line) are also shown for comparison. The M-R constraints for several NSs (magenta/green/lightblue/red contours), M-R curves from a few typical empirical dense matter EOSs (grey solid lines), and the constraints for the M-R relation based on events GW170817 (chocolate/pink bands) and GW190425 (cyan solid contour) are also shown.
See text for details.
}\label{fig_MR-C}
\end{figure}

The correlations of $M_{\rm{NS}}^{\max}$-$\Gamma_{\rm{c}}$ and $R_{\max}$-$\nu_{\rm{c}}$ also enable us to localize the causality boundary of the NS M-R curve.
Combining Eqs.\,(\ref{Rmax-n}) and (\ref{Mmax-G}) leads to,
\begin{equation}\label{rel-1}
\frac{M_{\rm{NS}}^{\max}}{M_{\odot}}\approx\frac{1.65\widehat{P}_{\rm{c}}}{1+3\widehat{P}_{\rm{c}}^2+4\widehat{P}_{\rm{c}}}\left(\frac{R_{\max}}{\rm{km}}-0.64\right)-0.106.
\end{equation}
The in-front coefficient $f\approx1.65\widehat{P}_{\rm{c}}/(1+3\widehat{P}_{\rm{c}}^2+4\widehat{P}_{\rm{c}})$ is a slow-varying function of $\widehat{P}_{\rm{c}}$, and it essentially explains the conventional quasi-linear correlation between $M_{\rm{NS}}^{\max}/M_{\odot}$ and $R_{\max}$ from model calculations.
In FIG.\,\ref{fig_MRmm}, we show the scatters of $M_{\rm{NS}}^{\max}$ and $R_{\max}$ using the RMF approaches and the non-relativistic EDFs of Gogny-like and Skyrme types together with a few microscopic EOSs\,\cite{CLZ23}.
The fitting lines of Eq.\,(\ref{rel-1}) adopt four fiducial values for $\widehat{P}_{\rm{c}}$ (captioned near the lines with the same color), from which one finds that for $\widehat{P}_{\rm{c}}\gtrsim0.3$ the expression (\ref{rel-1}) could reasonably describe the EOS samples.
However, obvious dispersions can be seen for these scatters.
In addition, certain EOSs are unfavored by the causal limit $\widehat{P}_{\rm{c}}\lesssim0.374$ (solid black line, corresponding to $s_{\rm{c}}^2\leq1$\,\cite{CLZ23}), see FIG.\,\ref{fig_MR-C} for further discussions.

Since the factor $(1+3\widehat{P}_{\rm{c}}^2+4\widehat{P}_{\rm{c}})/\widehat{P}_{\rm{c}}$ takes its minimum about 7.80 at $\widehat{P}_{\rm{c}}\approx0.374$, we have from Eq.\,(\ref{rel-1}) that,
\begin{equation}\label{rel-2}
{R_{\max}}/{\rm{km}}\gtrsim 4.73{M_{\rm{NS}}^{\max}}/{M_{\odot}}+1.14.
\end{equation}
The inequality (\ref{rel-2}) is obtained for the maximum-mass configuration via the causal condition $s_{\rm{c}}^2\leq1$,  it therefore gives a causality boundary for the NS M-R curve.
In FIG.\,\ref{fig_MR-C}, we show the causality boundary of (\ref{rel-2}) (black solid line) together with the M-R constraints for a few NSs including the J0740+6620 and J0030+0451\,\cite{Fon21,Riley21,Miller21,Salm22,Riley19,Miller19} (red/lightblue contours), the 4U\,1820-30 and EXO\,1745-248\,\cite{Oze16-a} (green/magenta contours), the recently reported (``black widow'') PSR J0952-0607 with a mass about 2.35$M_{\odot}$\,\cite{Romani22} (pink hatched band),  GW190814’s secondary component with a mass of $2.59^{+0.08}_{-0.09}M_{\odot}$\,\cite{Abbott2020} (cyan hatched band) and the M-R curves from several empirical dense matter EOSs\,\cite{MPA1,ENG,Mul96,Akmal1998,Glen85,SLy4,H4, ALF2,Prakash1995,Hem10,Chu14,BHB14,FSU,FSUd} (grey solid lines).
The violet  and the chocolate bands are obtained from the GW170817 event using the universal relations (URs) and the spectral EOS (specEOS) approaches \cite{Abbott2018}, respectively, while the cyan solid contour is based on the GW190425 event for a compact binary coalescence with total mass about 3.4M$_{\odot}$\,\cite{Abbott2020-a}. Additionally, the grey dotted line (marked as ``LPMY'') was empirically given by Ref.\,\cite{LP90} as $R_{\max}/\rm{km}\gtrsim  4.51M_{\rm{NS}}^{\max} /M_{\odot}$, by considering nuclear EOS effects.
Recently, the causality boundary for NS M-R curve was localized using an analysis of TA\,\cite{Fuji22} via the condition $\Delta\geq0$ (blue dashed line), where $\Delta\equiv 1/3-P/\varepsilon$ was suggested to measure the TA (see Section \ref{SEC_BRP} for more discussions on $\Delta$ and related issues).

At this point, it is necessary to discuss in some more details the fundamental difference between the apparent condition $P/\varepsilon\leq1$ from the principle of causality and $s^2\leq1$, with the latter being quite relevant in NSs. The physical origin of this distinction can be traced back to the nontrivial/nonlinear dependence of the $s_{\rm{c}}^2$ on $\widehat{P}_{\rm{c}}$, as shown clearly in Eq.\,(\ref{sc2}) which was revealed by the intrinsic structures of the TOV equations\,\cite{CLZ23} (similar clarifications were given recently in Ref.\,\cite{Fuji22} using the TA).
Actually, even an ``ultra-stiff'' EOS $P=\varepsilon$ may introduce probable tension/conflict with the observational data of NSs, e.g., see and compare the upper grey dashed line (marked by ``$P=\varepsilon$'') and the red solid contour for PSR J0740+6620 (at a 68\% confidence level).
Specifically, one can solve analytically the TOV equations adopting the linear EOS $P=w\varepsilon$ (for which $s^2=P/\varepsilon=w$ and therefore $P/\varepsilon\leq1$ and $s^2\leq1$ are totally equivalent) to give the M-R relation as (adopting $c=G=1$),
\begin{equation}
M_{\rm{NS}}^{(w)}=\frac{2w}{1+6w+w^2}R.
\end{equation}
Consequently, $M_{\rm{NS}}^{(w=1)}=R/4$ is obtained, which is smaller than both the Schwarzschild relation $M_{\rm{Schw}}=R/2$ and the Buchdahl's prediction $M_{\rm{Buch}}=4R/9$\,\cite{Buch59}, see the tan and deep-tan bands for the latter two relations in FIG.\,\ref{fig_MR-C}.
On the other hand, compared with the observational data, the ultra-relativistic Fermi gas EOS (with $P=\varepsilon/3$) leads to quite an inconsistent causality boundary (shown as the lower grey dashed line marked by ``$P=\varepsilon/3$''), i.e., $M_{\rm{NS}}^{(w=1/3)}=3R/14$\,\cite{Lightman1975}.
The central SSS is $7/9$ if $\widehat{P}_{\rm{c}}=1/3$ is inserted into Eq.\,(\ref{sc2}).

Interestingly, our result (\ref{rel-2}) puts a more stringent constraint on the M-R curve than those through the TA analysis\,\cite{Fuji22} and the one from Ref.\,\cite{LP90}. Moreover, it is very consistent with the observational data while excluding a few celebrated empirical dense matter EOSs as indicated in FIG.\,\ref{fig_MR-C}.
In particular, the observational boundary of M-R data for PSR J0740+6620\,\cite{Fon21,Riley21,Miller21,Salm22} and those from the events GW170817\,\cite{Abbott2018} and GW190425\,\cite{Abbott2020-a} are basically consistent with the relation (\ref{rel-2}).
Therefore, it is important to check the new causality boundary (\ref{rel-2}) as more observational data on NS masses/radii or GW events become available (see similar discussions given in Ref.\,\cite{Fuji22} and their FIG.3). As an illustration of applying the new causal boundary on the mass-radius curve, we give a lower limit for the radius as $R_{\max}\gtrsim12.25\,\rm{km}$ for the recently identified PSR J0952-0607\,\cite{Romani22} from (\ref{rel-2}),  while for the canonical NS the radius is bounded from below to about 7.77\,km (shown as the cyan solid diamond in FIG.\,\ref{fig_MR-C}).
Inversely, if the radii of NSs were known (constrained), e.g., $R_{\max}\approx12.39\,\rm{km}$ (for PSR J0740+6620\,\cite{Riley21} from NICER), using the causal limit $\widehat{P}_{\rm{c}}\lesssim0.374$ gives an upper limit $M_{\rm{NS}}^{\max}\lesssim2.38M_{\odot}$. Similarly if $R_{\max}\approx11.41\,\rm{km}$ or $13.69\,\rm{km}$ (the lower/upper constraints for radius of PSR J0740+6620\,\cite{Riley21}) is adopted, one has $M_{\rm{NS}}^{\max}\lesssim2.17M_{\odot}$ or $M_{\rm{NS}}^{\max}\lesssim2.65M_{\odot}$, respectively.

\section{Constraints on Neutron Star Compactness}\label{SEC_Compt}

An important quantity connected with the causality boundary is the NS compactness (parameter) defined as $\xi=M_{\rm{NS}}/R$ (adopting $c=G=1$).
For the maximum-mass configuration on the NS M-R curve, one obtains the upper bound for $\xi_{\max}={M_{\rm{NS}}^{\max}}/{R_{\max}}$ by neglecting the intercepts $0.106$ and $0.64$ in Eq.\,(\ref{Mmax-G}) and Eq.\,(\ref{Rmax-n}) respectively, as
\begin{equation}
\xi_{\max}\leq\frac{0.173\times10^4\Gamma_{\rm{c}}}{1.05\times 10^3\nu_{\rm{c}}}\left(\frac{M_{\odot}}{\rm{km}}\right)\approx\frac{2.44\widehat{P}_{\rm{c}}}{1+3\widehat{P}_{\rm{c}}^2+4\widehat{P}_{\rm{c}}}
\equiv \xi_{\max}^{\rm{(up)}}.
\label{C-ud}
\end{equation}
Therefore, $\xi_{\max}\lesssim0.313$ by taking $\widehat{P}_{\rm{c}}\approx0.374$, which is about $30\%$ smaller than the Buchdahl limit $4/9\approx0.444$.

\begin{figure}[h!]
\centering
\includegraphics[width=6.4cm]{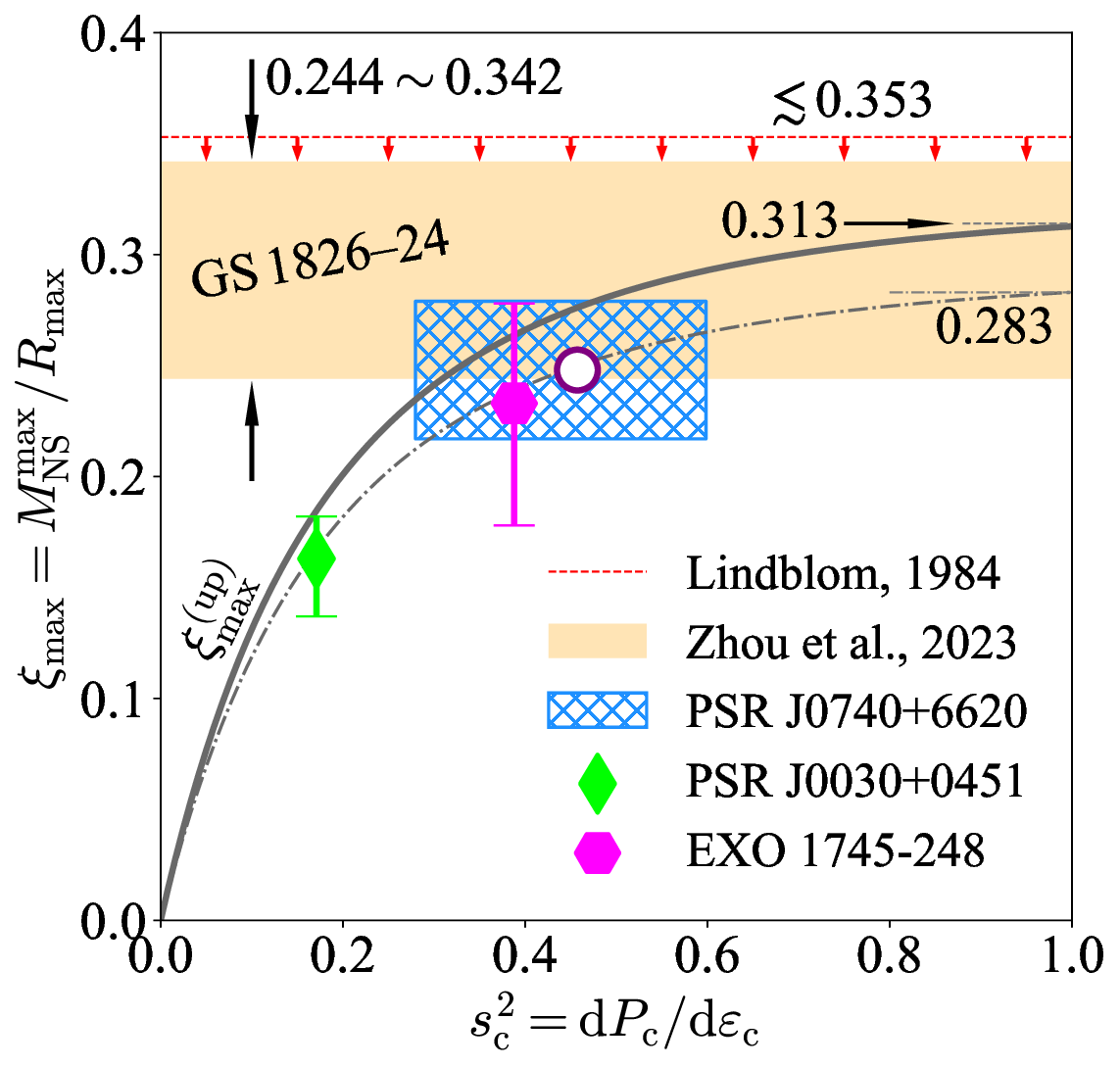}
\caption{Compactness parameter $\xi_{\max}$ for the maximum-mass configuration as a function of central SSS $s_{\rm{c}}^2$. Several other constraints/observational samples (see the text for details) are also shown for comparisons.}\label{fig_Compt}
\end{figure}

In FIG.\,\ref{fig_Compt}, we plot the compactness parameter $\xi_{\max}$ of Eq.\,(\ref{C-ud}) as a function of the central SSS $s_{\rm{c}}^2$ (which depends on the $\widehat{P}_{\rm{c}}$ in the form of Eq.\,(\ref{sc2})), where the compactness for PSR J0740-6620 about $0.217\mbox{$\sim$}0.279$ (with the central value about 0.248) is also shown by the lightblue hatched band (and the purple shallow circle) directly from its observational mass about $2.08_{-0.07}^{+0.07}M_{\odot}$\,\cite{Fon21} and radius about $12.39_{-0.98}^{+1.30}\,\rm{km}$\,\cite{Riley21}.
The compactness about $0.163_{-0.026}^{+0.019}$ for PSR J0030+0451\,\cite{Miller19} is shown by the light-green diamond (the SSS is obtained from its mass and radius using Eqs.\,(\ref{Mmax-G}), (\ref{Rmax-n}) and (\ref{sc2})).
A similar result for NS EXO 1745-248\,\cite{Oze16} is shown by the magenta hexagon.
We notice that neglecting the intercepts $0.106$ and $0.64$ in Eq.\,(\ref{Mmax-G}) and Eq.\,(\ref{Rmax-n}) slightly overestimates the compactness. Actually, by rewriting (\ref{rel-2}) we can obtain 
\begin{equation}
\xi_{\max}\lesssim0.313\cdot\left(1-{1.14\,\rm{km}}/{R_{\max}}\right),\end{equation} where $1.14/R_{\max}\ll1$ could be treated as a correction to Eq.\,(\ref{C-ud}).  Considering a typical NS radius $R_{\max}\approx12\,\rm{km}$ (with a reduction about 10\% on $\xi_{\max}$),  it then leads to $\xi_{\max}\lesssim0.283$ (grey dash-dotted line).
In this sense, the $\xi_{\max}\lesssim0.313$ from Eq.\ (\ref{C-ud}) provides an upper limit for the NS compactness.
An early constraint on the compactness $\xi$ about $\xi\lesssim0.353$\,\cite{Lind84} is indicated by the dashed red line (with a row of arrows).
Furthermore, the constraint on NS compactness can also be obtained from the gravitational redshift $z$ defined by $1+z=(1-2\xi)^{-1/2}$. Interestingly, a new constraint on $z$ was derived very recently from comparing X-ray burst model simulations with observational data for GS 1826-24 using newly measured atomic masses around the rp-process waiting-point nucleus $^{64}\rm{Ge}$\,\cite{Zhou23}. The resulting NS compactness is about $0.244\mbox{$\sim$}0.342$ with the most probable value 0.27 at 95\% confidence level (shown by the tan band). 
In addition, the lower limit of the tan band and the dash-dotted grey line together imply that $s_{\rm{c}}^2\gtrsim0.4$, implying the CB of 1/3 tends to break down.
Compared to all of the above constrains available on NS compactness, ours is probably the most stringent one so far.

We notice by passing that the compactness parameter is closely related to some bulk properties of NSs such as the moment of inertial $I$ and/or the NS binding energy $E_{\rm{b}}$\,\cite{LP01,LP04,LP07,XuJ,Bej03,Steiner16,LS05,Breu2016}. Therefore, the constraint on $\xi$ may naturally lead to some constraints on these quantities. For example, via the empirical expression for the moment of inertial as $I/M_{\rm{NS}}R^2\approx0.237(1+2.84\xi^2+18.9\xi^4)$ given by Ref.\,\cite{LS05}, one may obtain that $I/M_{\rm{NS}}R^2\lesssim0.49$ using $\xi\lesssim0.313$ or $I/M_{\rm{NS}}R^2\lesssim0.46$ using $\xi\lesssim0.283$.
Similarly, Ref.\,\cite{LP01} approximated the $E_{\rm{b}}$ as $E_{\rm{b}}/M_{\rm{NS}}\approx 0.6\xi/(1-\xi/2)$, and therefore $E_{\rm{b}}/M_{\rm{NS}}\lesssim0.22$ and $E_{\rm{b}}/M_{\rm{NS}}\lesssim0.20$ using $\xi\lesssim0.313$ and $\xi\lesssim0.283$, respectively.
We will not discuss these quantities further in the current work since they could be treated as direct corollaries of the constraints on $\xi$.

\section{The Ultimate Energy Density and Pressure in Neutron Stars}\label{SEC5}

Actually, the correlation $M_{\rm{NS}}^{\max}$-$\Gamma_{\rm{c}}$ alone is already useful for inferring the ultimate (maximum) energy density $\varepsilon_{\rm{ult}}$ as well as the ultimate pressure $P_{\rm{ult}}$ allowed in NSs before they collapse into black holes.
By rewriting  Eq.\,(\ref{Mmax-G}) and considering the casual limit $\widehat{P}_{\rm{c}}\lesssim0.374$, we obtain
\begin{equation}\label{eps_ult}
\varepsilon_{\rm{c}}\leq\varepsilon_{\rm{ult}}\equiv 6.32\left({M_{\rm{NS}}^{\max}}/{M_{\odot}}+0.106\right)^{-2}\,\rm{GeV}/\rm{fm}^3.
\end{equation}
Using the constraint $\widehat{P}_{\rm{c}}\lesssim0.374$ once again gives,
\begin{equation}
\label{P_ult}
P_{\rm{c}}\leq P_{\rm{ult}}\equiv 2.36\left({M_{\rm{NS}}^{\max}}/{M_{\odot}}+0.106\right)^{-2}\,\rm{GeV}/\rm{fm}^3.
\end{equation}
We note that the correlation between $R_{\max}$ and $\nu_{\rm{c}}$ of Eq.\,(\ref{Rmax-n}) is not used in establishing these limits.

\begin{figure}[h!]
\centering
\includegraphics[width=8.5cm]{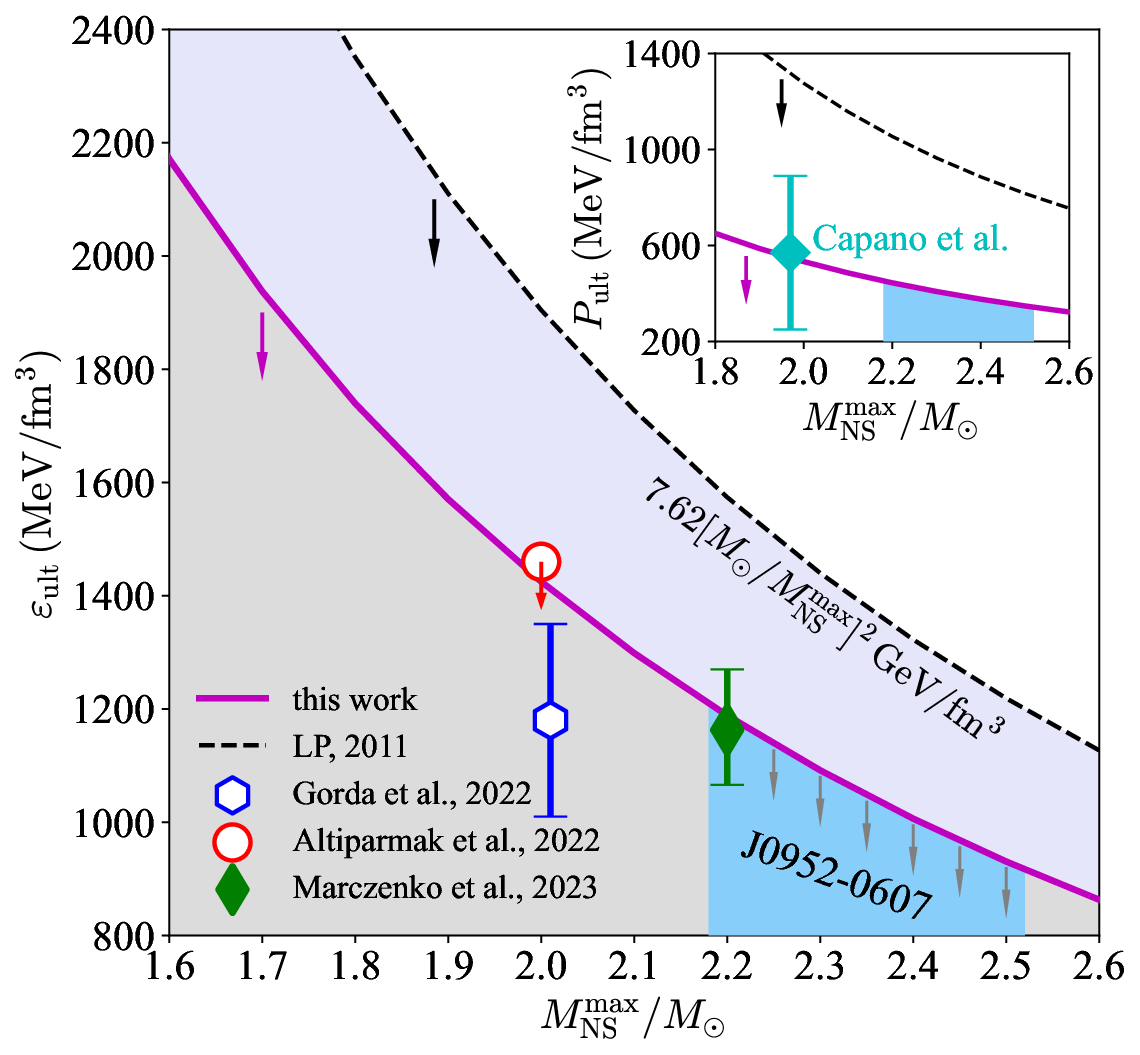}
\caption{The ultimate energy density $\varepsilon_{\rm{ult}}$ and the pressure $P_{\rm{ult}}$ (inset) allowed in NS cores as functions of the maximum-mass $M_{\rm{NS}}^{\max}$ of NS M-R relation.
The previous predictions on $\varepsilon_{\rm{ult}}$ and $P_{\rm{ult}}$ from Ref.\,\cite{LP11} are also shown for comparison (black dashed line). See text for details.
}\label{fig_ULTeps}
\end{figure}

\renewcommand*\tablename{\small TAB.}
\begin{table}[tbh!]
\centering{
\begin{tabular}{|c|c|c|c||c|c|} 
  \hline
$M_{\rm{NS}}^{\max}/M_{\odot}$&$\varepsilon_{\rm{ult}}$&$P_{\rm{ult}}$&Ref.&Eq.\,(\ref{eps_ult})&Eq.\,(\ref{P_ult})\\\hline\hline
       $\gtrsim1.97$&/&$570_{-320}^{+320}$&\cite{Capano20}&1.47&548\\\hline
       $\gtrsim2.0$&1.46&/&\cite{Alt22}&1.42&533\\\hline
       $\gtrsim2.01$&$1.18_{-0.17}^{+0.17}$&/&\cite{Gorda22}&1.41&528\\\hline
       $\gtrsim2.2$&$1.16_{-0.10}^{+0.11}$&/&\cite{Marc23}&1.19&444\\\hline
    \end{tabular}}
\caption{Summary of the constraints on $\varepsilon_{\rm{ult}}$ and $P_{\rm{ult}}$ existing in the literature and from Eqs.\,(\ref{eps_ult}) and (\ref{P_ult}), here $\varepsilon_{\rm{ult}}$ and $P_{\rm{ult}}$ are measured in $\rm{GeV}/\rm{fm}^3$ and $\rm{MeV}/\rm{fm}^3$, respectively.
}
\label{tab_ult}
\end{table}

Based on Eq.\  (\ref{eps_ult}) for $\varepsilon_{\rm{ult}}$ and Eq.\  (\ref{P_ult}) for $P_{\rm{ult}}$,  we find that the existence of a $2.08M_{\odot}$ ($1.97M_{\odot}$) NS leads to $\varepsilon_{\rm{c}}\lesssim1.32\,\rm{GeV}/\rm{fm}^3$ ($1.47\,\rm{GeV}/\rm{fm}^3$) and $P_{\rm{c}}\lesssim494\,\rm{MeV}/\rm{fm}^3$ ($548\,\rm{MeV}/\rm{fm}^3$), respectively.
It is necessary to point out that the upper limit $P_{\rm{c}}\lesssim548\,\rm{MeV}/\rm{fm}^3$ is quite consistent with $P_{\max}\lesssim570\,\rm{MeV}/\rm{fm}^3$ (shown as cyan diamond with error bars in the inset of FIG.\,\ref{fig_ULTeps}) obtained from Ref.\,\cite{Capano20} where the maximum masses are assumed to be larger than $1.97M_{\odot}$\,\cite{Dem10}.
The general $M_{\rm{NS}}^{\max}$-dependence of $\varepsilon_{\rm{ult}}$ and that of $P_{\rm{ult}}$ are shown by the magenta lines (together with the grey band). Also shown are the predictions on $\varepsilon_{\rm{ult}}$ and $P_{\rm{ult}}$ (black dashed lines with the lavender band) from Ref.\,\cite{LP11}, i.e., $\varepsilon_{\rm{ult}}\approx7.62[M_{\odot}/M_{\rm{NS}}^{\max}]^2\,\rm{GeV}/\rm{fm}^3$ and $P_{\rm{ult}}\approx5.12[M_{\odot}/M_{\rm{NS}}^{\max}]^2\,\rm{GeV}/\rm{fm}^3$ (cited as ``LP, 2011''), respectively.
Considering $M_{\rm{NS}}^{\max}/M_{\odot}\approx2.08$, these relations lead to the estimates $\varepsilon_{\rm{ult}}\lesssim1.76\,\rm{GeV}/\rm{fm}^3$ and $P_{\rm{ult}}\lesssim1.18\,\rm{GeV}/\rm{fm}^3$, which are about 33\% and 139\% larger than those from Eq.\ (\ref{eps_ult}) and Eq.\ (\ref{P_ult}), respectively.
In addition, we find the constraint $\varepsilon_{\rm{c}}\lesssim1.41\,\rm{GeV}/\rm{fm}^3$ (by using $M_{\rm{NS}}^{\max}/M_{\odot}\approx2.01$\,\cite{Ant13} in Eq.\,(\ref{eps_ult})) is close to $\varepsilon_{\rm{c}}\lesssim1.18_{-0.17}^{+0.17}\,\rm{GeV}/\rm{fm}^3$ from recent ab-initio QCD calculations\,\cite{Gorda22}, which also assumed the NS masses are greater than $2.01M_{\odot}$ as shown by the blue hexagon.
Similarly,  we have the limit $\varepsilon_{\rm{c}}\lesssim1.42\,\rm{GeV}/\rm{fm}^3$ (using $M_{\rm{NS}}^{\max}/M_{\odot}\approx2$ in Eq.\,(\ref{eps_ult})), which is close to the limit $1.46\,\rm{GeV}/\rm{fm}^3$ of Ref.\,\cite{Alt22} where the algorithm rejects masses $\leq 2M_{\odot}$ (red circle).
Furthermore, the maximum central energy density $\lesssim1.16_{-0.10}^{+0.11}\,\rm{GeV}/\rm{fm}^3$ from Ref.\,\cite{Marc23} is also shown using the green diamond, in which the maximum mass $M_{\rm{NS}}^{\max}$ is greater than about 2.2$M_{\odot}$.
Using our estimate of Eq.\,(\ref{eps_ult}), for such $M_{\rm{NS}}^{\max}\approx2.2M_{\odot}$ the central energy density is found to be about $1.19\,\rm{GeV}/\rm{fm}^3$.
We summarize in TAB.\,\ref{tab_ult} these constraints on $\varepsilon_{\rm{ult}}$ and $P_{\rm{ult}}$.
Finally, by considering the (black widow) PSR J0952-0607\,\cite{Romani22} with its mass about 2.35$M_{\odot}$, the ultimate energy density $\varepsilon_{\rm{ult}}$ and pressure $P_{\rm{ult}}$ are estimated to be about $1.05\,\rm{GeV}/\rm{fm}^3$ and $392\,\rm{MeV}/\rm{fm}^3$ (lightblue bands), respectively.

\section{Continuous Crossover of Sound Speed in Neutron Star Cores}\label{SEC3}

Up to now, we have investigated a few consequences of the $s_{\rm{c}}^2$ of Eq.\,(\ref{sc2}) holding only at NS centers.
In this section, we study the possible occurrence of continuous crossover signaled by a smooth reduction of $s^2$ at high densities in NS cores (as sketched in panel (b) of FIG.\,\ref{fig_sr2_sk}), by generalizing $s_{\rm{c}}^2$ of Eq.\,(\ref{sc2}) to finite distances away from NS centers.

\begin{figure}[h!]
\centering
\includegraphics[width=8.5cm]{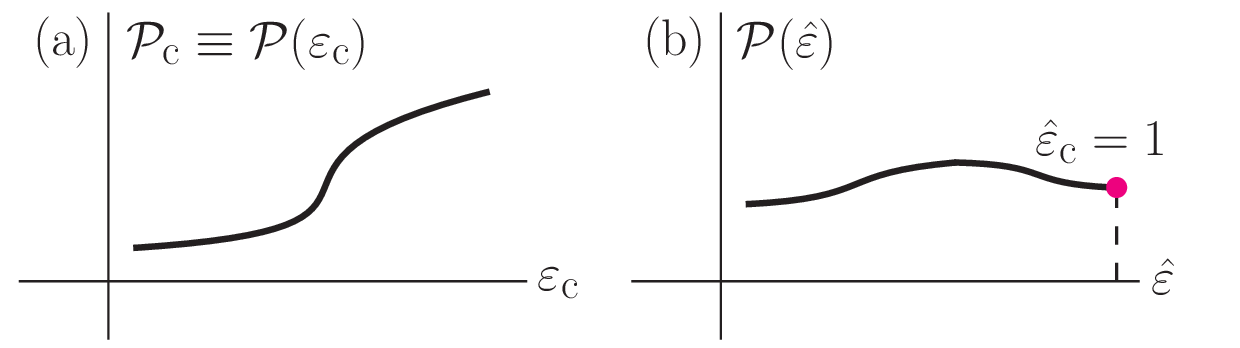}
\caption{Sketch for two types of dependence of certain physical quantities on energy density.
}\label{fig_Pec}
\end{figure}

In this and next sections, we shall encounter two types of dependence of a certain quantity $\mathcal{P}$ on the energy density, one is the dependence of (central) $\mathcal{P}_{\rm{c}}\equiv\mathcal{P}(\varepsilon_{\rm{c}})$ on $\varepsilon_{\rm{c}}$ (panel (a) of FIG.\,\ref{fig_Pec}) or equivalently on $\widehat{P}_{\rm{c}}$, e.g., FIG.\,\ref{fig_NEG-K3}, FIG.\,\ref{fig_tc}, etc.
The other is the dependence of $\mathcal{P}$ on the reduced energy density $\widehat{\varepsilon}=\varepsilon/\varepsilon_{\rm{c}}$, e.g., Eq.(\ref{smu}) and Eq.\,(\ref{Pmu3}), etc., see panel (b) of FIG.\,\ref{fig_Pec}.
The latter actually encapsulates the radial-dependence of the quantity, i.e., a finite $\widehat{r}$ corresponds to a finite $\widehat{\varepsilon}<1$.
We may write explicitly the subscripts to avoid potential confusions.

\subsection{1st-order in $\widehat{\varepsilon}-1$: Sign of $s_{\textmd{c}}^2-s^2(\widehat{\varepsilon})$ for $\widehat{\varepsilon}<1$}

One can straightforwardly obtain the SSS at distance $\widehat{r}$ from the center $\widehat{r}=0$ by considering the perturbative expressions for the reduced pressure $\widehat{P}\approx \widehat{P}_{\rm{c}}+b_2\widehat{r}^2+b_4\widehat{r}^4+\cdots$ and the reduced energy density $\widehat{\varepsilon}\approx1+a_2\widehat{r}^2+a_4\widehat{r}^4+\cdots$\,\cite{CLZ23},
\begin{equation}
s^2(\widehat{r})=\frac{\d\widehat{P}}{\d\widehat{\varepsilon}}=\frac{\d\widehat{P}}{\d\widehat{r}}\frac{\d\widehat{r}}{\d\widehat{\varepsilon}}
=\frac{b_2+2b_4\widehat{r}^2+\cdots}{a_2+2a_4\widehat{r}^2+\cdots}.\label{io-2}
\end{equation}
Taking $\widehat{r}=0$ in Eq.\,(\ref{io-2}) gives us $s_{\rm{c}}^2=b_2/a_2$ and therefore $a_2=b_2/s_{\rm{c}}^2<0$ since $b_2=-6^{-1}(1+3\widehat{P}_{\rm{c}}^2+4\widehat{P}_{\rm{c}})<0$\,\cite{CLZ23}, i.e., the energy density at the center is larger than its surroundings. However, it does not necessarily mean that the center is stiffer than its surroundings in terms of the SSS.
We can expand Eq.\,(\ref{io-2}) as an even power series of $\widehat{r}$ (which becomes accurate as $\widehat{r}\to0$),
\begin{align}
s^2(\widehat{r})\approx&s_{\rm{c}}^2\left[1+\frac{2}{b_2}\left(b_4-s_{\rm{c}}^2a_4\right)\widehat{r}^2\right]+\mathcal{O}\left(\widehat{r}^4\right).
\label{io-3}
\end{align}

The expression for $a_4$ could be obtained by using the expansion of $\widehat{P}$ over $\widehat{\varepsilon}$. Specifically, by collecting the terms in the expansion of $\widehat{P}=\sum_{k=1}d_k\widehat{\varepsilon}^k$ in front of $\widehat{r}^4$ and comparing it with $b_4\widehat{r}^4$ in the expansion of $\widehat{P}$ over the coefficients $\{b_k\}$'s, one finds that,
\begin{equation}\label{def-a4}
a_4=\frac{1}{s_{\rm{c}}^2}\left(b_4-a_2^2\sum_{k=1}^K\frac{k(k-1)}{2}d_k\right),
\end{equation}
where the following general sum rules are used,
\begin{align}
\widehat{P}_{\rm{c}}=&\sum_{k=1}^Kd_k,~~
s_{\rm{c}}^2=\left.\frac{\d\widehat{P}}{\d\widehat{\varepsilon}}\right|_{\widehat{\varepsilon}=\widehat{\varepsilon}_{\rm{c}}=1}=\sum_{k=1}^Kkd_k\label{io-4-s2}.
\end{align}
Here $K$ is the effective truncation order of the expansions.
Putting the expression of $a_4$ of Eq.\,(\ref{def-a4}) into Eq.\,(\ref{io-3}) leads to the final expression for $s^2$ as,
\begin{align}\label{oo-2}
s^2(\widehat{r})\approx s_{\rm{c}}^2+2a_2D\widehat{r}^2,~~D=\sum_{k=1}^{K}\frac{k(k-1)}{2}d_k.
\end{align}

Furthermore, since we have for the $\widehat{\varepsilon}$ away from the center perturbatively as $\widehat{\varepsilon}(\widehat{r})\approx1+a_2\widehat{r}^2$, the dependence of $s^2$ on $\widehat{r}^2$ could be transformed into that on $\widehat{\varepsilon}$ as,
\begin{equation}\label{s2eps}
s^2(\widehat{\varepsilon})\approx s_{\rm{c}}^2+2D\left(\widehat{\varepsilon}-1\right).
\end{equation}
The above approximation (\ref{s2eps}) is expected to be valid only near the NS centers (nevertheless, the sign of $D$ is sufficient for our purpose to deduce a relative relation between $s^2(\widehat{\varepsilon}$) and $s_{\rm{c}}^2$).
According to Eq.\,(\ref{s2eps}),  we then have
\begin{align}
D<0\;\leftrightarrow&\; s_{\rm{c}}^2<s^2(\widehat{\varepsilon})\notag\\
\leftrightarrow&\;\rm{``reduction of $s^2$\;toward\;NS\;centers}\mbox{''},
\end{align}
since $\widehat{\varepsilon}-1<0$, see panel (b) of FIG.\,\ref{fig_sr2_sk}.
In both Eq.\,(\ref{oo-2}) and Eq.\,(\ref{s2eps}) the zeroth-order term is the SSS at center,  what we investigate is the possible crossover at zero temperature (cold NSs) and high densities $\gtrsim5\rho_{\rm{sat}}$\,\cite{Brandes23,Fuku20,Fuji23,ZLi22}, with a peaked behavior in $s^2$ (see next subsection).

Besides (\ref{io-4-s2}),  the stability/causality condition for any $\widehat{\varepsilon}$, i.e., $0\leq s^2\leq 1$, can be written as,
\begin{equation}\label{oo-1}
0\leq d_1+2d_2\widehat{\varepsilon}+3d_3\widehat{\varepsilon}^2+\cdots\leq1,
\end{equation}
which guarantees the pressure $\widehat{P}$ never becomes negative. 
The analyses given up to now are general and not limited to the maximum-mass configuration $M_{\rm{NS}}^{\max}$.

For the maximum-mass configuration $M_{\rm{NS}}^{\max}$ on the NS M-R curve, the inequality $\d^2M_{\rm{NS}}^{\max}/\d\varepsilon_{\rm{c}}^2<0$ gives
\begin{align}\label{io-7}
\left.\frac{\d s_{\rm{c}}^2}{\d\widehat{P}_{\rm{c}}}\right|_{M_{\rm{NS}}^{\max}}<\sigma_{\rm{c}}^2\equiv
\frac{\d}{\d\widehat{P}_{\rm{c}}}s_{\rm{c}}^2=
\frac{2}{3}\frac{9\widehat{P}_{\rm{c}}^4-3\widehat{P}_{\rm{c}}^2+4\widehat{P}_{\rm{c}}+2}{(1-3\widehat{P}_{\rm{c}}^2)^2},
\end{align}
here $s_{\rm{c}}^2$ is given by Eq.\,(\ref{sc2}).
Inequality (\ref{io-7}) implies, e.g.,  for PSR J0740+6620 that $\d s_{\rm{c}}^2/\d\widehat{P}_{\rm{c}}\lesssim2.74$ using $\widehat{P}_{\rm{c}}\approx0.24$\,\cite{CLZ23}, which means if $\widehat{P}_{\rm{c}}$ increases by about 0.1, the increasing of $s_{\rm{c}}^2$ should be smaller than 0.274.
Moreover, we obtain using the relation $\d s^2/\d\widehat{P}=\d s^2/\d\widehat{\varepsilon}\cdot\d\widehat{\varepsilon}/\d\widehat{P}=\d^2\widehat{P}/\d\widehat{\varepsilon}^2\cdot\d\widehat{\varepsilon}/\d\widehat{P}$ an equivalent form of (\ref{io-7}),
\begin{equation}\label{io-5}
\left.\frac{\d^2\widehat{P}}{\d\widehat{\varepsilon}^2}\right|_{\widehat{\varepsilon}=\widehat{\varepsilon}_{\rm{c}}=1}=\sum_{k=1}^Kk(k-1)d_k=2D<\sigma_{\rm{c}}^2s_{\rm{c}}^2.
\end{equation}
The coefficients $d_1$ and $d_2$ (via conditions of (\ref{io-4-s2}))  and the correction $D$ are given respectively by,
\begin{align}
d_1=&2\widehat{P}_{\rm{c}}-s_{\rm{c}}^2
+\sum_{k=3}^{K}(k-2)d_k,\label{ref-d1}\\
d_2=&-\widehat{P}_{\rm{c}}+s_{\rm{c}}^2
-\sum_{k=3}^{K}(k-1)d_k,\label{ref-d2}\\
D=&-3\widehat{P}_{\rm{c}}+2s_{\rm{c}}^2
+\sum_{k=3}^{K}\frac{(k-2)(k-3)}{2}d_{k}.\label{ref-D}
\end{align}
The first two terms in $D$ are deterministic while the last one has certain randomness (reflecting the uncertainties of the dense matter EOS).

As an example, we use the above scheme to determine the correction coefficient $D$ for $K=5$.
The sum rules of (\ref{io-4-s2}) (or equivalently Eqs.\,(\ref{ref-d1}) and (\ref{ref-d2})) can be used to express the coefficients $d_1$ and $d_2$ as $d_1=2\widehat{P}_{\rm{c}}-s_{\rm{c}}^2+d_3+2d_4+3d_5$ and $d_2=-\widehat{P}_{\rm{c}}+s_{\rm{c}}^2-2d_3-3d_4-4d_5$, respectively.
Then, the vanishing of $s^2$ when $\widehat{\varepsilon}=0$ or equivalently $d_1=0$ enables us to write $d_3=-2\widehat{P}_{\rm{c}}+s_{\rm{c}}^2-2d_4-3d_5$.
Therefore, according to the basic definition of $D$ in Eq.\,(\ref{oo-2}), we obtain $D=d_2+3d_3+6d_4+10d_5=d_4+3d_5+2s_{\rm{c}}^2-3\widehat{P}_{\rm{c}}$, which is just the content of Eq.\,(\ref{ref-D}).
See Appendix \ref{app0} for a proof of Eq.\,(\ref{ref-D}) for general $K$.

\begin{figure}[h!]
\centering
\includegraphics[width=6.4cm]{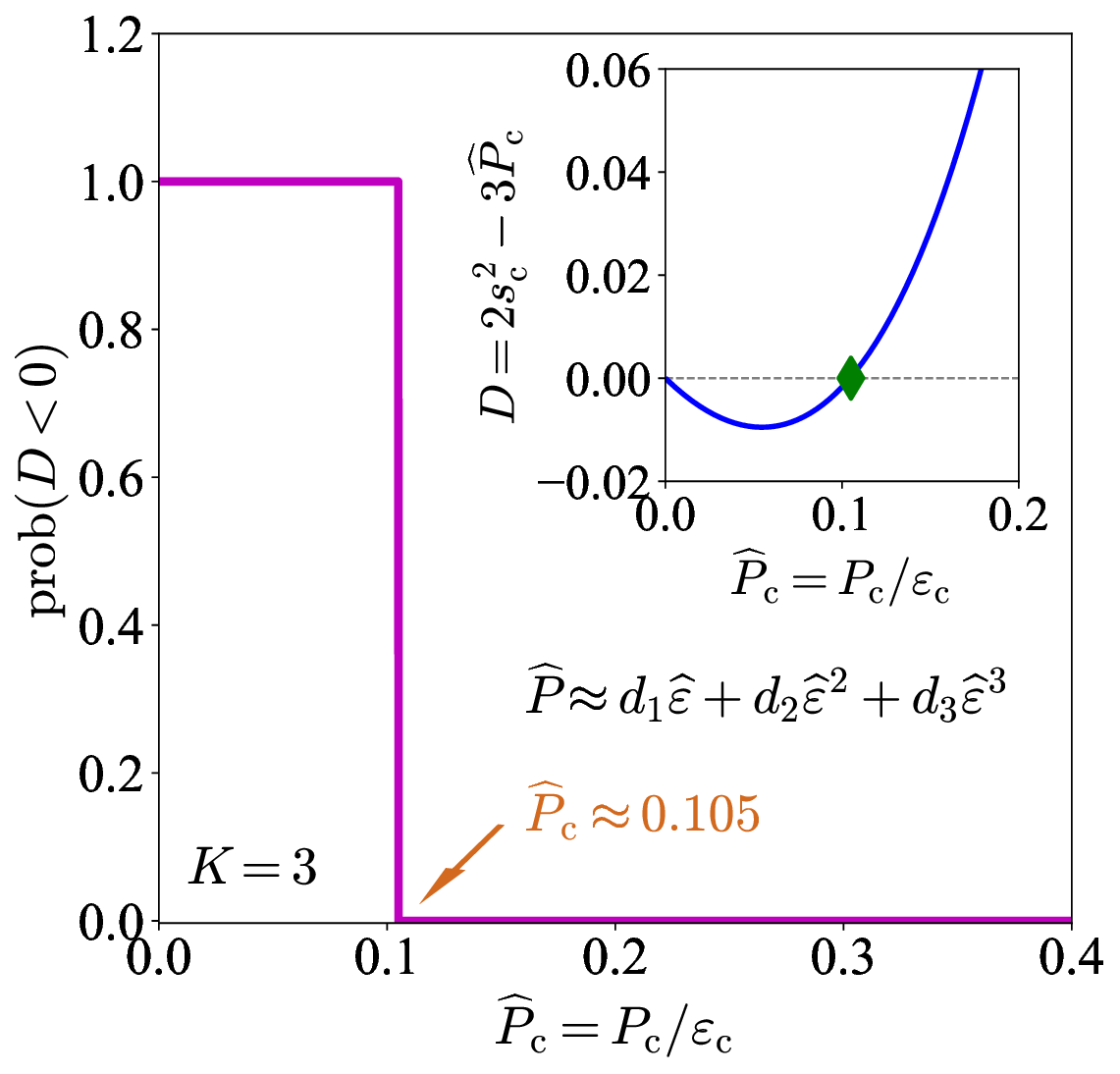}
\caption{The probability of $D<0$ for $K=3$ with $K$ the truncation order of expansion $\widehat{P}=\sum_{k=1}^Kd_k\widehat{\varepsilon}^k=d_1\widehat{\varepsilon}+d_2\widehat{\varepsilon}^2+d_3\widehat{\varepsilon}^3$.}\label{fig_NEG-K3}
\end{figure}

\begin{figure*}
\centering
\includegraphics[width=4.2cm]{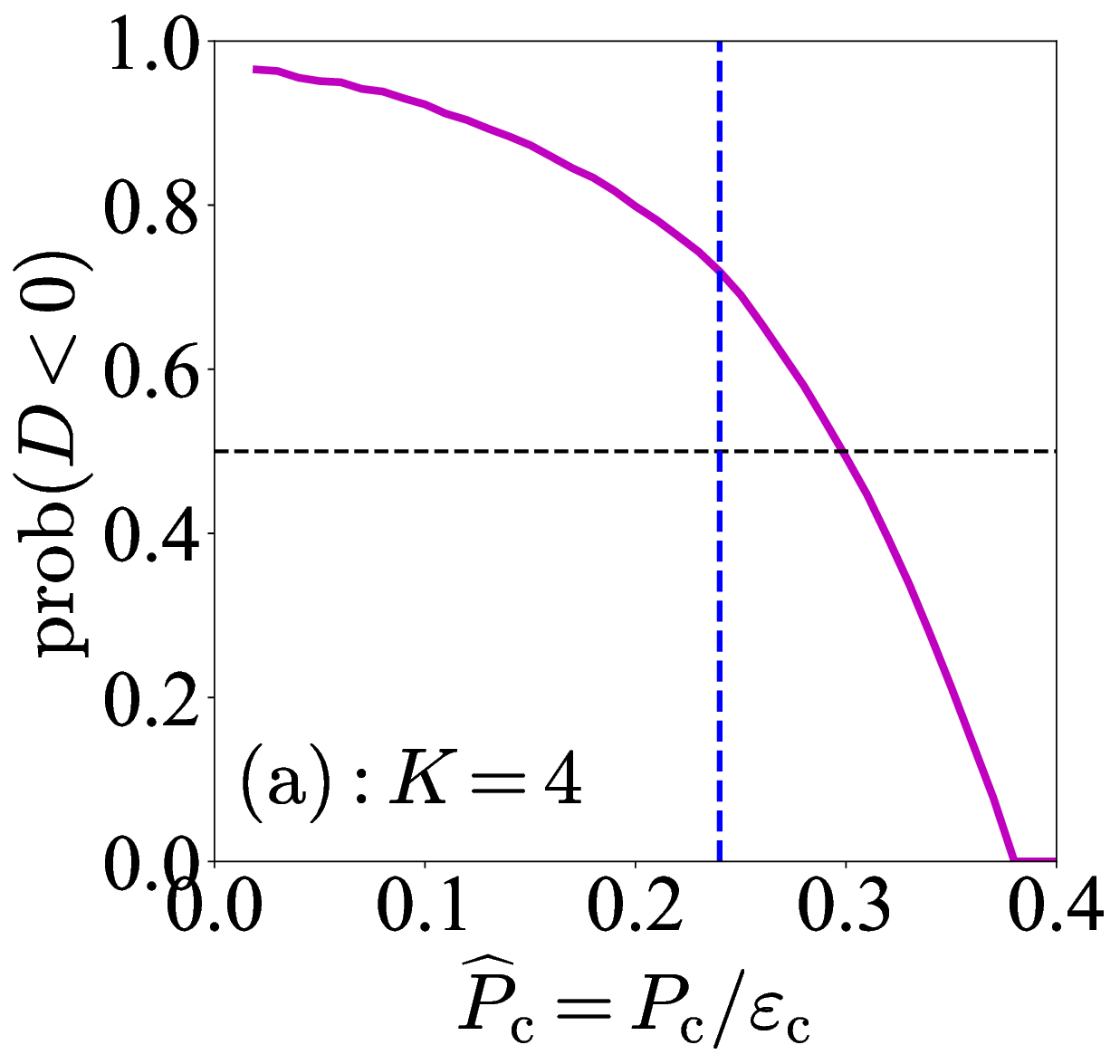}\quad
\includegraphics[width=4.2cm]{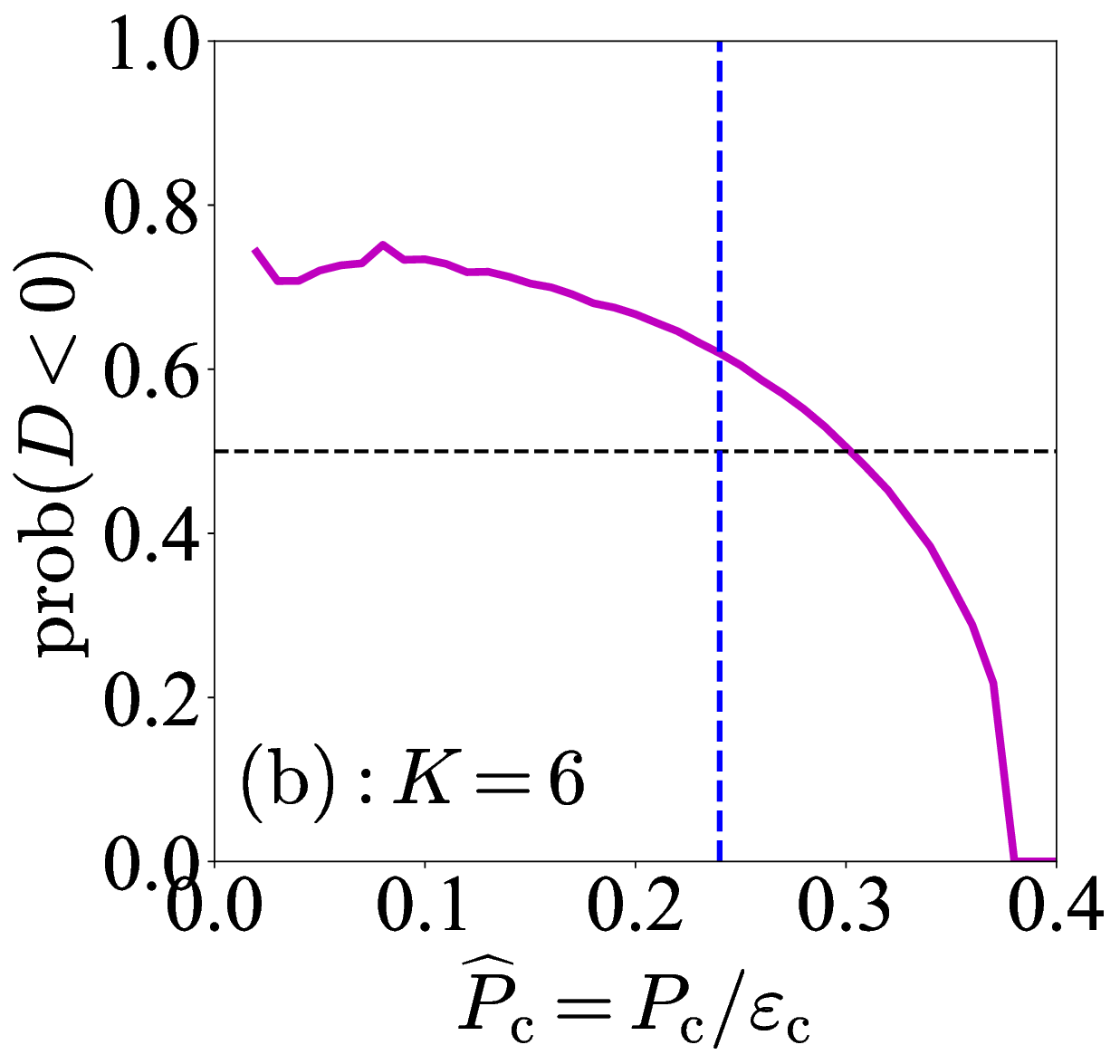}\quad
\includegraphics[width=4.2cm]{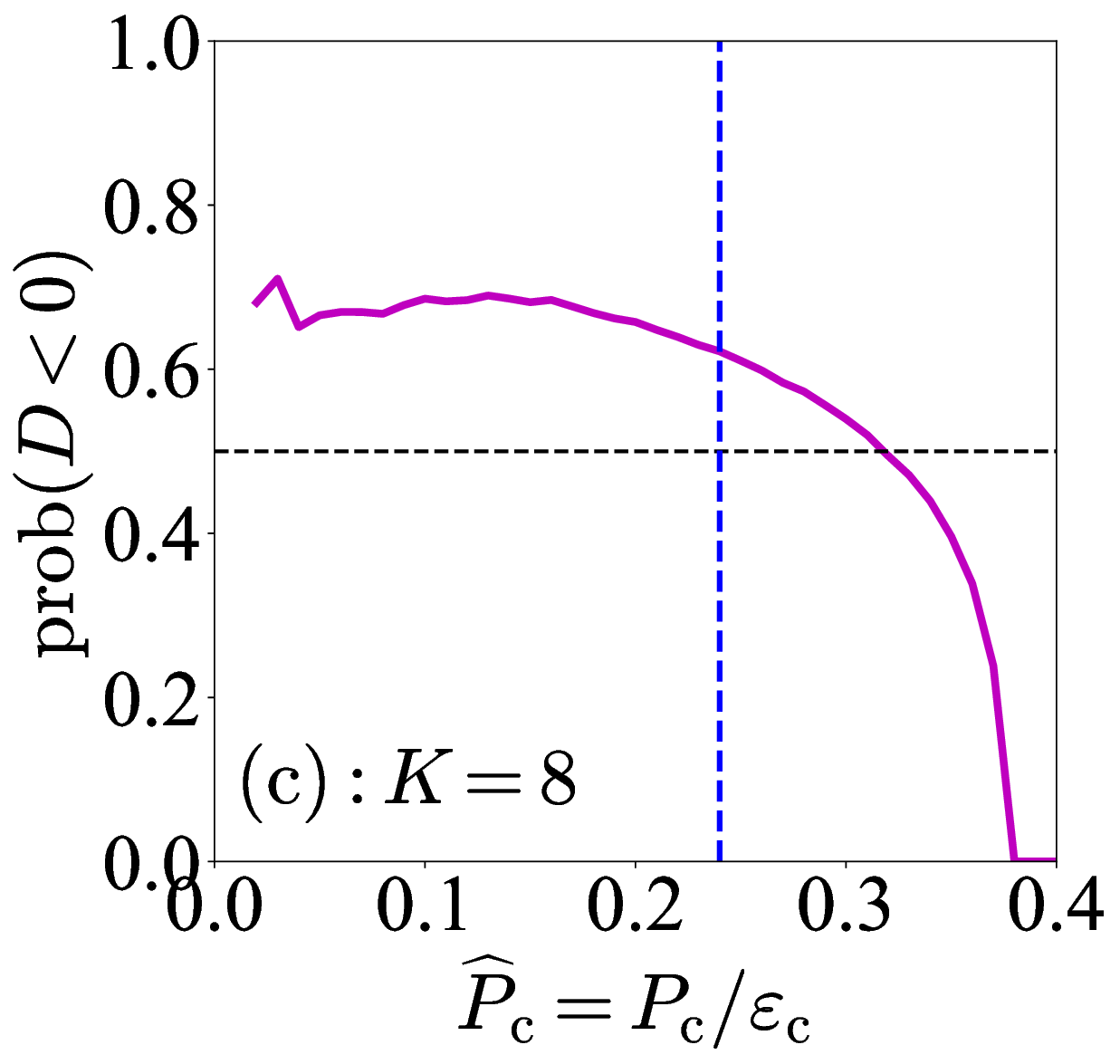}\quad
\includegraphics[width=4.2cm]{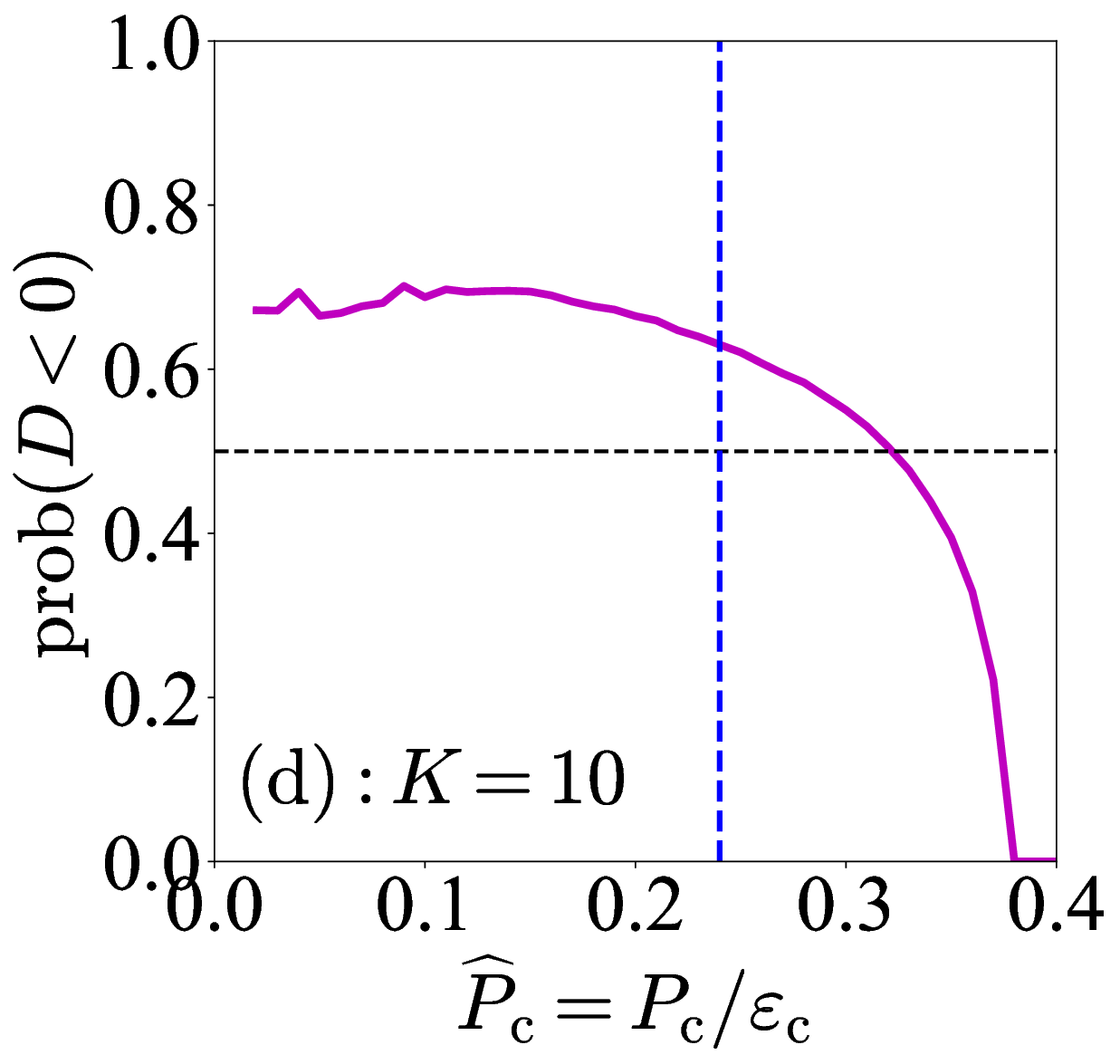}\\
\includegraphics[width=4.2cm]{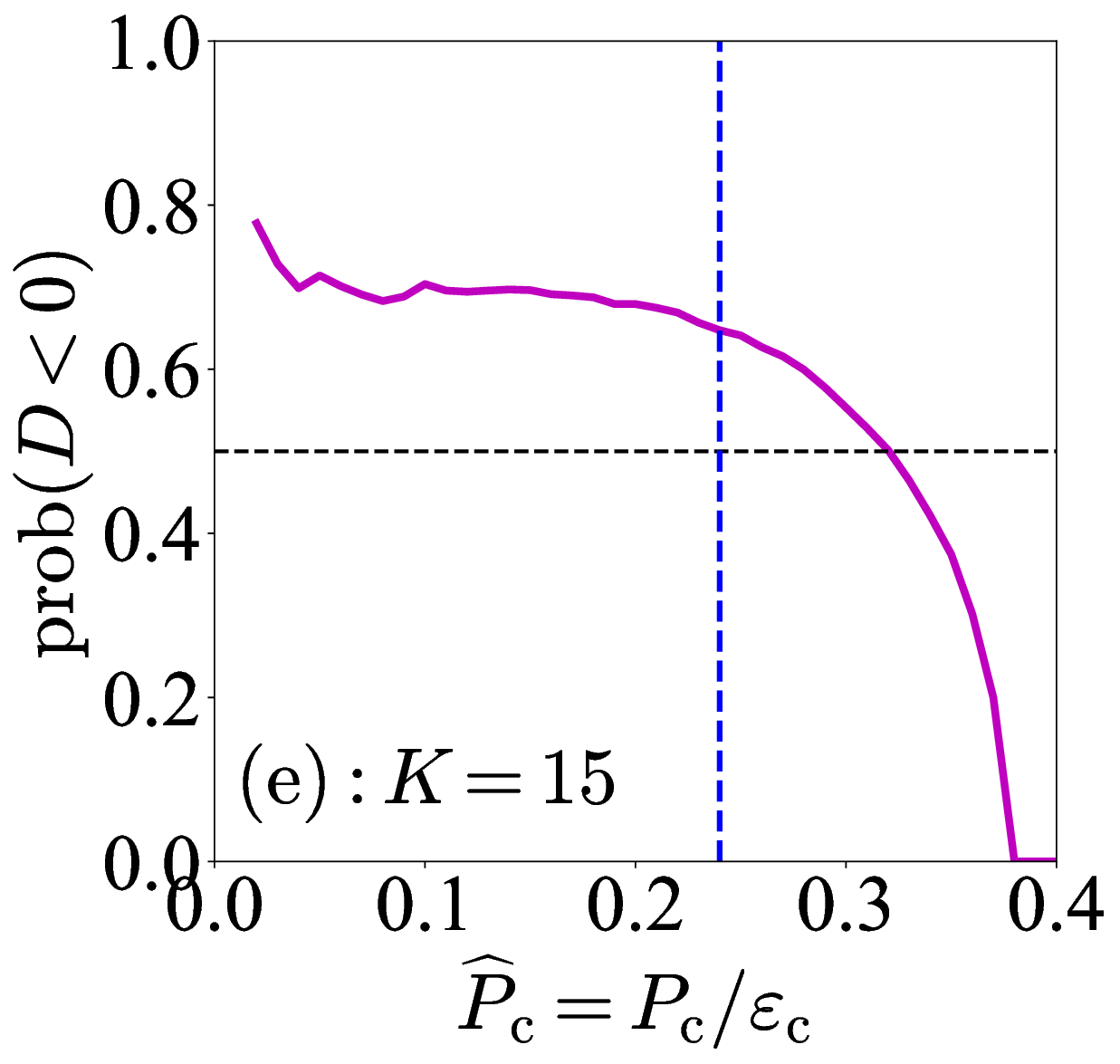}\quad
\includegraphics[width=4.2cm]{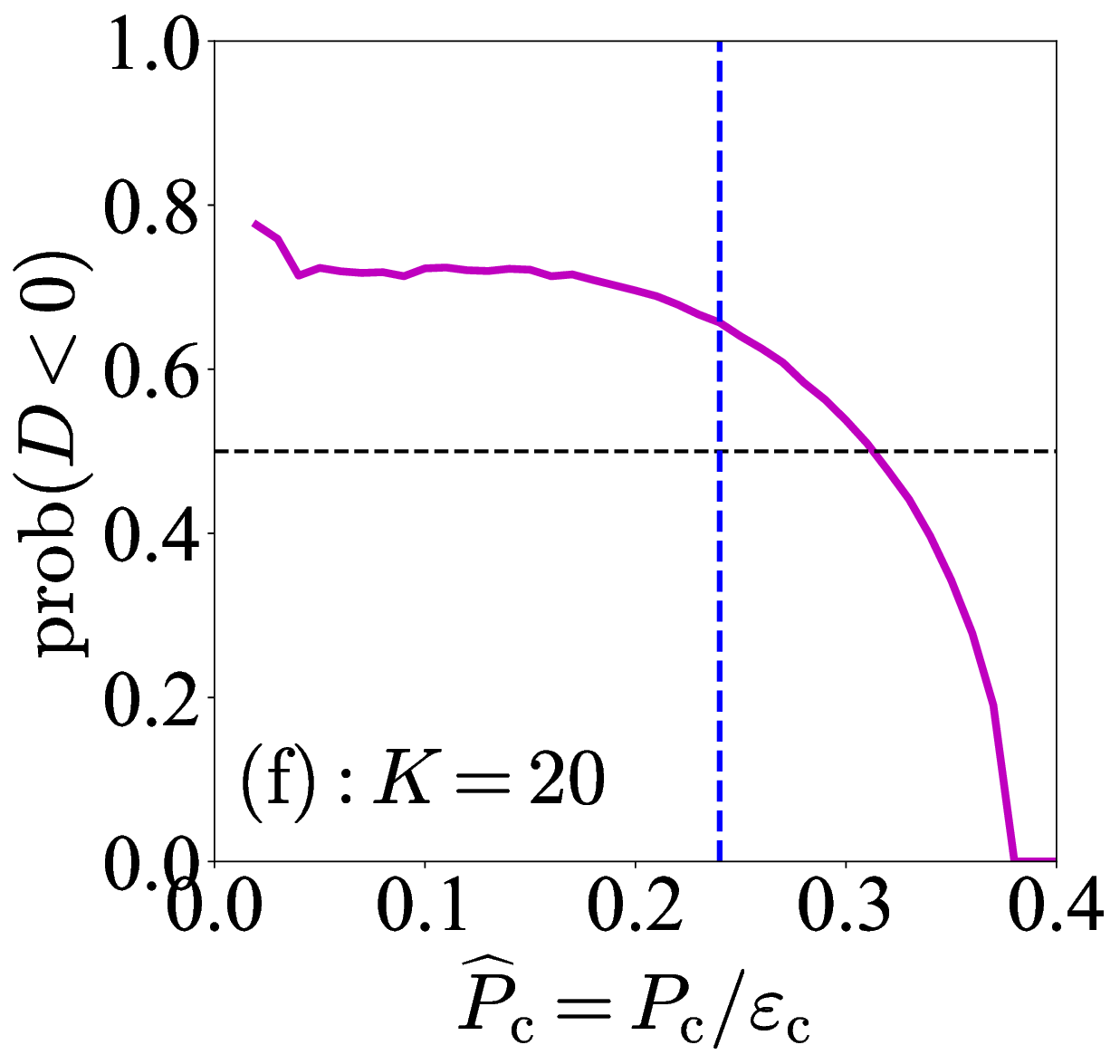}\quad
\includegraphics[width=4.2cm]{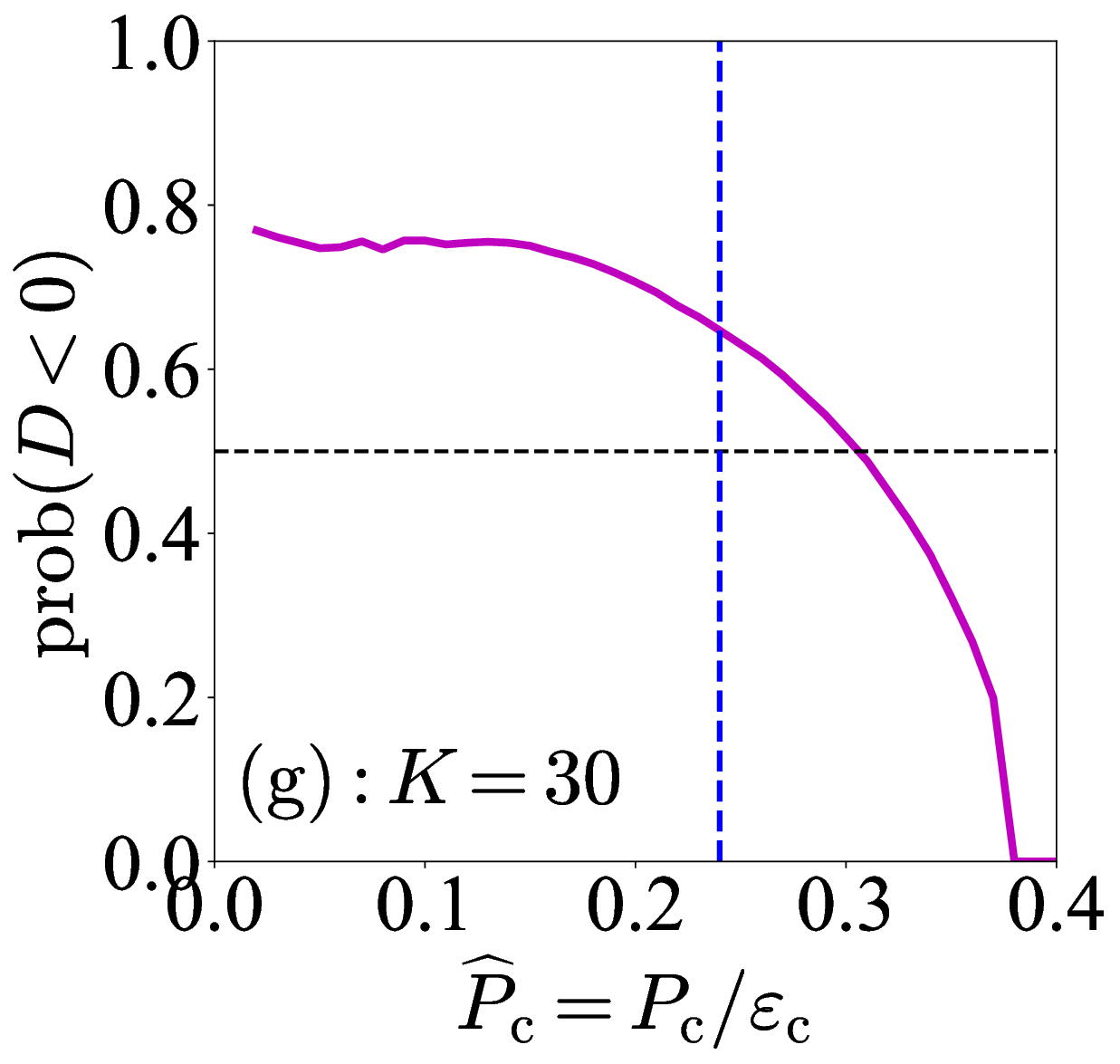}\quad
\includegraphics[width=4.2cm]{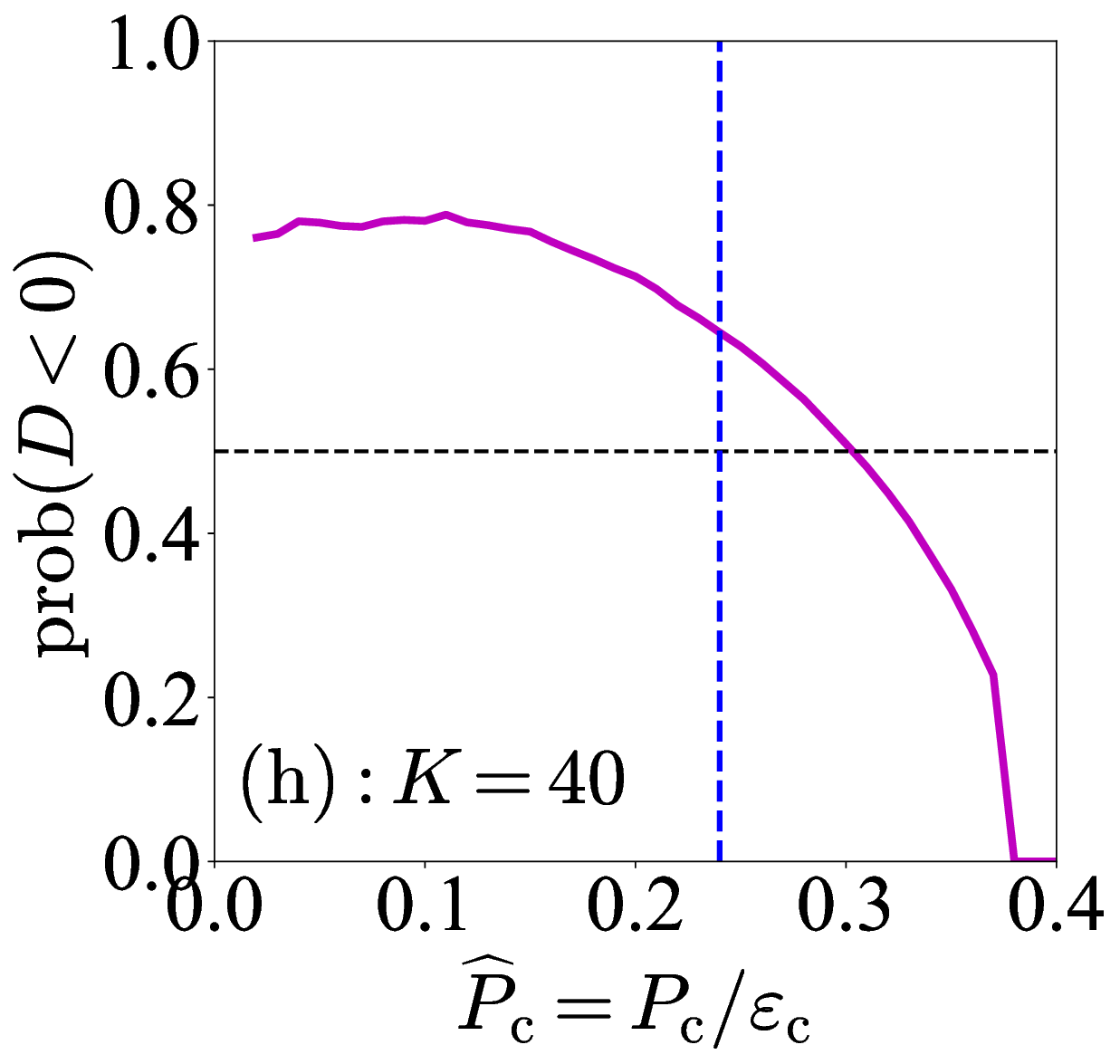}
\caption{The same as FIG.\,\ref{fig_NEG-K3} but for more cases of the truncation order $K$. Dashed vertical blue line in each panel is for $\widehat{P}_{\rm{c}}\approx0.24$ (PSR J0740+6620) extracted in Ref.\,\cite{CLZ23}, and the black horizontal dashed line is plotted at $\rm{prob}(D<0)=50\%$.
}\label{fig_NEG}
\end{figure*}

The probability of $D<0$ could be estimated by combining the condition (\ref{oo-1}) of $0\leq s^2\leq1$,  inequality (\ref{io-5}) together with the general sum rules of (\ref{io-4-s2}), i.e.,
\begin{align}\label{io-6}
&\rm{prob}(D<0)
\approx\frac{\#[0\leq s^2\leq1\;\rm{and}\;2D<\sigma_{\rm{c}}^2s_{\rm{c}}^2\;\rm{and}\;D<0]}{\#[0\leq s^2\leq1\;\rm{and}\;2D<\sigma_{\rm{c}}^2s_{\rm{c}}^2]}, 
\end{align}
for the maximum-mass configuration $M_{\rm{NS}}^{\max}$ of the NS M-R curve,  here $\#$ counts for the number of samples of [$\dots$].
In the following, we uniformly sample the coefficients $d_k$ for $k\geq4$ within certain ranges fulfilling the requirements $0\leq s^2\leq 1$ and $2D<\sigma_{\rm{c}}^2s_{\rm{c}}^2$ and count the events of $D<0$ to estimate the probability of $D<0$.  The value of $d_2$ could be obtained correspondingly using Eq.\,(\ref{ref-d2}) and $d_3$ via Eq.\,(\ref{ref-d1}), i.e., the sum rules of (\ref{io-4-s2}).
For a certain truncation order $K$,  different empirical ranges for the parameters $d_k$ are adopted to boost the sampling efficiency.

The simplest situation is $K=3$, i.e., $\widehat{P}\approx d_1\widehat{\varepsilon}+d_2\widehat{\varepsilon}^2+d_3\widehat{\varepsilon}^3\approx d_2\widehat{\varepsilon}^2+d_3\widehat{\varepsilon}^3$.
Correspondingly, we shall obtain $d_2=-s_{\rm{c}}^2+3\widehat{P}_{\rm{c}}$ and $d_3=s_{\rm{c}}^2-2\widehat{P}_{\rm{c}}$,
and the condition $D<0$ (see Eq.\,(\ref{ref-D})) is now equivalently to (with the approximation holding for small $\widehat{P}_{\rm{c}}$ as we have $s_{\rm{c}}^2\approx4\widehat{P}_{\rm{c}}/3$),
\begin{equation}\label{2cc}
u_{\rm{c}}\equiv 2s_{\rm{c}}^2-3\widehat{P}_{\rm{c}}\approx-{\widehat{P}_{\rm{c}}}/{3}
<0.
\end{equation}
Considering Eq.\,(\ref{sc2}), we then obtain $\widehat{P}_{\rm{c}}\lesssim0.105$ in order to make $D<0$ (see the inset of FIG.\,\ref{fig_NEG-K3} where the diamond is at $\widehat{P}_{\rm{c}}\approx0.105$).
For $\widehat{P}_{\rm{c}}\gtrsim0.105$, there is no space for the occurrence of $D<0$, and the transition on the probability of $D<0$ from $\widehat{P}_{\rm{c}}\lesssim0.105$ to $\widehat{P}_{\rm{c}}\gtrsim0.105$ is sharp (FIG.\,\ref{fig_NEG-K3}). Equivalently, this means $s_{\rm{c}}^2$ is definitely larger than its surroundings for $\widehat{P}_{\rm{c}}\gtrsim0.105$.
Due to the importance of $D$ for our analysis, we explain/illustrate in Appendix \ref{appK} why the $D$ tends to be negative for small $\widehat{P}_{\rm{c}}$ using $K=4$.

The above example with $K=3$ involving the cubic polynomial of $\widehat{P}$ over $\widehat{\varepsilon}$ is a very special case for demonstration. In order to analyze the 
probability of $D<0$ more generally, we study more cases of varying $K$ as shown in FIG.\,\ref{fig_NEG} (each panel simulates $10^8$ uniform samples). 
It is seen that as the truncation order $K$ increases, the probability of $D<0$ eventually stabilizes (see panels (a) to (h) of FIG.\,\ref{fig_NEG} where $K$ varies from 4 to 40).
One finds from the curves that for $\widehat{P}_{\rm{c}}\lesssim0.3$ the probability of $D<0$ is larger than about 50\%, indicating that the continuous crossover probably occurs near the center.
For PSR J0740+6620 with its mass $\approx2.08\pm 0.07M_{\odot}$\,\cite{Fon21,Riley21,Miller21,Salm22} which is close to the theoretical predicted maximum NS mass about $2.01\mbox{$\sim$}2.16M_{\odot}$\,\cite{Rezzolla2018} (Ref.\,\cite{Mar17} showed that it should be smaller than about $2.17M_{\odot}$ while Ref.\,\cite{Ruiz18} predicted $M_{\rm{NS}}^{\max}\lesssim2.16\mbox{$\sim$}2.28M_{\odot}$ via general relativistic magnetohydrodynamics simulations) as well as $\widehat{P}_{\rm{c}}\approx0.24$ and $s_{\rm{c}}^2\approx0.45$\,\cite{CLZ23}, this probability is found to be larger than about 63\%. As $\widehat{P}_{\rm{c}}$ increases even further, the probability of $D<0$ decreases.
In the limiting case of $s_{\rm{c}}^2\to1$ (or equivalently $\widehat{P}_{\rm{c}}\to0.374$), this probability is extremely small.

The physical reason for these behaviors could be traced back to the (deterministic) term $2s_{\rm{c}}^2-3\widehat{P}_{\rm{c}}$ in $D$ of Eq.\,(\ref{ref-D}) which is negative for small $\widehat{P}_{\rm{c}}$ (see Eq.\,(\ref{2cc})), though the summation term of Eq.\,(\ref{ref-D}) maybe either positive or negative (EOS uncertainties).
It  indicates that $\widehat{P}_{\rm{c}}$ is the relevant quantity for the onset of the continuous crossover ($s^2(\widehat{\varepsilon})>s_{\rm{c}}^2$) at zero temperature and high densities (i.e., in the cores of cold NSs)\,\cite{Brandes23,Fuku20,Fuji23,ZLi22}: It is likely to happen for large ($\widehat{P}_{\rm{c}}$ is small and the core is denser) and massive NSs (so they are near the maximum-mass configuration on the M-R curve). 
At the Newtonian limit $s_{\rm{c}}\approx4\widehat{P}_{\rm{c}}/3$, since $u_{\rm{c}}=-\widehat{P}_{\rm{c}}/3$ the coefficient $D$ tends always to be negative (e.g., see relevant discussion/illustration given in Appendix \ref{appK} and FIG.\,\ref{fig_WhyDNL}). In this case,  we expect that the probability of $s^2(\widehat{\varepsilon})>s_{\rm{c}}^2$ is large. Therefore the general-relativistic effects reduce the probability for $s^2(\widehat{\varepsilon})>s_{\rm{c}}^2$.

\begin{figure}[h!]
\centering
\includegraphics[height=3.8cm]{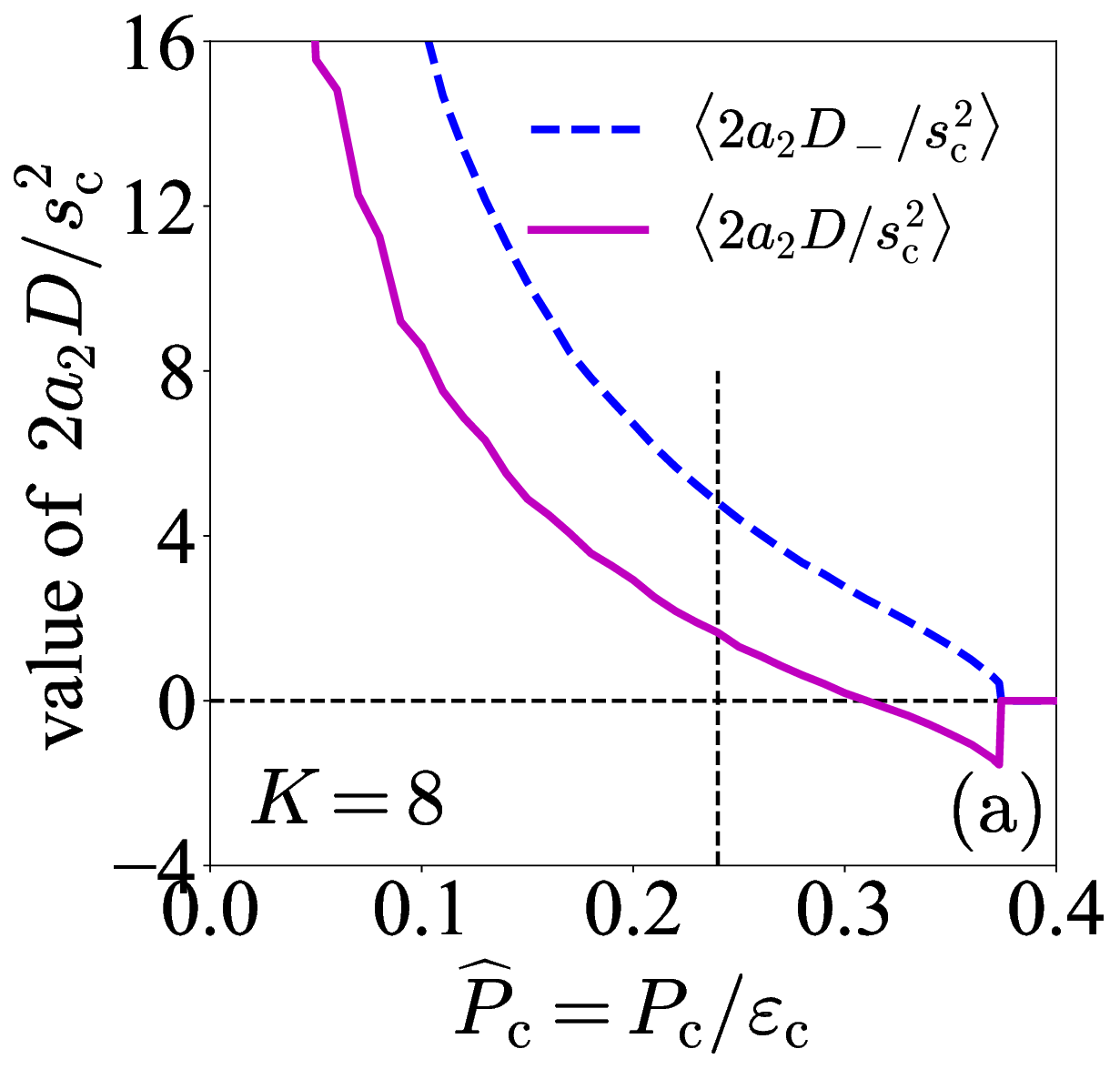}\quad
\includegraphics[height=3.8cm]{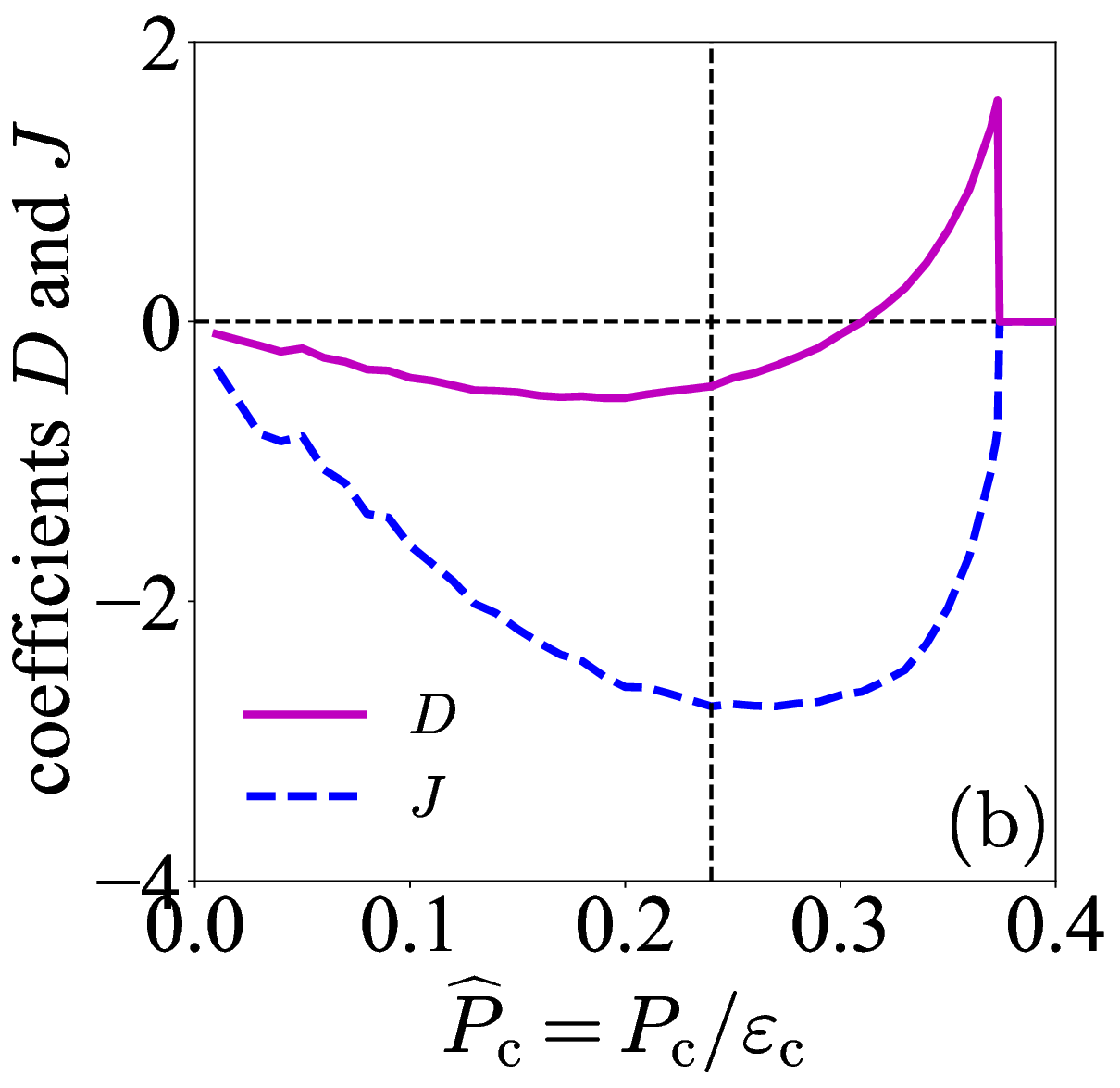}
\caption{Averages of $2a_2D/s_{\rm{c}}^2$ and $2a_2D_-/s_{\rm{c}}^2$ as functions of $\widehat{P}_{\rm{c}}$ (panel (a)) and coefficients $D$ and $J$ (panel (b)) defined in Eq.\,(\ref{oo-2}) and Eq.\,(\ref{ref-J}), respectively, the vertical dashed line is at $\widehat{P}_{\rm{c}}=0.24$.
}\label{fig_AD}
\end{figure}

According to Eq.\ (\ref{oo-2}), the deviation of SSS from its central value is proportional to  $2a_2D\widehat{r}^2$. To evaluate the relative changing rate of SSS from its central value, 
we show in panel (a) of FIG.\,\ref{fig_AD} the $\langle 2a_2D/s_{\rm{c}}^2\rangle$ defined as
\begin{equation}
\left\langle {2a_2D}/{s_{\rm{c}}^2}\right\rangle=\left({2a_2}/{s_{\rm{c}}^2}\right)\sum_{k=\pm}\rm{prob}(D_k)D_k,
\end{equation}
here $D_k=D_{\pm}$ denotes the value of $D>0$ or $D<0$. For a comparison, the average value of $2a_2D_-/s_{\rm{c}}^2$ is also shown (blue dashed line).
Considering $\widehat{P}_{\rm{c}}\approx0.24$\,\cite{CLZ23} for PSR J0740+6620\,\cite{Fon21,Riley21,Miller21,Salm22},  we obtain $
\langle 2a_2D/s_{\rm{c}}^2\rangle\approx1.6$,
i.e.,  $s^2(\widehat{r})/s_{\rm{c}}^2\approx1+1.6\widehat{r}^2$ for small $\widehat{r}$.
Transforming it back using Eq.\,(\ref{s2eps}) gives us $s^2(\widehat{\varepsilon})/s_{\rm{c}}^2\approx 1+2(1-\widehat{\varepsilon})\approx 3-2\widehat{\varepsilon}$, here $\widehat{\varepsilon}$ being very close to 1 is assumed.
Equivalently, we can rewrite the radial-dependence of the SSS by using Eq.\,(\ref{Rmax-n}) approximately as (by neglecting the intercept $0.64$),
\begin{equation}
{s^2}/{s_{\rm{c}}^2}\approx1+\frac{9.4\widehat{P}_{\rm{c}}\langle a_2D/s_{\rm{c}}^2\rangle}{1+3\widehat{P}_{\rm{c}}^2+4\widehat{P}_{\rm{c}}}\left(\frac{r}{R_{\max}}\right)^2,
\end{equation}
here $r=\widehat{r}Q$ is the physical radius and $R_{\max}$ is measured in km.
Taking $\widehat{P}_{\rm{c}}\approx0.24$, we obtain $ s^2/s_{\rm{c}}^2\approx1+0.85(r/R_{\max})^2$. 

The consistency between Eq.\,(\ref{io-3}) and Eq.\,(\ref{oo-2}) implies $b_4-s_{\rm{c}}^2a_4<0$ if $D<0$ due to the negativeness of $a_2$ and $b_2$.
Consequently,  we obtain $a_4>{b_4}/{s_{\rm{c}}^2}>0$ since $b_4>0$\,\cite{CLZ23}.
As a rough (order of magnitude) estimate for $a_4$, e.g., we take $\widehat{P}_{\rm{c}}\gtrsim0.16$ (for massive NSs), therefore $a_4>b_4/s_{\rm{c}}^2\gtrsim0.86$, i.e., $a_4\sim\mathcal{O}(1)$ (see Appendix \ref{appB} for another order of magnitude estimate for $a_4$).
Since all of our perturbative expansions including $\widehat{\varepsilon}$ and $\widehat{\rho}=\rho/\rho_{\rm{c}}$ (see Eq.\,(\ref{pk-4})) over $\widehat{r}$ are dimensionless, it is reasonable to expect  the magnitude of the expansion coefficients is natural, i.e., it is $\sim\mathcal{O}(1)$.
On the other hand, if $D>0$, then the consistency between Eq.\,(\ref{io-3}) and Eq.\,(\ref{oo-2}) implies $a_4<b_4/s_{\rm{c}}^2$.
Combining these two cases leads to
\begin{equation}\label{Da4}
D\left({b_4}/{s_{\rm{c}}^2}-a_4\right)>0,
\end{equation}
i.e., $D$ and $b_4/s_{\rm{c}}^2-a_4$ have the same sign.
The inequality (\ref{Da4}) is a general constraint on the expansion coefficients.

\subsection{2nd-order in $\widehat{\varepsilon}-1$: Estimates on the Peak of $s^2$}

The $s^2$ approaches zero at the core-crust transition density below which the dynamical fluctuations in uniform matter will be unstable indicating the onset of forming clusters in the crust.  It implies that there should exist a peak in $s^2$ somewhere along the path away from the center if $D<0$. In fact, Refs.\,\cite{Ecker22,Ecker23} predicted that the peak of $s^2$ for NSs with masses $M_{\rm{NS}}^{\max}\gtrsim2M_{\odot}$ is located at about 2/3 of its radius. As $\varepsilon$ is a decreasing function of $\widehat{r}$, the peak of $s^2$ is located at a energy density $\varepsilon_{\rm{pk}}<\varepsilon_{\rm{c}}$, e.g., Ref.\,\cite{Tak23} predicted that $\varepsilon_{\rm{pk}}\approx565\,\rm{MeV}/\rm{fm}^3$ for massive NSs.
Take our over-simplified EOS with truncation order $K=3$ (FIG.\,\ref{fig_NEG-K3}) as an example,  the peak $\widehat{\varepsilon}_{\rm{pk}}$ of $s^2(\widehat{\varepsilon})$ can be obtained as $\widehat{\varepsilon}_{\rm{pk}}\approx5/6+\widehat{P}_{\rm{c}}+7\widehat{P}_{\rm{c}}^2/2+13\widehat{P}_{\rm{c}}^3\gtrsim5/6$ and $s^2(\widehat{\varepsilon}_{\rm{pk}})\approx25\widehat{P}_{\rm{c}}/18+5\widehat{P}_{\rm{c}}^2/9+17\widehat{P}_{\rm{c}}^3/6$ (with $\widehat{P}_{\rm{c}}$ being small $\lesssim0.105$) using $s^2\approx2(3\widehat{P}_{\rm{c}}-s_{\rm{c}}^2)\widehat{\varepsilon}+3(s_{\rm{c}}^2-2\widehat{P}_{\rm{c}})\widehat{\varepsilon}^3$.
In addition, the peak of $s^2$ could not be estimated through Eq.\,(\ref{s2eps}) since it is linear in $\widehat{\varepsilon}$ and the higher-order terms, e.g., $a_4$, $b_4$ and $a_6$, etc., are necessary for such estimate.

We now use our perturbative expansions as a tool and include these higher-order contributions to explore whether there would exist a plateau for $\widehat{P}$ for a finite range of $\widehat{\varepsilon}$ (relevant for sharp PTs) and estimate the peak of $s^2$ (near NS centers) if it exists.
When considering terms including order $b_6$-coefficient in the expansion of $\widehat{P}$ and the $a_6$-term in the expansion of $\widehat{\varepsilon}$, 
namely $\widehat{P}\approx \widehat{P}_{\rm{c}}+b_2\widehat{r}^2+b_4\widehat{r}^4+b_6\widehat{r}^6$ and $\widehat{\varepsilon}\approx 1+a_2\widehat{r}^2+a_4\widehat{r}^4+a_6\widehat{r}^6$,
we derive the SSS like Eq.\,(\ref{io-3}),
\begin{align}\label{ef-1}
s^2/s_{\rm{c}}^2\approx&
1+\frac{2}{b_2}\left(b_4-s_{\rm{c}}^2a_4\right)\widehat{r}^2\notag\\
&+\frac{3}{b_2}\left[
\left(b_6-s_{\rm{c}}^2a_6\right)-\frac{4}{3}\frac{a_4}{a_2}\left(b_4-s_{\rm{c}}^2a_4\right)
\right]\widehat{r}^4.
\end{align}
The coefficient $a_6$ could be eliminated as $a_4$ in Eq.\,(\ref{io-3}): Expanding the reduced pressure $\widehat{P}=\sum_{k=1}^Kd_k\widehat{\varepsilon}^k$ with $\widehat{\varepsilon}=1+\sum_{k=1}^Ka_k\widehat{r}^k$ and comparing the coefficient of $\widehat{r}^6$ in the expansion $\widehat{P}=\widehat{P}_{\rm{c}}+\sum_{k=1}^Kb_k\widehat{r}^k$.
The result is,
\begin{equation}\label{a6b6}
b_6-s_{\rm{c}}^2a_6=a_2^3J+2a_2a_4D,
\end{equation}
where $J$ is a quantity constructed from $d_k$'s defined in the expansion $\widehat{P}=\sum_{k=1}^Kd_k\widehat{\varepsilon}^k$,
\begin{equation}\label{ref-J}
J=\sum_{k=1}^K\frac{k(k-1)(k-2)}{6}d_k
=d_3+4d_4+10d_5+\cdots.
\end{equation}
The coefficient $J$ has certain randomness like the coefficient $D$ (characterizing the uncertainties of the dense matter EOS).
Using this $J$, we can rewrite the radial variation of $s^2$ as,
\begin{equation}\label{ef-2}
s^2(\widehat{r})\approx s_{\rm{c}}^2+2a_2D\widehat{r}^2+\left(3a_2^2J+2a_4D\right)\widehat{r}^4.
\end{equation}

As discussed above, the peak of $s^2$ exists in the situation of $D<0$ and where $a_4>b_4/s_{\rm{c}}^2>0$ (see inequality (\ref{Da4})).
Taking the derivative $\d s^2/\d \widehat{r}$ and setting it to zero locates the peak of $s^2$ to
\begin{equation}\label{da-1}
\widehat{r}_{\rm{pk}}=\left(-\frac{a_2D}{3a_2^2J+2a_4D}\right)^{1/2}.
\end{equation}
Here $a_2D>0$, so the expression under square root is positive if $J<-2a_4D/3a_2^2$ (with the latter being positive since $a_4D<0$).
A further quantitative estimate for $\widehat{r}_{\rm{pk}}$ depends strongly on the coefficients $a_4$, $D$ and $J$.
Inverting $\widehat{\varepsilon}\approx 1+a_2\widehat{r}^2+a_4\widehat{r}^4+a_6\widehat{r}^6$ gives us $\widehat{r}^2\approx(\mu/a_2)[1-(a_4/a_2^2)\mu+(2a_4^2/a_2^4-a_6/a_2^3)\mu^2]$, where $\mu\equiv\widehat{\varepsilon}-1<0$.
Putting this $\widehat{r}^2$ into the full expression for $s^2(\widehat{r})$ to order $\widehat{r}^4$ of Eq.\,(\ref{ef-2}) enables us to write the $s^2(\widehat{\varepsilon})=s^2(\mu)$ as,
\begin{align}\label{smu}
s^2(\mu)\approx& s_{\rm{c}}^2+2D\mu
+3J\mu^2
-\frac{2}{a_2^2}\left(
{3a_4}J+\frac{a_6}{a_2}D
\right)\mu^3,
\end{align} 
which generalizes (\ref{s2eps}).
Thus we have to order $\mathcal{O}(\mu^3)$,
\begin{align}\label{Pmu3}
\widehat{P}(\mu)\approx&\widehat{P}_{\rm{c}}
+s_{\rm{c}}^2\mu+\left.\frac{1}{2}\frac{\d s^2}{\d\mu}\right|_{\mu=0}\mu^2
+\left.\frac{1}{6}\frac{\d^2s^2}{\d\mu^2}\right|_{\mu=0}\mu^3\notag\\
\approx&\widehat{P}_{\rm{c}}+s_{\rm{c}}^2\mu+D\mu^2+J\mu^3.
\end{align}

In fact, Eq.\,(\ref{Pmu3}) could be obtained straightforwardly via the expansion $\widehat{P}=\sum_{k=1}^Kd_k\widehat{\varepsilon}^k$.
However, the coefficient $J$ (Eq.\,(\ref{ref-J})) now like the coefficient $D$ (Eq.\,(\ref{ref-D})) is not totally random, but is constrained through the general requirements $0\leq s^2\leq 1$ and $\sigma_{\rm{c}}^2s_{\rm{c}}^2>2D$ used in the algorithm of Eq.\,(\ref{io-6}), i.e., the randomness of $J$ and $D$ is effectively reduced by the physical constraints.
Therefore, we can estimate $J$ in a similar manner as for $D$.
Taking $\widehat{P}_{\rm{c}}\approx0.24$ (for PSR J0740+6620) we obtain an estimate $J\approx-2.7$ (panel (b) of FIG.\,\ref{fig_AD}).
Combining $s_{\rm{c}}^2\approx0.45$ and $D\approx-0.45$ (panel (b) of FIG.\,\ref{fig_AD}) for $\widehat{P}_{\rm{c}}\approx0.24$, the $\widehat{P}(\widehat{\varepsilon})$ is found to be an increasing function of $\widehat{\varepsilon}$ for $\widehat{\varepsilon}\gtrsim0.7$. It basically excludes the possibility of having a plateau in $\widehat{P}$ at high densities near the center.
Including the last term in Eq.\,(\ref{smu}) may slightly shift $\widehat{\varepsilon}\gtrsim0.7$ to a higher value, here $a_6$ could be estimated via Eq.\,(\ref{a6b6}) with the expression for coefficient $b_6$ given in Ref.\,\cite{CLZ23}.
Moreover, the $D$-term and $J$-term contribute to $s^2(\mu)$ (and also $\widehat{P}(\mu)$) with different signs (since $\mu<0$) for $\widehat{P}_{\rm{c}}\lesssim0.3$ (panel (b) of FIG.\,\ref{fig_AD}), explaining the existence of a peak in $s^2(\mu)$.
One finds the peak of $s^2(\mu)$ from Eq.\,(\ref{smu}) (neglecting the last term) as,
\begin{equation}\label{s2pk}
\mu_{\rm{pk}}\approx-D/3J.
\end{equation}  
We can see that the possible peak in the derivative part of $s^2$ is shifted to a slightly lower value (see Eq.\,(\ref{tpk})).

\begin{figure}[h!]
\centering
\includegraphics[width=6.4cm]{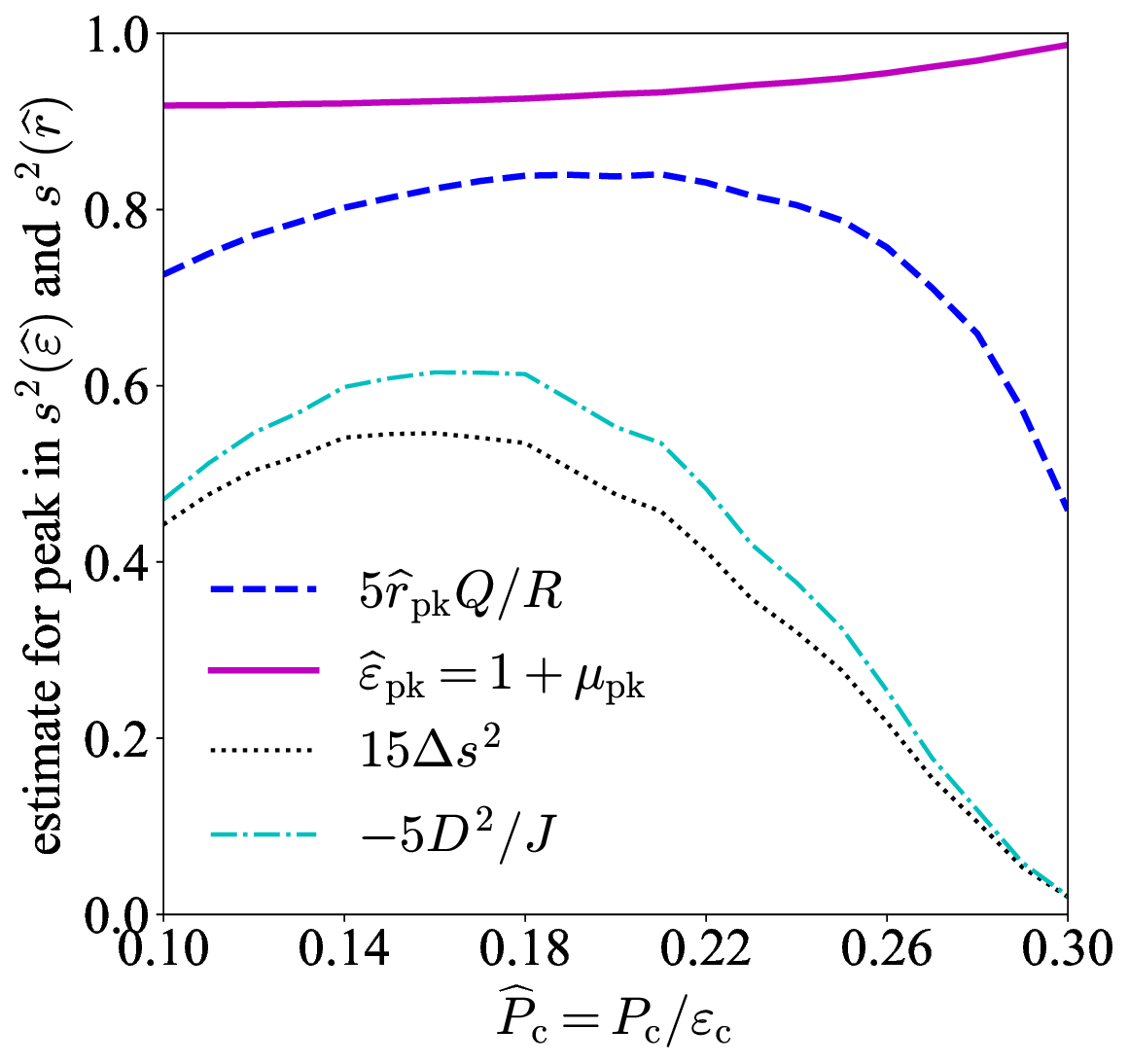}
\caption{Estimate for the peak in $s^2$ either for $s^2(\widehat{\varepsilon})$ or $s^2(\widehat{r})$,
the enhancement on $s^2$, i.e., $\Delta s^2=s^2(\widehat{r}_{\rm{pk}})-s_{\rm{c}}^2=-D^2/(3J+2a_4D/a_2^2)$ and its approximation $15\cdot(-D^2/3J)=-5D^2/J$ are shown.  The coefficient $D$ becomes averagely negative when $\widehat{P}_{\rm{c}}\gtrsim0.3$ (see FIG.\,\ref{fig_AD}).
In the simulation, the effective truncation order of the expansions is set at $K=8$.}
\label{fig_s2peak}
\end{figure}

As an illustration, we take $\varepsilon_{\rm{c}}\approx901\,\rm{MeV}/\rm{fm}^3$ (which is the central energy density for PSR J0740+6620)\,\cite{CLZ23} and $R\approx 12.39\,\rm{km}$\,\cite{Riley21} while  let $\widehat{P}_{\rm{c}}$ be a free parameter to demonstrate the numerical results.  The radius length is obtained as $Q=(4\pi G\varepsilon_{\rm{c}})^{-1/2}\approx8.7\,\rm{km}$. Shown in FIG.\,\ref{fig_s2peak} are our estimates for the peak in $s^2$ for either $s^2(\widehat{\varepsilon})$ or $s^2(\widehat{r})$.
Specifically, $\widehat{\varepsilon}_{\rm{pk}}=1+\mu_{\rm{pk}}\approx1-D/3J$ and $\widehat{r}_{\rm{pk}}$ is given by Eq.\,(\ref{da-1}).
The enhancement on $s^2$, i.e., $\Delta s^2=s^2(\widehat{r}_{\rm{pk}})-s_{\rm{c}}^2=-D^2/(3J+2a_4D/a_2^2)$ from Eq.\,(\ref{ef-2}) is also shown, where $a_4\approx1$ is adopted.
Considering that $|2a_4D/a_2^2|\ll|3J|$, we have $\Delta s^2\approx-D^2/3J$ using Eq.\,(\ref{s2pk}) (see the cyan dash-dotted line in FIG.\,\ref{fig_s2peak} for $15\cdot(-D^2/3J)=-5D^2/J$).
The difference between $\Delta s^2$ using Eq.\,(\ref{ef-2}) and Eq.\,(\ref{smu}) is that 
Eq.\,(\ref{ef-2}) is full to order $\widehat{r}^4$ while Eq.\,(\ref{smu}) is an approximation (even the $\mu^3$-term is included).
It is seen that as the $\widehat{P}_{\rm{c}}$ increases, the $\widehat{\varepsilon}_{\rm{pk}}$ eventually approaches 1, implying that the peak $\widehat{\varepsilon}_{\rm{pk}}$ eventually moves to the center, and it may even disappear if the $\widehat{P}_{\rm{c}}$ increases further. 
Actually for $\widehat{P}_{\rm{c}}\gtrsim0.3$ (see panel (b) of FIG.\,\ref{fig_AD}), the $D$ changes from being negative to positive and therefore both terms $2D\mu$ and $3J\mu^2$ in $s^2(\mu)$ are negative,
i.e., $s_{\rm{c}}^2$ is larger than its surroundings.
 Similar phenomena occur for the $\Delta s^2$ and $\widehat{r}_{\rm{pk}}Q/R$. For example, we have for $\widehat{P}_{\rm{c}}\to0.374$ that $s_{\rm{c}}^2\to1$ (Eq.\,(\ref{sc2})) and therefore $\Delta s^2\to0$ (no space under this limit for $s^2$ to be enhanced when going outward from the center).
In our illustration,  we find $(\Delta s^2)_{\max}\approx0.038$ occurring for $\widehat{P}_{\rm{c}}\approx0.16$, therefore $(\Delta s^2)_{\max}/s_{\rm{c}}^2\approx14\%$ using $s_{\rm{c}}^2\approx0.26$.
Similarly, we have $\Delta s^2/s_{\rm{c}}^2\approx5\%$ for $\widehat{P}_{\rm{c}}\approx0.24$.

For situations with positive $D$ (thus $a_2D<0$), the SSS may firstly decrease as finite $\widehat{r}$ develops and then increase further if $3a_2^2J+2a_4D>0$.
There exists then a valley $\widehat{r}_{\rm{va}}$ in $s^2$, which may still be estimated by using Eq.\,(\ref{da-1}).
The self-consistent requirement between expansions (\ref{ef-1}) and (\ref{ef-2}) considering the coefficients of $\widehat{r}^4$ gives us (see the similar inequality (\ref{Da4}) from the preceding order),
\begin{equation}\label{ef-3}
\left[\frac{4a_4}{3a_2}\left(\frac{b_4}{s_{\rm{c}}^2}-a_4\right)
-\left(\frac{b_6}{s_{\rm{c}}^2}-a_6\right)\right]\left(J+\frac{2a_4}{3a_2^2}D\right)>0,
\end{equation}
where $b_4$ and $b_6$ were given in the Appendix of Ref.\,\cite{CLZ23} and $a_4$ as well as $a_6$ are expected to be $\mathcal{O}(1)$.

The above results/estimates on the peak of $s^2$ as well as the qualitative predictions for the $\widehat{P}(\widehat{\varepsilon})=\widehat{P}(\mu)$ are expected to be (nearly) model independent, because we rely only on (without using extra assumptions) the general conditions/requirements (including the inequality $0\leq s^2\leq1$ of (\ref{oo-1}),  inequality (\ref{io-5}) and the sum rules of (\ref{io-4-s2})) together with formula (\ref{sc2}) and also bound the high-order coefficients such as $a_4$ (Appendix \ref{appB}) and $a_6$, etc.

It is necessary to emphasize that due to its perturbative nature, however, the above analysis on the $\widehat{\varepsilon}$-dependence of $s^2$ (or the signature of the continuous crossover) is only effective and qualitatively reasonable at finite distances $\widehat{r}$ being very close to NS centers ($\widehat{r}=0$).
Whether there exist sharp PTs at some large distances away from the centers (or at certain intermediate energy density $\varepsilon$) could not be excluded immediately and should be in principle analyzed by including more higher-order terms in the expansions,  and the results may have certain model dependence.

Furthermore, $\widehat{P}(\mu)<\widehat{P}_{\rm{c}}$ is definite for $\mu\approx0^-$ considering Eq.\,(\ref{Pmu3}) which is fundamentally different from $s^2(\mu)$ of Eq.\,(\ref{smu}), since $D$ still has sizable probabilities to be positive even for small $\widehat{P}_{\rm{c}}$ and thus $s^2(\mu)<s_{\rm{c}}^2$ (panel (c) of FIG.\,\ref{fig_sr2_sk}).
We discuss in further details in the next section on the possible bound on $P/\varepsilon$ in NS cores.

\section{Upper Bound for $P/\varepsilon$ and Conformal Anomaly near Neutron Star Centers}\label{SEC_BRP}

In this section, we discuss more on $P/\varepsilon$ in NS cores.
Considering the trace anomaly (or the conformality measure) $\Delta=1/3-P/\varepsilon$ and compared with the analysis given in Ref.\,\cite{Fuji22}, our formula (\ref{sc2}) predicts a slightly different bound for $\Delta_{\rm{c}}$ as $\Delta_{\rm{c}}\gtrsim-0.04$, and for $\widehat{P}_{\rm{c}}\approx0.24$ (PSR J0740+6620) we found $\Delta_{\rm{c}}\approx0.09$.  
Although our $\Delta_{\rm{c}}$ and that from TA analysis are close to each other, the physical origins seem different. While the TA is based on properties of pQCD, ours could be traced back to the structures encapsulated in the (general-relativistic) TOV equations. Thus, whether there exists deep connection between them needs further investigations. 

Firstly, we mention a few similar evaluations of $\Delta$ in the recent literature. An analysis in Ref.\,\cite{Ecker23} using an agnostic EOS showed that $\Delta$ is very close to zero for $M_{\rm{NS}}^{\max}\gtrsim2.18\mbox{$\sim$}2.35M_{\odot}$ and may be slightly negative for even more massive NSs (e.g., $\Delta\gtrsim-0.02$ for $M_{\rm{NS}}^{\max}\gtrsim2.52M_{\odot}$).
Moreover, in Ref.\,\cite{Tak23}, the central minimum value of $\Delta$ is found to be about $\Delta_{\min}\approx0.04$ using the NICER data together with the tidal deformability from GW170817.
A value of $\Delta_{\min}\approx-0.05$ was inferred considering additionally the second component of GW190814 as a neutron star with mass about $2.59M_{\odot}$\,\cite{Abbott2020} using two hadronic EOS models\,\cite{Tak23}. See also the recent works on the similar issue in Refs.\,\cite{Mus23,Pro23}, e.g., Ref.\,\cite{Mus23} predicted that $\Delta_{\rm{c}}\gtrsim-0.046$ if $M_{\rm{NS}}^{\max}\geq2.2M_{\odot}$.
Another analysis within the Bayesian framework considering the state-of-the-art theoretical calculations showed that $\Delta\gtrsim-0.01$ (where $M_{\rm{NS}}^{\max}\approx2.27_{-0.11}^{+0.11}M_{\odot}$) and for a $2M_{\odot}$ NS the polytropic index $\gamma\equiv\d\ln P/\d\ln\varepsilon=s^2/(P/\varepsilon)$ is found to be about 2\,\cite{Ann23}.
The $\gamma$ parameter in NS centers, actually,  can not be unity\,\cite{Fuji22,Ann23} since $\Delta_{\rm{c}}$ ($\widehat{P}_{\rm{c}}$) and $s_{\rm{c}}^2$ could not approach zero (1/3) and 1/3 simultaneously, as shown by Eq.\,(\ref{sc2}).
Physically, due to the nonlinear nature of the central EOS in NSs, it is likely that $\gamma\neq1$ (actually $\gamma=1$ if and only if $P\propto\varepsilon$),  see the relevant discussions in detail in Section \ref{SEC4} and FIG.\,\ref{fig_MR-C}.
Using our formula (\ref{sc2}), it is straightforward to find that $4/3\leq\gamma_{\rm{c}}\lesssim2.67$ and considering $\widehat{P}_{\rm{c}}\approx0.24_{-0.07}^{+0.05}$ for PSR J0740+6620\,\cite{CLZ23}, we obtain $\gamma_{\rm{c}}\approx1.86_{-0.21}^{+0.22}$.
Moreover, for $\Delta_{\rm{c}}=0$ the $\gamma_{\rm{c}}$ index is $7/3\approx2.33$.
Furthermore, Ref.\,\cite{Ann23} also defined a quantity $\Theta\equiv [\Delta^2+(P/\varepsilon-s^2)^2]^{1/2}$ to measure the conformality and found it should be $\lesssim0.65$ for all NSs.
The criterion $\Theta\lesssim0.2$ (besides $\gamma\lesssim1.75$) was adopted\,\cite{Ann23} to identify the nearly-conformality at a given density.
Based on our formula (\ref{sc2}), we can find $0.19\lesssim\Theta_{\rm{c}}\lesssim0.63$ with the minimum/maximum value obtained at $\widehat{P}_{\rm{c}}\approx0.18$/$\widehat{P}_{\rm{c}}\approx0.374$ and $\Theta_{\rm{c}}\approx0.22_{-0.03}^{+0.09}$ for $\widehat{P}_{\rm{c}}\approx0.24_{-0.07}^{+0.05}$ (PSR J0740+6620).

\renewcommand*\tablename{\small TAB.}
\begin{table}[tbh!]
\centering{
\begin{tabular}{|c|c|c|c|c|} 
  \hline
Quantity&Range&$\widehat{P}_{\rm{c}}\approx0.18$&$0.24$&$0.374$\\\hline\hline
       $\Delta_{\rm{c}}=1/3-\widehat{P}_{\rm{c}}$ &$-0.041\lesssim\Delta_{\rm{c}}\leq1/3$&$0.15$&$0.09$&$-0.041$\\\hline
       $s_{\rm{c}}^2=\d{P}_{\rm{c}}/\d\varepsilon_{\rm{c}}$&$0\leq s_{\rm{c}}^2\leq1$&$0.31$&$0.45$&$1$\\\hline
       $\gamma_{\rm{c}}=s_{\rm{c}}^2/\widehat{P}_{\rm{c}}$&$4/3\leq\gamma_{\rm{c}}\lesssim2.67$&$1.68$&1.86&$2.67$\\\hline
            $t_{\rm{c}}=\widehat{P}_{\rm{c}}- s_{\rm{c}}^2$&$-0.63\lesssim t_{\rm{c}}\leq0$&$-0.13$&$-0.21$&$-0.63$\\\hline
       $\Theta_{\rm{c}}=(\Delta_{\rm{c}}^2+t_{\rm{c}}^2)^{1/2}$&$0.19\lesssim\Theta_{\rm{c}}\lesssim0.63$&$0.19$&0.22&$0.63$\\\hline
       $u_{\rm{c}}=2s_{\rm{c}}^2-3\widehat{P}_{\rm{c}}$&$-0.01\lesssim u_{\rm{c}} \lesssim0.88$&$0.07$&$0.17$&$0.88$\\\hline
    \end{tabular}}
\caption{Ranges of several quantities relevant for the trace anomaly as defined in the text, the values for them at three reference $\widehat{P}_{\rm{c}}$ are also shown (last three columns), $t_{\rm{c}}=\d\Delta_{\rm{c}}/\d\ln\varepsilon_{\rm{c}}=\widehat{P}_{\rm{c}}-s_{\rm{c}}^2$ is the logarithmic
derivative of $\Delta_{\rm{c}}$, and $u_{\rm{c}}= 2s_{\rm{c}}^2-3\widehat{P}_{\rm{c}}$ is the quantity closely related to the possible crossover occurring near the NS centers (see Eq.\,(\ref{ref-D})).
}
\label{sstab}
\end{table}

\begin{figure}[h!]
\centering
\includegraphics[width=6.4cm]{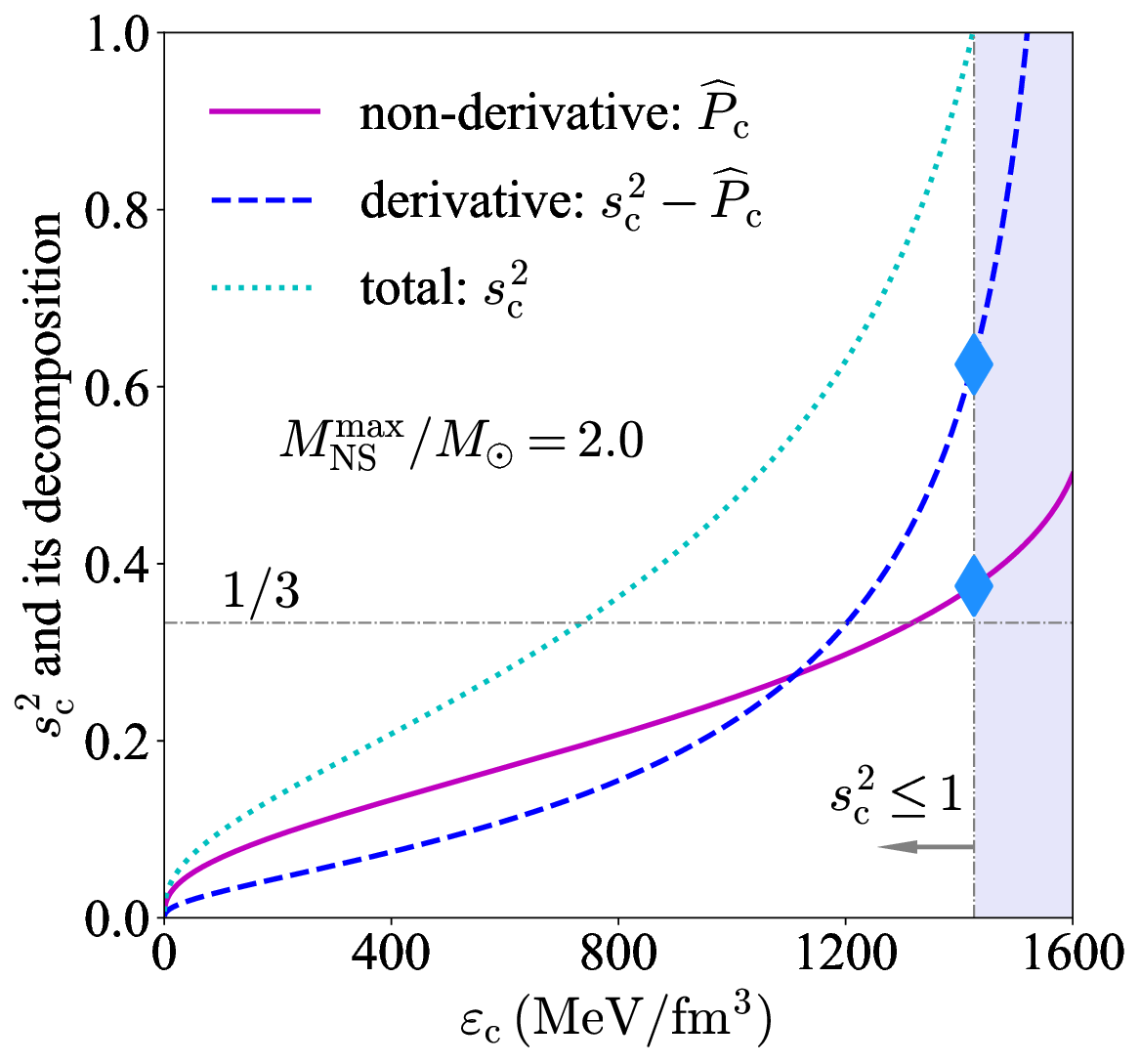}
\caption{Decomposition of $s_{\rm{c}}^2$ into its non-derivative part ($\widehat{P}_{\rm{c}}$) and derivative part ($s_{\rm{c}}^2-\widehat{P}_{\rm{c}}$) for $M_{\rm{NS}}^{\max}/M_{\odot}=2$.
Solid diamonds characterize the boundary of $s_{\rm{c}}^2\leq1$ (left of the lavender band).
}\label{fig_tc}
\end{figure}

We summarize in TAB.\,\ref{sstab} these values at three reference $\widehat{P}_{\rm{c}}$ (0.18, 0.24 and 0.374), where $t_{\rm{c}}\equiv\d\Delta_{\rm{c}}/\d\ln\varepsilon_{\rm{c}}=\widehat{P}_{\rm{c}}-s_{\rm{c}}^2$ as the logarithmic
derivative of $\Delta_{\rm{c}}$\,\cite{Ann23} (with respect to $\varepsilon_{\rm{c}}$) is also given. We have $t_{\rm{c}}\to0$ if $\gamma_{\rm{c}}\to1$ (the so-called conformal limit of matter), therefore $-t_{\rm{c}}$ characterizes the deviation from the conformal limit.  In Ref.\,\cite{Fuji22}, $-t=s^2-P/\varepsilon$ and $1/3-\Delta=P/\varepsilon$ are decomposed as the derivative and non-derivative parts of $s^2$, respectively.
Equivalently,  one has $\Theta=(\Delta^2+t^2)^{1/2}$.
See FIG.\,\ref{fig_tc} for an example where $M_{\rm{NS}}^{\max}/M_{\odot}=2$ is adopted for the illustration (using Eq.\,(\ref{Mmax-G}) to solve for $\widehat{P}_{\rm{c}}$ and $s_{\rm{c}}^2$).
According to Eq.\,(\ref{sc2}), we find both the non-derivative and derivative parts of $s_{\rm{c}}^2$ are increasing functions of $\varepsilon_{\rm{c}}$, and take their maximum values at $\widehat{P}_{\rm{c}}\approx0.374$.
The solid diamonds in FIG.\,\ref{fig_tc} characterize the boundary $\widehat{P}_{\rm{c}}\lesssim0.374$,  i.e., the curves above $\varepsilon_{\rm{c}}\gtrsim1.42\,\rm{GeV}/\rm{fm}^3$ (for $M_{\rm{NS}}^{\max}/M_{\odot}=2$, see TAB.\,\ref{tab_ult}) violate the causality condition $s_{\rm{c}}^2\leq1$.
For other values of $M_{\rm{NS}}^{\max}$ the shapes of the curves in FIG.\,\ref{fig_tc} are similar.
Shortly after, we may show that there exists a peak in the derivative part of $s^2$ when considering positions having finite distance from the center (see Eq.\,(\ref{tpk})).

\begin{figure}[h!]
\centering
\includegraphics[width=6.4cm]{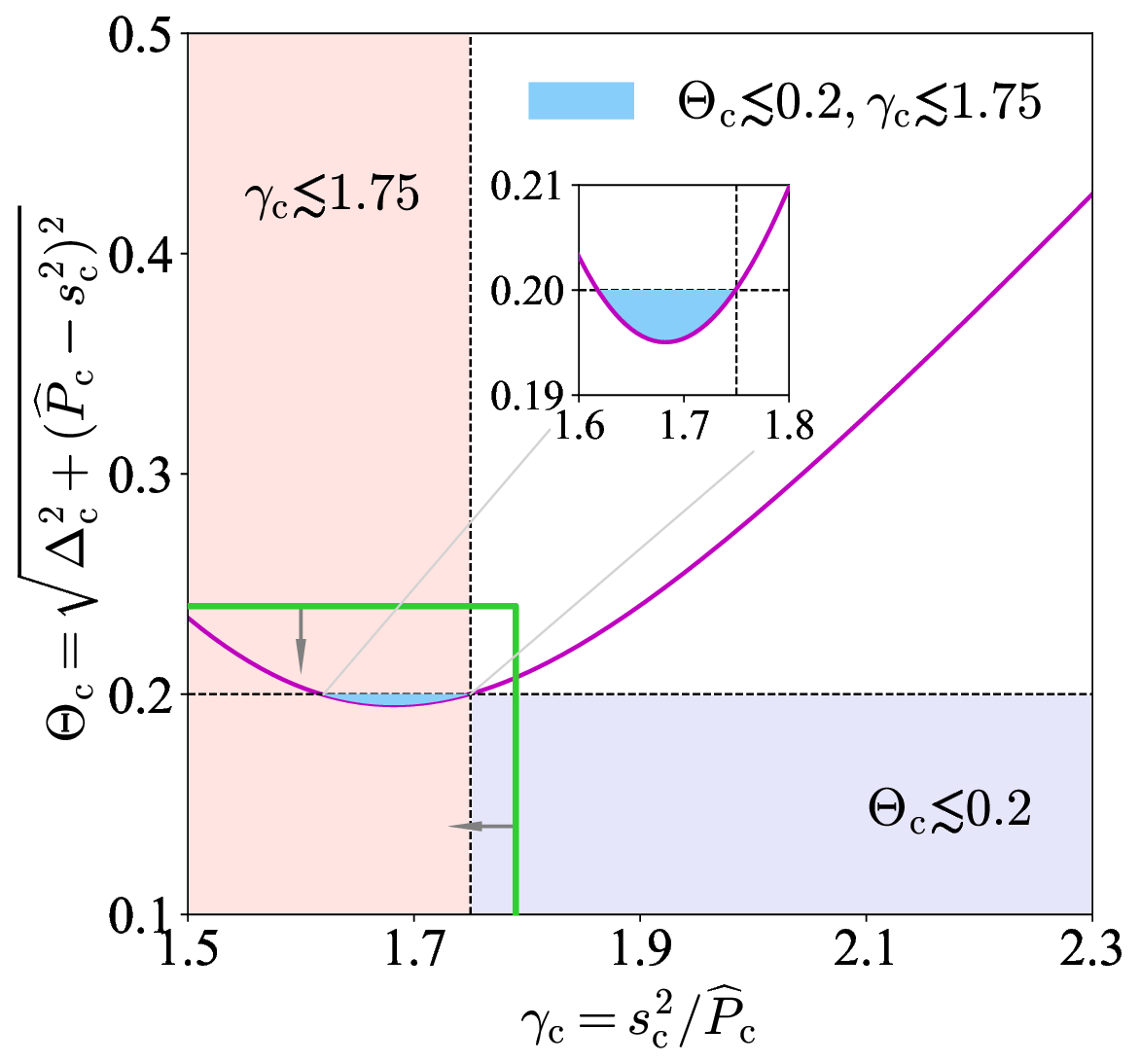}
\caption{Dependence of $\Theta_{\rm{c}}$ on the polytropic index $\gamma_{\rm{c}}$ for the maximum-mass configuration $M_{\rm{NS}}^{\max}$, where in the lightblue region two criteria $\Theta_{\rm{c}}\lesssim0.2$ and $\gamma_{\rm{c}}\lesssim1.75$ hold simultaneously.
The constraints from Ref.\,\cite{Ann23} on $\gamma$ and $\Theta$ are also shown (green solid lines associated with arrows) for comparison.
The inset amplifies the lightblue region.
}\label{fig_Th-g}
\end{figure}

In TAB.\,\ref{sstab}, the column with $\widehat{P}_{\rm{c}}\approx0.18$ is also shown as the measure $\Theta_{\rm{c}}$ takes its minimum at this $\widehat{P}_{\rm{c}}$.
Interestingly, if one requires $\Theta_{\rm{c}}\approx0.2$, then two solutions $\widehat{P}_{\rm{c}}\approx0.21$ and $\widehat{P}_{\rm{c}}\approx0.16$ should be obtained,  the corresponding $\gamma_{\rm{c}}$ is found to be about 1.75 and 1.62, respectively.
We have approximately $\Theta_{\rm{c}}\approx1/3-\eta+2\eta^2/3+13\eta^3/6+41\eta^4/8+\cdots$, where $0.5\gtrsim\eta\equiv1-4/3\gamma_{\rm{c}}\geq0$ acts as a small-expansion quantity (since $2.67\gtrsim\gamma_{\rm{c}}\geq4/3$).
These numerical values indicate that empirically the near-conformality may be possible in the core of very massive NSs (with masses $\gtrsim2M_{\odot}$). 
We notice that $\widehat{P}_{\rm{c}}\approx0.189$\,\cite{CLZ23} and correspondingly $\Theta_{\rm{c}}\approx0.195$ and $\gamma_{\rm{c}}\approx1.70$ are obtained if the radius about $13.7\,\rm{km}$ was adopted for PSR J0740+6620\,\cite{Miller21}.
However, the likelihood for realizing conformality near the centers is small if the criteria $\Theta_{\rm{c}}\lesssim0.2$ and $\gamma_{\rm{c}}\lesssim1.75$\,\cite{Ann23} were adopted simultaneously, as shown in FIG.\,\ref{fig_Th-g}, where the constraints $\gamma\lesssim1.79$ and $\Theta\lesssim0.24$ from Ref.\,\cite{Ann23} are shown (green solid lines associated with arrows) for comparison.
Moreover, Eq.\,(\ref{gr-1})$\sim$Eq.\,(\ref{TGG}) given in the following together may extend this conclusion to situations where $\widehat{\varepsilon}\lesssim1$ (or equivalently at some finite distances away from the center).
Furthermore, FIG.\,\ref{fig_Th-g} also supports the conclusion that the criterion $\Theta_{\rm{c}}\lesssim0.2$ is more restrictive than $\gamma_{\rm{c}}\lesssim1.75$\,\cite{Ann23}.

\begin{figure}[h!]
\centering
\includegraphics[width=6.4cm]{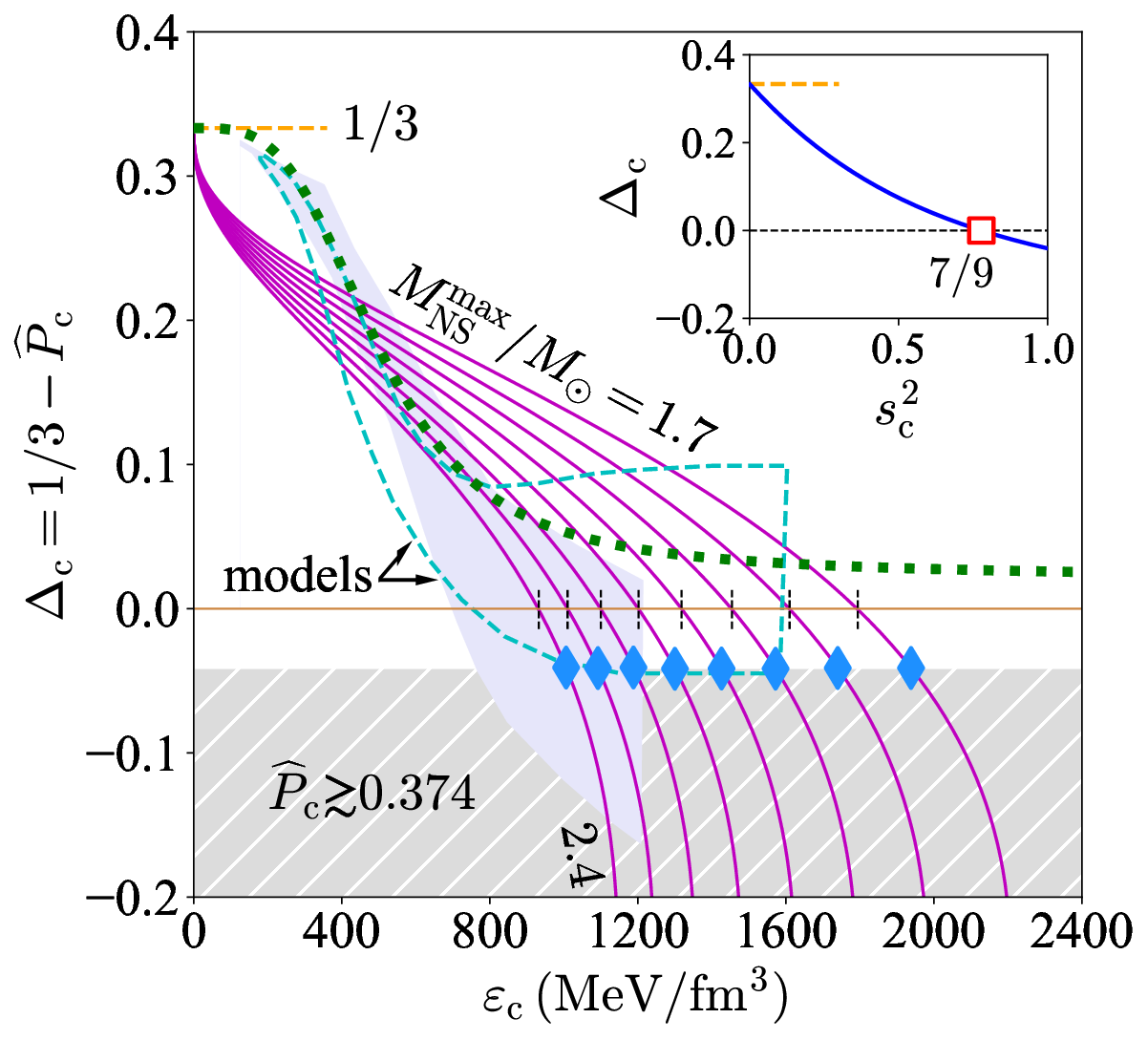}
\caption{Conformality measure $\Delta_{\rm{c}}=1/3-\widehat{P}_{\rm{c}}$ as a function of $\varepsilon_{\rm{c}}$ for different $M_{\rm{NS}}^{\max}/M_{\odot}$ (from 1.7 to 2.4), where lightblue solid diamonds mark the points set by $\widehat{P}_{\rm{c}}\lesssim0.374$ while the vertical black dashed lines are for the vanishing points of $\Delta_{\rm{c}}$ for each $M_{\rm{NS}}^{\max}$.
Two model calculations (lavender/cyan)\,\cite{Fuji22,Gorda22} and a parametrization for $\Delta$ (green dotted)\,\cite{Fuji22} are also shown for comparison.
Inset gives the dependence of $\Delta_{\rm{c}}$ on $s_{\rm{c}}^2$.
}\label{fig_TAc}
\end{figure}

In FIG.\,\ref{fig_TAc}, we show the $\Delta_{\rm{c}}=1/3-\widehat{P}_{\rm{c}}$ as a function of $\varepsilon_{\rm{c}}$ for different values of $M_{\rm{NS}}^{\max}$ (varying from 1.7$M_{\odot}$ to $2.4M_{\odot}$) using the mass correlation (\ref{Rmax-n}). The parametrization for $\Delta$ suggested by Ref.\,\cite{Fuji22} is also shown (green dotted line), which is positive-definite by construction.
In addition, two model predictions for $\Delta$ are represented by lavender\,\cite{Fuji22} and cyan dashed\,\cite{Gorda22} bands using machine-learning algorithms and ab-initio QCD calculations, respectively.
For relatively light NSs, the condition $\Delta_{\rm{c}}\geq0$ generally holds since the central energy density $\varepsilon_{\rm{c}}$ is relatively low, e.g., for $M_{\rm{NS}}^{\max}/M_{\odot}=1.7$ the condition $\varepsilon_{\rm{c}}\lesssim1.8\,\rm{GeV}/\rm{fm}^3$ is safely satisfied. On the other hand,  the $\varepsilon_{\rm{c}}$ may exceed the value set by $\Delta_{\rm{c}}\approx0$ as $M_{\rm{NS}}^{\max}$ increases. 
Our constraint on the lower bound of $\Delta_{\rm{c}}$ is quite consistent with those from Refs.\,\cite{Fuji22,Gorda22}.
The inset of FIG.\,\ref{fig_TAc} shows the dependence of $\Delta_{\rm{c}}$ on $s_{\rm{c}}^2$. It is seen that for $s_{\rm{c}}^2\geq7/9\approx0.78$ (see Eq.\,(\ref{sc2})) the $\Delta_{\rm{c}}$ becomes negative. In short, FIG.\,\ref{fig_TAc} clearly explains why the criterion $\Delta\geq0$ tends to break down for massive NSs with increasing $M_{\rm{NS}}^{\max}$\,\cite{Ecker23,Tak23,Ann23,Brandes23-a}, instead of for light NSs.

We have studied above the consequences of $\widehat{P}_{\rm{c}}$ on a few quantities.
We now investigate to which extent the limit $\widehat{P}_{\rm{c}}\lesssim0.374$ holds.
In order to bound $P/\varepsilon=\widehat{P}/\widehat{\varepsilon}$ (notice that the hatted pressure and energy density are both scaled with $\varepsilon_{\rm{c}}$), we need to take three generalizations of $\widehat{P}_{\rm{c}}\lesssim0.374$ obtained from Eq.\,(\ref{sc2}) by studying the following questions:
\begin{enumerate}[label=(\alph*),leftmargin=*]
\item How does the $\widehat{P}/\widehat{\varepsilon}$ behave at a finite $\widehat{r}$ for the maximum-mass configuration $M_{\rm{NS}}^{\max}$?
\item How does the limit $\widehat{P}_{\rm{c}}\lesssim0.374$ modify when we consider stable NSs on the M-R curve away from the maximum-mass configuration?
\item By combining (a) and (b), how does the $P/\varepsilon$ behave for stable NSs at finite distances $
\widehat{r}$ away from their centers?
\end{enumerate}

For the first question,  since the pressure $\widehat{P}$ and $\widehat{\varepsilon}$ are decreasing functions of $\widehat{r}$, i.e.,
\begin{align}
\widehat{P}\approx& \widehat{P}_{\rm{c}}+b_2\widehat{r}^2<\widehat{P}_{\rm{c}},\label{pk-1}\\
\widehat{\varepsilon}\approx&\widehat{\varepsilon}_{\rm{c}}+a_2\widehat{r}^2=1+s_{\rm{c}}^{-2}b_2\widehat{r}^2<1=\widehat{\varepsilon}_{\rm{c}},\label{pk-2}
\end{align}
we obtain by taking their ratio
\begin{align}
{P}/{\varepsilon}=
{\widehat{P}}/{\widehat{\varepsilon}}\approx&{\widehat{P}_{\rm{c}}}/{\widehat{\varepsilon}_{\rm{c}}}+\left(1-\frac{\widehat{P}_{\rm{c}}}{s_{\rm{c}}^2}\right)b_2\widehat{r}^2
\approx
\widehat{P}_{\rm{c}}+\frac{1}{4}b_2\widehat{r}^2<\widehat{P}_{\rm{c}}.\label{pk-3}
\end{align}
Generally, $1-\widehat{P}_{\rm{c}}/s_{\rm{c}}^2>0$ and the approximation $s_{\rm{c}}^2\approx4\widehat{P}_{\rm{c}}/3$ is used for small $\widehat{P}_{\rm{c}}$ in the last step.
This means not only $\widehat{P}$ and $\widehat{\varepsilon}$ decrease for finite $\widehat{r}$,  but also does their ratio $\widehat{P}/\widehat{\varepsilon}$.
Thus for NSs at the maximum-mass configuration of the M-R curves, we have $\widehat{P}/\widehat{\varepsilon}\leq\widehat{P}_{\rm{c}}\lesssim0.374$.
Considering the second question and for stable NSs on the M-R curve, one has $\Psi\sim\d M_{\rm{NS}}/\d\varepsilon_{\rm{c}}>0$ and Eq.\,(\ref{sc2}) should be modified to
\begin{equation}\label{io-11}
s_{\rm{c}}^2=\widehat{P}_{\rm{c}}\left(1+\frac{1+\Psi}{3}\frac{1+3\widehat{P}_{\rm{c}}^2+4\widehat{P}_{\rm{c}}}{1-3\widehat{P}_{\rm{c}}^2}\right).
\end{equation}
Consequently,  the upper limit for $\widehat{P}_{\rm{c}}$ constrained by $s_{\rm{c}}^2\leq1$ should be smaller than $0.374$ due to the positiveness of $\Psi$.
Furthermore, for the last question (c),  the inequality (\ref{pk-3}) still holds and is modified for small $\widehat{P}_{\rm{c}}$,
\begin{equation}\label{pk-3a}
 \widehat{P}/\widehat{\varepsilon}\approx\widehat{P}_{\rm{c}}+\frac{1+\Psi}{4+\Psi}b_2\widehat{r}^2<\widehat{P}_{\rm{c}}.
 \end{equation}

Combining the above three aspects, we find that $P/\varepsilon=\widehat{P}/\widehat{\varepsilon}\leq\widehat{P}_{\rm{c}}\lesssim0.374$ holds for all stable NSs along the M-R curve either at or being close to their centers. Nevertheless, the validity of this conclusion is still limited to small $\widehat{r}$ (like the SSS) due to the perturbtive nature of the expansions of $\widehat{P}(\widehat{r})$ and $\widehat{\varepsilon}(\widehat{r})$. Whether $P/\varepsilon$ could exceed such upper limit at even larger distances away from the centers depends on the joint analysis of $s^2$ and $P/\varepsilon$,  by including more higher order contributions of the expansions.

The upper bound $P/\varepsilon\lesssim0.374$ (at least near the NS centers) is an intrinsic property of the TOV equations, which embody the strong-field aspects of gravity (GR).
In this sense, there is no guarantee {\it a prior} that this bound is consistent with all microscopic nuclear EOSs. This is mainly because the latter were conventionally constructed without considering the strong-field ingredients of gravity,  see, e.g.,  the sizable tension between our causality boundary on the M-R diagram of FIG.\,\ref{fig_MR-C} (solid black line) and predictions of a few nuclear EOSs.
The robustness of such upper bound for $P/\varepsilon$ can be checked only by observable astrophysical quantities/processes involving strong-field aspects of gravity such as NS M-R data,  NS-NS mergers and/or NS-black hole mergers\,\cite{Shibata2015}.

We use the perturbative expansion of $P/\varepsilon$ and Eq.\,(\ref{smu}) to explain the existence of a peak in the derivative part $-t$ of the SSS\,\cite{Fuji22}.
Dividing Eq.\,(\ref{Pmu3}) by $\mu=\widehat{\varepsilon}-1$ gives the ratio $P/\varepsilon$ to order $\mu^2$ as $P/\varepsilon\approx\widehat{P}_{\rm{c}}-t_{\rm{c}}\mu+(D+t_{\rm{c}})\mu^2$ where $t_{\rm{c}}=\widehat{P}_{\rm{c}}-s_{\rm{c}}^2$, this expansion generalizes Eq.\,(\ref{pk-3}).
Consequently, the derivative part $-t$ of $s^2$ is obtained as $-t(\widehat{\varepsilon})=-t(\mu)\approx-t_{\rm{c}}+(2D+t_{\rm{c}})\mu+(3J-D-t_{\rm{c}})\mu^2$ (to order $\mu^2$), from which the peak could be found,
\begin{equation}\label{tpk}
\mu_{\rm{pk}}^{(-t)}=\widehat{\varepsilon}_{\rm{pk}}^{(-t)}-1
=\frac{1}{2}\frac{2D+t_{\rm{c}}}{D+t_{\rm{c}}-3J}.
\end{equation}
Taking for example $\widehat{P}_{\rm{c}}\approx0.16\mbox{$\sim$}0.30$, this peak is shown to be located at about $-\mu_{\rm{pk}}^{(-t)}\approx10\%\mbox{$\sim$}5\%$ (see discussions given after Eq.\,(\ref{Pmu3}) for the estimate on $D$ and $J$).
On the other hand, using the expansion $P/\varepsilon$ just given here, we deduce that $t_{\rm{c}}/2(D+t_{\rm{c}})>0$, i.e., there would be no peak in the non-derivative part of $s^2$ or this part is an increasing function of $\varepsilon$\,\cite{Fuji22}.
The value of $-t$ at $\mu_{\rm{pk}}^{(-t)}$ is,
\begin{equation}
-t_{\rm{pk}}\equiv
-t(\mu_{\rm{pk}}^{(-t)})=\frac{4D^2+12Jt_{\rm{c}}-3t_{\rm{c}}^2}{4D-12J+4t_{\rm{c}}}.
\end{equation}
Then, we have $-t_{\rm{pk}}\approx0.15\mbox{$\sim$}0.35$ and the non-derivative part (of $s^2$), namely $P/\varepsilon\approx0.14\mbox{$\sim$}0.28$ for $\widehat{P}_{\rm{c}}\approx0.16\mbox{$\sim$}0.30$. Moreover, comparing $\mu_{\rm{pk}}^{(-t)}$ of Eq.\,(\ref{tpk}) and $\mu_{\rm{pk}}=-D/3J$ of $s^2(\mu)$ (see Eq.\,(\ref{s2pk})) leads us to 
\begin{equation}\label{tpk1}
\mu_{\rm{pk}}-\mu_{\rm{pk}}^{(-t)}=\frac{2D^2}{9J^2}
\frac{1+t_{\rm{c}}/D+3Jt_{\rm{c}}/2D^2}{1-D/3J-t_{\rm{c}}/3J}
>0.
\end{equation}
This means the location of the peak in $s^2$ (i.e., $\mu_{\rm{pk}}$) occurs at a higher energy density than that of the peak in $-t$ (i.e., $\mu_{\rm{pk}}^{(-t)}$), see also FIG.\,2 of Ref.\,\cite{Fuji22}.  Equivalently, the peak in $s^2$ is closer to the NS center than the peak in $-t$.
When higher order terms of $\mu$ are included in the expansions, the specific location of $\mu_{\rm{pk}}$ and $\mu_{\rm{pk}}^{(-t)}$ may vary, however the relation $\mu_{\rm{pk}}-\mu_{\rm{pk}}^{(-t)}>0$ would not be changed.

\begin{figure}[h!]
\centering
\includegraphics[width=6.4cm]{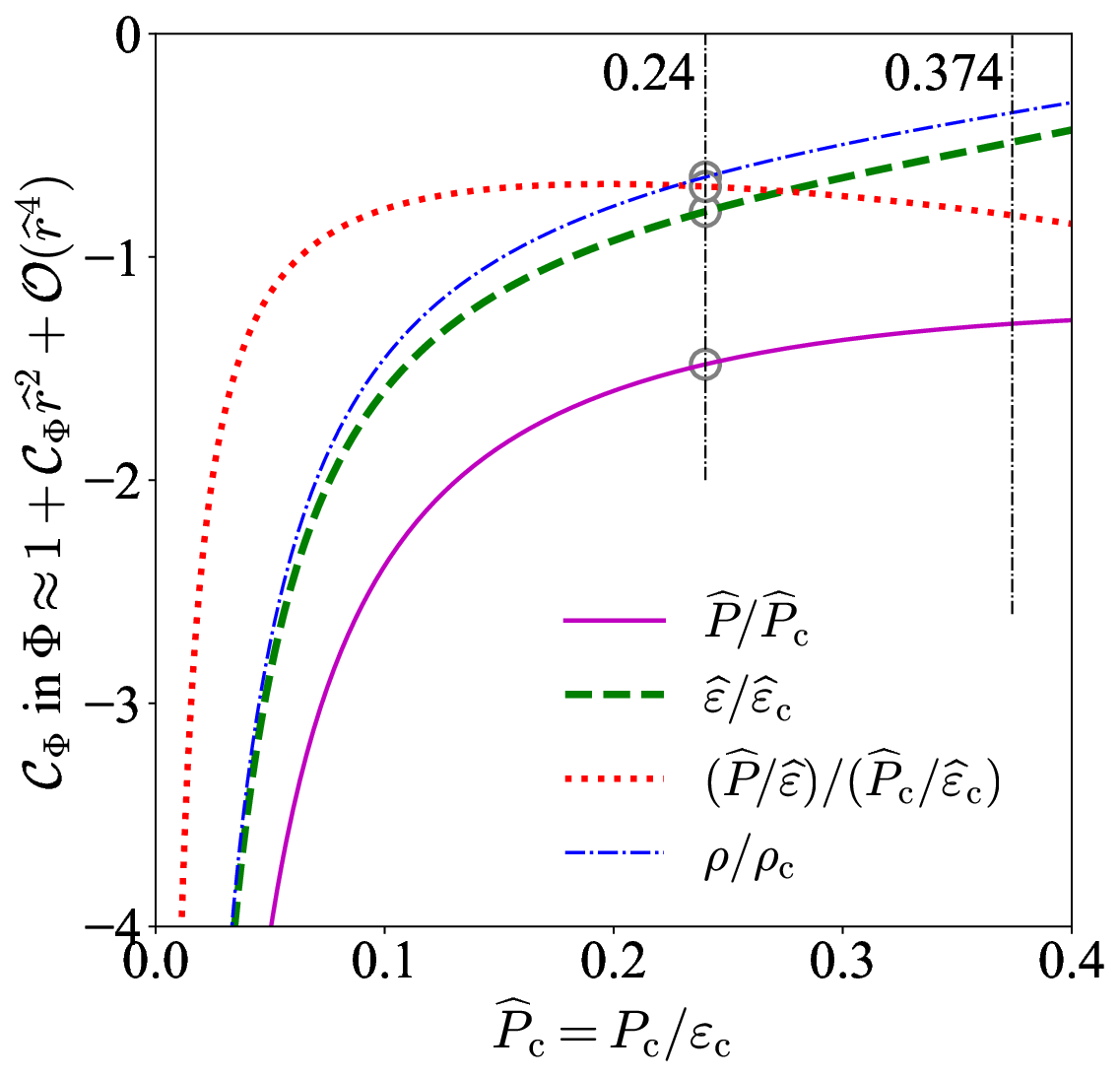}
\caption{The $\widehat{P}_{\rm{c}}$-dependence of coefficients $\mathcal{C}_{\Phi}$ appeared in $\Phi\approx 1+\mathcal{C}_{\Phi}\widehat{r}^2+\mathcal{O}(\widehat{r}^4)$ for $\Phi=\widehat{P}/\widehat{P}_{\rm{c}}$, $\widehat{\varepsilon}/\widehat{\varepsilon}_{\rm{c}}$, $(\widehat{P}/\widehat{\varepsilon})/(\widehat{P}_{\rm{c}}/\widehat{\varepsilon}_{\rm{c}})$ and $\rho/\rho_{\rm{c}}$.
}\label{fig_eps_P_rho}
\end{figure}

In FIG.\,\ref{fig_eps_P_rho}, we show the $\widehat{P}_{\rm{c}}$-dependence of coefficient $\mathcal{C}_{\Phi}$ appeared in the approximation $\Phi\approx 1+\mathcal{C}_{\Phi}\widehat{r}^2+\mathcal{O}(\widehat{r}^4)$ for $\Phi=\widehat{P}/\widehat{P}_{\rm{c}},~ \widehat{\varepsilon}/\widehat{\varepsilon}_{\rm{c}},~ (\widehat{P}/\widehat{\varepsilon})/(\widehat{P}_{\rm{c}}/\widehat{\varepsilon}_{\rm{c}})$ and $\rho/\rho_{\rm{c}}$, for the $M_{\rm{NS}}^{\max}$ configuration.
Here besides the relations for $\widehat{P}$, $\widehat{\varepsilon}$ and $\widehat{P}/\widehat{\varepsilon}$, we also have the corresponding relation for baryon density $\rho$ as (to order $\widehat{r}^4)$,
\begin{equation}
\widehat{\rho}\equiv
\rho/\rho_{\rm{c}}\approx1+\left(\frac{b_2/s_{\rm{c}}^2}{1+\widehat{P}_{\rm{c}}}\right)\widehat{r}^2+\frac{1}{1+\widehat{P}_{\rm{c}}}\left(a_4-\frac{{b_2^2}/{2s_{\rm{c}}^2}}{1+\widehat{P}_{\rm{c}}}\right)\widehat{r}^4
,\label{pk-4}
\end{equation}
obtained using the thermodynamic relation $\rho\partial\varepsilon/\partial\rho=P+\varepsilon$ with $\rho_{\rm{c}}$ the baryon number density at NS center.
Here, $a_4$ characterizes the model dependence of $\rho/\rho_{\rm{c}}$.
Without any surprise, $\rho<\rho_{\rm{c}}$ for finite $\widehat{r}\neq0$.
By taking $\widehat{P}_{\rm{c}}\approx0.24$ for PSR J0740+6620\,\cite{CLZ23}, we then obtain numerically $\widehat{P}/\widehat{P}_{\rm{c}}\approx 1-1.5\widehat{r}^2$, $\widehat{\varepsilon}/\widehat{\varepsilon}_{\rm{c}}\approx1-0.8\widehat{r}^2$, $(\widehat{P}/\widehat{\varepsilon})/(\widehat{P}_{\rm{c}}/\widehat{\varepsilon}_{\rm{c}})\approx1-0.7\widehat{r}^2$, and $\rho/\rho_{\rm{c}}\approx1-0.6\widehat{r}^2$, here $\widehat{\varepsilon}_{\rm{c}}=\varepsilon_{\rm{c}}/\varepsilon_{\rm{c}}=1$.
Two reference values for $\widehat{P}_{\rm{c}}$ at 0.24 and 0.374 are marked using the black dash-dotted lines.
It is clearly shown that all these $\mathcal{C}_{\Phi}$'s are definitely negative for $0\leq\widehat{P}_{\rm{c}}\lesssim0.374$,  which are different from the coefficient $\mathcal{C}_{\Phi}$ for $\Phi=s^2/s_{\rm{c}}^2$.
The latter may be either positive or negative (FIG.\,\ref{fig_NEG} and FIG.\,\ref{fig_AD}).
By recovering the physical radius $r$, one can find from Eq.\,(\ref{pk-1}) the (reduced) pressure $P/P_{\rm{c}}\approx1-0.78(r/R_{\max})^2$ (neglecting the intercept 0.64 in Eq.\,(\ref{Rmax-n})), holding for perturbatively small $r/R_{\max}$.
Combing (\ref{pk-2}) and (\ref{pk-4}) gives (to order $\mu^2$ and $\widehat{P}_{\rm{c}}^2$),
\begin{align}\label{kkk}
\rho/\rho_{\rm{c}}\approx&1+ \chi\mu-s_{\rm{c}}^2\chi^2\mu^2
\approx
\widehat{\varepsilon}-\mu\left(1+\frac{4}{3}\mu\right)\widehat{P}_{\rm{c}}\left(1-\widehat{P}_{\rm{c}}\right),
\end{align} where $\chi=1/(1+\widehat{P}_{\rm{c}})$ and $\mu=\widehat{\varepsilon}-1<0$, i.e., 
$\widehat{\rho}\approx\widehat{\varepsilon}$ to leading-order near the center and $\widehat{\rho}\gtrsim\widehat{\varepsilon}$ considering finite $\widehat{P}_{\rm{c}}$.
Relation (\ref{kkk}) gives us an estimate on the location of the possible peak in $s^2(\rho)$ as a function of $\rho$, e.g.,
by considering $\widehat{\varepsilon}_{\rm{pk}}-1=\mu_{\rm{pk}}=-D/3J$ of Eq.\,(\ref{s2pk}) and FIG.\,\ref{fig_s2peak} we find that the peak is at $\widehat{\rho}_{\rm{pk}}=\rho_{\rm{pk}}/\rho_{\rm{c}}\approx95\%$ by taking $D\approx-0.45$ and $J\approx-2.7$ (relevant for PSR J0740+6620), i.e., it is very close to the center (the finite-$\widehat{P}_{\rm{c}}$ correction of (\ref{kkk}) is $\lesssim1\%$).
The expansion (\ref{pk-4}) is useful and we give in Appendix \ref{appB} an order of magnitude estimate for the coefficient $a_4$ via the decreasing feature of $\widehat{\rho}$ as a function of $\widehat{r}$.

In addition, since $P/\varepsilon$ is a decreasing function for small $\widehat{r}$ as given in Eq.\,(\ref{pk-3}),  the conformal anomaly $\Delta=1/3-P/\varepsilon$ essentially increases with $\widehat{r}$ (i.e., when going outward) not so far ($\widehat{r}$ being small).
This means that it is disadvantage for the NS matter to become (nearly) conformal when going away from the center not so far.
This is also reflected in the behaviors of $\gamma=s^2/(P/\varepsilon)$ and $\Theta=[\Delta^2+(P/\varepsilon-s^2)^2]^{1/2}$ at finite $\widehat{r}$ or finite $\widehat{\varepsilon}$ (or equivalently $\mu$).
In fact, we can show straightforwardly that,
\begin{align}
\gamma/\gamma_{\rm{c}}\approx&1
+\frac{b_2}{s_{\rm{c}}^2}\left(1+\frac{2D}{s_{\rm{c}}^2}-\frac{s_{\rm{c}}^2}{\widehat{P}_{\rm{c}}}\right)\widehat{r}^2\notag\\
\approx&1-\frac{3D}{16\widehat{P}_{\rm{c}}^2}\widehat{r}^2
,\label{gr-1}\\
\Theta/\Theta_{\rm{c}}\approx&1
+\frac{b_2}{s_{\rm{c}}^2}\frac{3t_{\rm{c}}(1+3s_{\rm{c}}^2-6\widehat{P}_{\rm{c}}-6D)}{1+9s_{\rm{c}}^2-6\widehat{P}_{\rm{c}}(1+3s_{\rm{c}}^2)+18\widehat{P}_{\rm{c}}^2}\widehat{r}^2\notag\\
\approx&1+\frac{1-6D}{8}\widehat{r}^2
,\label{Tr-1}
\end{align}
where $b_2=-6^{-1}(1+3\widehat{P}_{\rm{c}}^2+4\widehat{P}_{\rm{c}})$ and the coefficients in front of $\widehat{r}^2$ in $\gamma/\gamma_{\rm{c}}$ and $\Theta/\Theta_{\rm{c}}$ are both positive, the second line for each quantity keeps only the leading-order term in $\widehat{P}_{\rm{c}}$.
Numerically, then $\Theta/\Theta_{\rm{c}}\approx1+3.8\widehat{r}^2$ and $\gamma/\gamma_{\rm{c}}\approx1+2.3\widehat{r}^2$ for $\widehat{P}_{\rm{c}}\approx0.24$ using $D\approx-0.45$ (panel (b) of FIG.\,\ref{fig_AD}).
For small $\widehat{P}_{\rm{c}}\to0$, the second line of Eq.\,(\ref{Tr-1}) gives approximately $\Theta/\Theta_{\rm{c}}\approx1+\widehat{r}^2/8$ since $D\sim\widehat{P}_{\rm{c}}^2\to0$ indicated by Eq.\,(\ref{gr-1}) (see also panel (b) of FIG.\,\ref{fig_AD}).
Similarly, the $\mu$-dependence of $\gamma$ and $\Theta$ could be obtained,
\begin{align}
\gamma/\gamma_{\rm{c}}\approx&1+\left(\frac{t_{\rm{c}}}{\widehat{P}_{\rm{c}}}+\frac{2D}{s_{\rm{c}}^2}\right)\mu\notag\\
\approx&1+\frac{3D}{2\widehat{P}_{\rm{c}}}\mu,\label{gm-1}\\
\Theta/\Theta_{\rm{c}}\approx&1+
\frac{3t_{\rm{c}}+9s_{\rm{c}}^2(t_{\rm{c}}+2D)-18\widehat{P}_{\rm{c}}(t_{\rm{c}}+D)}{1+9s_{\rm{c}}^2-6\widehat{P}_{\rm{c}}(1+3s_{\rm{c}}^2)+18\widehat{P}_{\rm{c}}^2}\mu\notag\\
\approx&1+(6D-1)\widehat{P}_{\rm{c}}\mu.
\label{Tm-1}
\end{align}
The coefficients in front of $\mu$ in Eqs.\,(\ref{gm-1}) and (\ref{Tm-1}) are both negative, e.g.,  we have $\gamma/\gamma_{\rm{c}}\approx1-2.9\mu$ and $\Theta/\Theta_{\rm{c}}\approx1-4.8\mu$ for $\widehat{P}_{\rm{c}}\approx0.24$. 
Therefore, Eq.\,(\ref{gr-1})$\sim$Eq.\,(\ref{Tm-1}) together generalize the conclusion of FIG.\,\ref{fig_Th-g}: If the matter at centers of massive NSs was not conformal,  it is likely its nearby surroundings are also  not conformal.
Combining the leading-order terms of Eq.\,(\ref{gm-1}) and Eq.\,(\ref{Tm-1}) or those of Eq.\,(\ref{gr-1}) and Eq.\,(\ref{Tr-1}) leads us to the correlation between the $\gamma$ and $\Theta$ parameters as,
\begin{equation}\label{TGG}
\Delta\Theta\approx\left(4-{2D}/{3}\right)\widehat{P}_{\rm{c}}^2\cdot\Delta\gamma,
\end{equation}
which holds near NS centers.
Eq.\,(\ref{TGG}) tells that two positive quantities $\Delta\Theta\equiv\Theta/\Theta_{\rm{c}}-1$ and $\Delta\gamma\equiv\gamma/\gamma_{\rm{c}}-1$ is positively correlated ($4-2D/3>0$ according to panel (b) of FIG.\,\ref{fig_AD}).

Another feature of Eq.\,(\ref{io-11}) is that for NSs close to the maximum-mass point $M_{\rm{NS}}^{\max}$ along the M-R curve, one has $\d^2M_{\rm{NS}}/\d\varepsilon_{\rm{c}}^2<0$ and therefore the criterion (\ref{io-6}) still holds.
However, the deterministic term $u_{\rm{c}}=2s_{\rm{c}}^2-3\widehat{P}_{\rm{c}}$ in the coefficient $D$ of Eq.\,(\ref{ref-D}) should be modified to (for small $\widehat{P}_{\rm{c}}$),
\begin{equation}
u_{\rm{c}}=2s_{\rm{c}}^2-3\widehat{P}_{\rm{c}}\approx-\frac{1-2\Psi}{3}\widehat{P}_{\rm{c}},~~\Psi>0,\end{equation}
by expanding Eq.\,(\ref{io-11}) over $\widehat{P}_{\rm{c}}$ as $s_{\rm{c}}^2\approx(4+\Psi)\widehat{P}_{\rm{c}}/3$ (Newtonian limit).
A positive $\Psi$ tends to make the correction $D$ positive and therefore reduce the probability of $s_{\rm{c}}^2<s^2(\widehat{\varepsilon})$ (see Eq.\,(\ref{s2eps})).
For example, a canonical NS is less likely to have the crossover in the core than a massive NS (assuming they have similar radii), since the former is more likely (than the latter) to be on climbing stage of the M-R curve.
Our analysis is consistent with Ref.\,\cite{Ecker23} which predicted the increasing of $s^2$ is reduced even to disappear when going out as the NS mass decreases from $M_{\rm{NS}}^{\max}$ to $1.4M_{\odot}$.
On the other hand,  for massive NSs with similar radii of PSR J0740+6620 (thus the $\widehat{P}_{\rm{c}}$'s are also similar), the reduction on the probability of $D<0$ due to a (small) positive $\Psi$ is expected to be small as they are near the maximum-mass configuration\,\cite{Rezzolla2018,Mar17,Ruiz18}.

\section{Distinguishing Compactness from Stiffness in Neutron Stars, and pQCD prediction from General Relativity requirement on Trace Anomaly}\label{SEC_CSD}
With both analytical analyses and numerical examples, we have investigated separately in detail in the previous two sections the radial variation of SSS and the various bounds on the trace anomaly. 
To this end, it is useful to summarize the main differences between the compactness and stiffness of NSs. It is also important to emphasize the respective applicabilities of the pQCD prediction and GR limit on trace anomaly in NSs.  

\begin{figure}[h!]
\centering
\includegraphics[width=8.cm]{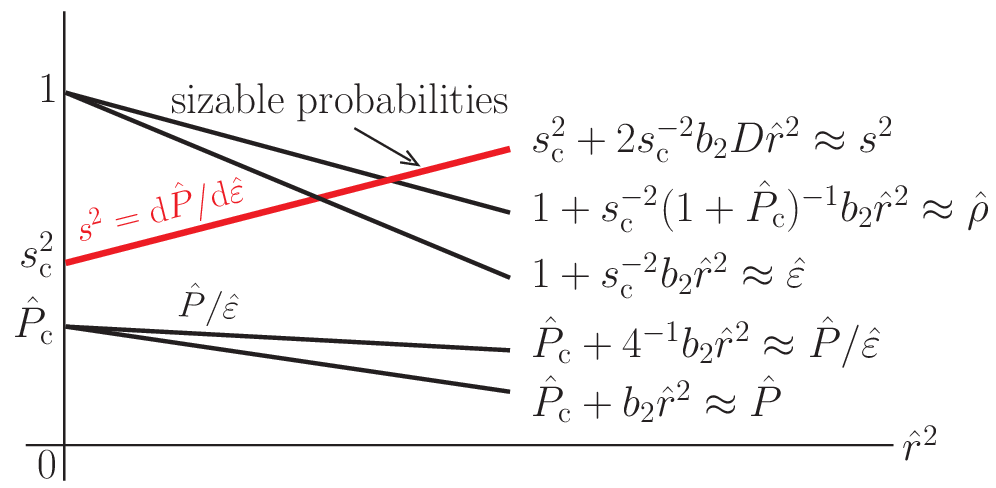}
\caption{Qualitative sketch of the reduction of $\widehat{P}$, $\widehat{\varepsilon}$,  $\widehat{P}/\widehat{\varepsilon}$ and $\widehat{\rho}$ (these quantities characterize the compactness/denseness of the matter) to order $\widehat{r}^2$ and the probable increasing of $s^2$ (characterizing the stiffness of the matter) near NS centers.
Here $b_2=-6^{-1}(1+3\widehat{P}_{\rm{c}}^2+4\widehat{P}_{\rm{c}})$ and the coefficient $D$ depends on $\widehat{P}_{\rm{c}}$ (see panel (b) of FIG.\,\ref{fig_AD}).
}\label{fig_re_pe}
\end{figure}

According to the perturbative expressions (\ref{pk-1})$\sim$(\ref{pk-3}) (as well as FIG.\,\ref{fig_eps_P_rho}),  the energy density $\varepsilon$, the pressure $P$, their ratio $P/\varepsilon$ and the baryon density $\rho$ (see Eq.\,(\ref{pk-4})) are all decreasing functions of $\widehat{r}$ definitely.
While on the other hand, the SSS $s^2$ has sizable probabilities to be enhanced when going outward from  centers of NSs, see sketches shown in FIG.\,\ref{fig_re_pe}.
This demonstrates the fundamental difference between the stiffness (characterized effectively by $s^2$) and the denseness/compactness (characterized effectively by $\rho$, $P$, $\varepsilon$ or $P/\varepsilon$), although they are closely related to each other (see, e.g., FIG.\,\ref{fig_Compt} or the $s_{\rm{c}}^2$ of Eq.\,(\ref{sc2}) itself).
An illustrative/useful example is that $s_{\rm{c}}^2$ (of Eq.\,(\ref{sc2})) approaches $1/3$ (stiffness) earlier than $\widehat{P}_{\rm{c}}\to1/3$ (compactness).
Physically, it is because the EOS of NS matter (especially near the centers) is nonlinear (otherwise $s^2=P/\varepsilon$, $\gamma=1$ and $\Theta=[\Delta^2+(P/\varepsilon-s^2)^2]^{1/2}=|\Delta|$).

\begin{figure}[h!]
\centering
\includegraphics[width=8.6cm]{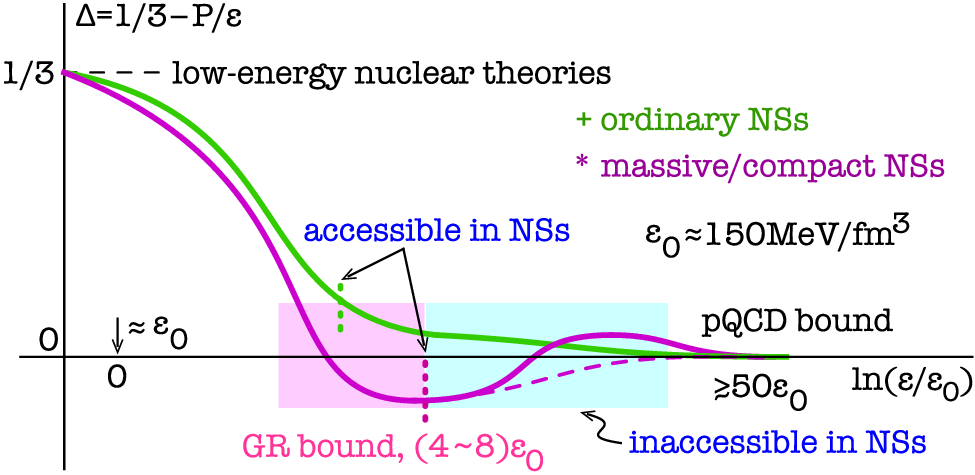}
\caption{Sketch of the patterns for $\Delta=1/3-P/\varepsilon$ in NSs. The $\Delta$ is well constrained around the fiducial density $\varepsilon_0\approx150\,\rm{MeV}/\rm{fm}^3$ by low-energy nuclear theories and is predicted to vanish due to conformality of the matter at $\varepsilon\gtrsim50\varepsilon_0$ using pQCD theories.
Massive and compact NSs provide a unique opportunity to probe the negativeness of $\Delta$ in certain energy density regions, where the GR bound on $\Delta$ (with $\varepsilon$ being around ($4\mbox{$\sim$}8)\varepsilon_0$) is expected to be more relevant than the pQCD bound ($\varepsilon\gtrsim50\varepsilon_0$) in these NSs.
}\label{fig_Dp_sk}
\end{figure}

Closely related is the trace anomaly (or the conformality measure) $\Delta=1/3-P/\varepsilon$\,\cite{Fuji22,Bjorken83}.
We sketch in FIG.\,\ref{fig_Dp_sk} the $\Delta$ as a function of energy density $\varepsilon$, here $\varepsilon_0\approx150\,\rm{MeV}/\rm{fm}^3$ is the fiducial energy density at $\rho_{\rm{sat}}$,
around which the low-energy nuclear theories constrain the $\Delta$ quite well.
The pQCD bound $\Delta=0$ applies effectively at very large energy densities $\varepsilon\gtrsim50\varepsilon_0\approx7.5\,\rm{GeV}/\rm{fm}^3$\,\cite{Fuji22,Kur10}, which is far larger than the energy density reacheable in NSs. Thus, it is possibly relevant but not fundamental for explaining the observed $P/\varepsilon\gtrsim1/3$ in massive NSs (using microscopic/phenomenological models).
On the other hand, we have demonstrated that a GR bound on $\widehat{P}_{\rm{c}}=P_{\rm{c}}/\varepsilon_{\rm{c}}$ and $P/\varepsilon$ (near NS centers) naturally emerges when dissecting perturbatively the TOV equations without using any specific EOS model\,\cite{CLZ23}. In this sense, the GR bound on $\Delta$ (with the $\varepsilon$ being roughly around (4$\sim$8$)\varepsilon_0$) is likely more relevant/fundamental than the pQCD prediction for nuclear EOS in NSs, although the latter may influence the extraction of $\Delta$\,\cite{Mus23}. 
Whether these two specific bounds (being very close to each other) are inner-related may deepen our understanding on the connection between GR and the microscopic theories of elementary particles.
Our above discussions and findings are not out of bounds as ultimately properties of compact objects are determined by the Hamilton's principle using the total action of the whole system including gravity, matter (including nuclear matter, dark matter and energy) and their couplings\,\cite{NAP2011}.

As a negative $\Delta$ is unlikely to be observed in ordinary NSs (e.g., NSs with masses $\sim 1.7M_{\odot}$), see FIG.\,\ref{fig_TAc}, the evolution of $\Delta$ is probably like the green line in FIG.\,\ref{fig_Dp_sk}. An (unconventional) exception may come from light but very compact NSs, e.g., a $1.7M_{\odot}$ NS with radius about 9.3\,km (therefore $\varepsilon_{\rm{c}}\approx1.86\,\rm{GeV}/\rm{fm}^3$ and $P_{\rm{c}}\approx654\,\rm{MeV}/\rm{fm}^3$ and $\widehat{P}_{\rm{c}}\approx0.351$) has its $\Delta_{\rm{c}}\approx-0.02$ (FIG.\,\ref{fig_TAc}).
On the other hand,  massive and compact NSs (masses $\gtrsim2M_{\odot}$) have the most relevance to observe a negative $\Delta$ (as indicated by the magenta line in FIG.\,\ref{fig_Dp_sk}), and how the negative $\Delta$ evolves to the pQCD bound may tell more on the properties of supradense matter.
Unfortunately, the region with $\varepsilon\gtrsim8\varepsilon_0$ is largely inaccessible in NSs (sketched by the light-cyan rectangle band) due to their self-gravitating nature (see FIG.\,\ref{fig_ULTeps} which indicates that observations of massive NSs should inevitably put an upper bound on the central energy density as $\varepsilon_{\rm{c}}\lesssim\varepsilon_{\rm{ult}}$).

\section{Summary}\label{SEC6}

Using the central SSS $s_{\rm{c}}^2$ and NS mass/radius scaling obtained from analyzing perturbatively structures of the scaled TOV equations\,\cite{CLZ23}, we studied the radial variation of SSS, 
trace anomaly and several closely related properties of NSs in an EOS-model independent manner.
We found that the reduced pressure $\widehat{P}_{\rm{c}}=P_{\rm{c}}/\varepsilon_{\rm{c}}$ is the most relevant quantity determining the onset of a continuous crossover of SSS  in the cores of massive NSs.
With sizable probabilities the crossover occurs near the centers of massive NSs if
$\widehat{P}_{\rm{c}}\lesssim0.3$, e.g., with a probability $\gtrsim63\%$ for
$\widehat{P}_{\rm{c}} \approx 0.24$ (PSR J0740+6620). The resulting peak of $s^2$ or its derivative part (defined as $s^2-P/\varepsilon$) near NS centers is generally demonstrated.

With the help of the universal correlations of $M_{\rm{NS}}^{\max}$-$\Gamma_{\rm{c}}$ and $R_{\max}$-$\nu_{\rm{c}}$\,\cite{CLZ23} and the nonlinear dependence of $s_{\rm{c}}^2$ on $\widehat{P}_{\rm{c}}$ (of Eq.\,(\ref{sc2})), a new causality boundary for NS M-R curve as $R_{\max}/\rm{km}\gtrsim 4.73M_{\rm{NS}}^{\max}/M_{\odot}+1.14$ was obtained. It is shown to be excellently consistent with several NS mass/radius observations and puts a more stringent constraint on the dense matter EOS compared to the ones available in the literature.
Moreover, the NS maximum compactness parameter is limited to $\lesssim0.313\cdot(1-1.14\,\rm{km}/R_{\max})\lesssim0.283$ (using $R_{\max}\approx12\,\rm{km}$).
The difference between $s^2\leq1$ (stiffness) and $P/\varepsilon\leq1$ (compactness) in NSs is clarified,
with its physical origin being traced back to the nonlinear characteristic of the core EOS in NSs. 
While $s^2\leq1$ is more relevant than $P/\varepsilon\leq1$ in NSs,  the latter is shown to be $\lesssim 0.374$ or equivalently $\Delta=1/3-P/\varepsilon\gtrsim-0.041$ around the centers of stable NSs.
Such bound is a direct consequence of GR (encapsulated in the TOV equations), and is expected to be more fundamental than the pQCD prediction of dense matter EOS for NSs.
Closely related, the matter in cores of massive NSs is unlikely to be conformal if the criteria $\Theta=[\Delta^2+(P/\varepsilon-s^2)^2]^{1/2}\lesssim0.2$ and $\gamma=\d\ln P/\d\ln\varepsilon\lesssim1.75$ were adopted. Moreover,  the violation of conformal bound is found to basically depend on $\widehat{P}_{\rm{c}}$, e.g.,  the bound is broken if $M_{\rm{NS}}^{\max}\gtrsim1.9M_{\odot}$ as $s_{\rm{c}}^2$ increases with $M_{\rm{NS}}^{\max}/M_{\odot}$ (under the assumption NSs with masses about $1.3\mbox{$\sim$}2.3M_{\odot}$ have similar radii $\approx$12\,km).
Furthermore, the ultimate energy density $\varepsilon_{\rm{ult}}$ and pressure $P_{\rm{ult}}$ allowed in NSs are estimated. The observations of massive NSs may induce corresponding upper limits for $\varepsilon_{\rm{ult}}$ and $P_{\rm{ult}}$, e.g., the existence of a 2.08$M_{\odot}$ NS leads to $\varepsilon_{\rm{ult}}\lesssim1.32\,\rm{GeV}/\rm{fm}^3$ and $P_{\rm{ult}}\lesssim494\,\rm{MeV}/\rm{fm}^3$, respectively. This improves previous constraints on the same quantities.

Finally, unlike most existing studies of NSs in the literature, our analyses are carried out without using any model EOS for NS matter thus largely free of the uncertainties in modeling the dense matter EOS. We have demonstrated that some properties of the underlying NS EOS (pressure versus energy density) can be inferred directly from the observational data alone. This in turn can help constrain predictions of dense matter EOS based on nuclear many-body theories. Overall, our work contributes to realizing the ultimate goal of understanding properties of strong-field gravity, dense matter and their couplings in compact objects.

\section*{Acknowledgements}

We would like to thank Peng-Cheng Chu for helpful discussions.
This work is supported in part by the U.S. Department of Energy, Office of Science, under Award No. DE-SC0013702, the CUSTIPEN (China-U.S. Theory Institute for Physics with Exotic Nuclei) under
US Department of Energy Grant No. DE-SC0009971, the National Natural Science Foundation of China under Grant No.12235010.

\appendix
\renewcommand\theequation{\Alph{section}\arabic{equation}}
\renewcommand\thefigure{\alph{section}\arabic{figure}}

\setcounter{figure}{0}

\section{ Proof of Eq.\,(\ref{ref-D}) for a General $K$}\label{app0}

From Eq.\,(\ref{ref-d1}), we can solve for $d_3$ in terms of other $d_k$'s (using the property that $s^2=0$ for $\widehat{\varepsilon}=0$),
\begin{equation}
d_3=-2\widehat{P}_{\rm{c}}+ s_{\rm{c}}^2-\sum_{k=4}^K(k-2)d_k,
\end{equation}
where we split the summation $\sum_{k=3}^K(k-2)d_k$ in Eq.\,(\ref{ref-d1}) into $d_3+\sum_{k=4}^K(k-2)d_k$.
Therefore, according to the basic definition of $D$ of Eq.\,(\ref{oo-2}), we obtain,
\begin{align}
D=&\sum_{k=1}^K\frac{k(k-1)}{2}d_k
=d_2+3d_3+\sum_{k=4}^K\frac{k(k-1)}{2}d_k\notag\\
=&2s_{\rm{c}}^2-3\widehat{P}_{\rm{c}}+\sum_{k=4}^K\left(\frac{k(k-1)}{2}-(k-1)-(k-2)\right)d_k
\notag\\
=&2s_{\rm{c}}^2-3\widehat{P}_{\rm{c}}+\sum_{k=4}^K\frac{(k-2)(k-3)}{2}d_k,
\end{align}
here expression for $d_2$ of Eq.\,(\ref{ref-d2}) is used.
In summary,  condition $d_1=0$ and the sum rules of (\ref{io-4-s2}) together enables us to rewrite $D$ in the form of Eq.\,(\ref{ref-D}), and for the special case of $K=3$ the $D$ becomes deterministic.

\section{Explaining the Sign of $D$ and $J$}\label{appK}

In the main text, we analytically study the case of $K=3$ (see FIG.\,\ref{fig_NEG-K3}), where a sharp change for the probability of $D<0$ to $D>0$ occurs at $\widehat{P}_{\rm{c}}\approx0.105$.
For more general values of $K$, similar phenomenon may happen.
In this appendix, we try to explain why the coefficient $D$ tends to be negative (positive) for small (large) $\widehat{P}_{\rm{c}}$, using the expansions with $K=4$.
We have two relations, i.e., $d_2+d_3+d_4=\widehat{P}_{\rm{c}}$ and $2d_2+3d_3+4d_4=s_{\rm{c}}^2$, from which we can solve for $d_2$ and $d_3$ to be expressed in terms of $d_4$.
The general condition $0\leq s^2\leq1$ now reads $0\leq2d_2\widehat{\varepsilon}+3d_3\widehat{\varepsilon}^2+4d_4\widehat{\varepsilon}^3\leq1$.
Putting the expressions of $d_2$ and $d_3$ into it gives the upper and lower limit for $d_4$, i.e., $d_4^{\rm{(l)}}\leq d_4\leq d_4^{\rm{(u)}}$, where
\begin{align}\label{def_d4ul}
d_4^{\rm{(l)}}=&\frac{\widehat{\varepsilon}(\widehat{P}_{\rm{c}}-s_{\rm{c}}^2)-\widehat{P}_{\rm{c}}+\widehat{\varepsilon}^{-1}}{4\widehat{\varepsilon}^2-3\widehat{\varepsilon}-1},~~
d_4^{\rm{(u)}}=\frac{\widehat{\varepsilon}(\widehat{P}_{\rm{c}}-s_{\rm{c}}^2)-\widehat{P}_{\rm{c}}}{4\widehat{\varepsilon}^2-3\widehat{\varepsilon}-1}.
\end{align}
Adding the deterministic term $u_{\rm{c}}=-3\widehat{P}_{\rm{c}}+2s_{\rm{c}}^2$ of Eq.\,(\ref{2cc}), we obtain correspondingly $D^{\rm{(u/l)}}=u_{\rm{c}}+d_4^{\rm{(u/l)}}$ (see Eq.\,(\ref{ref-D})).

\begin{figure}[h!]
\centering
\includegraphics[width=4.2cm]{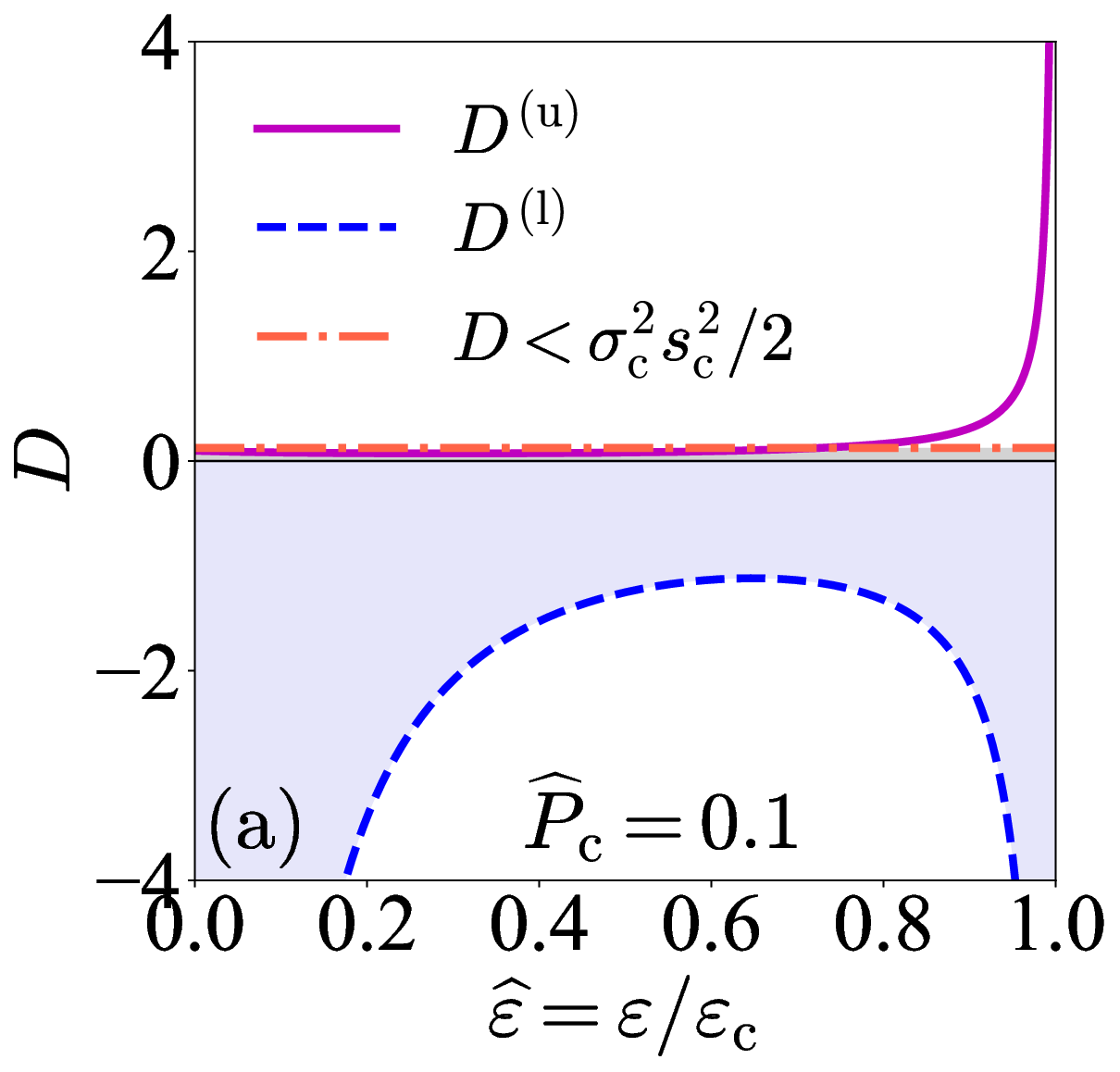}
\includegraphics[width=4.2cm]{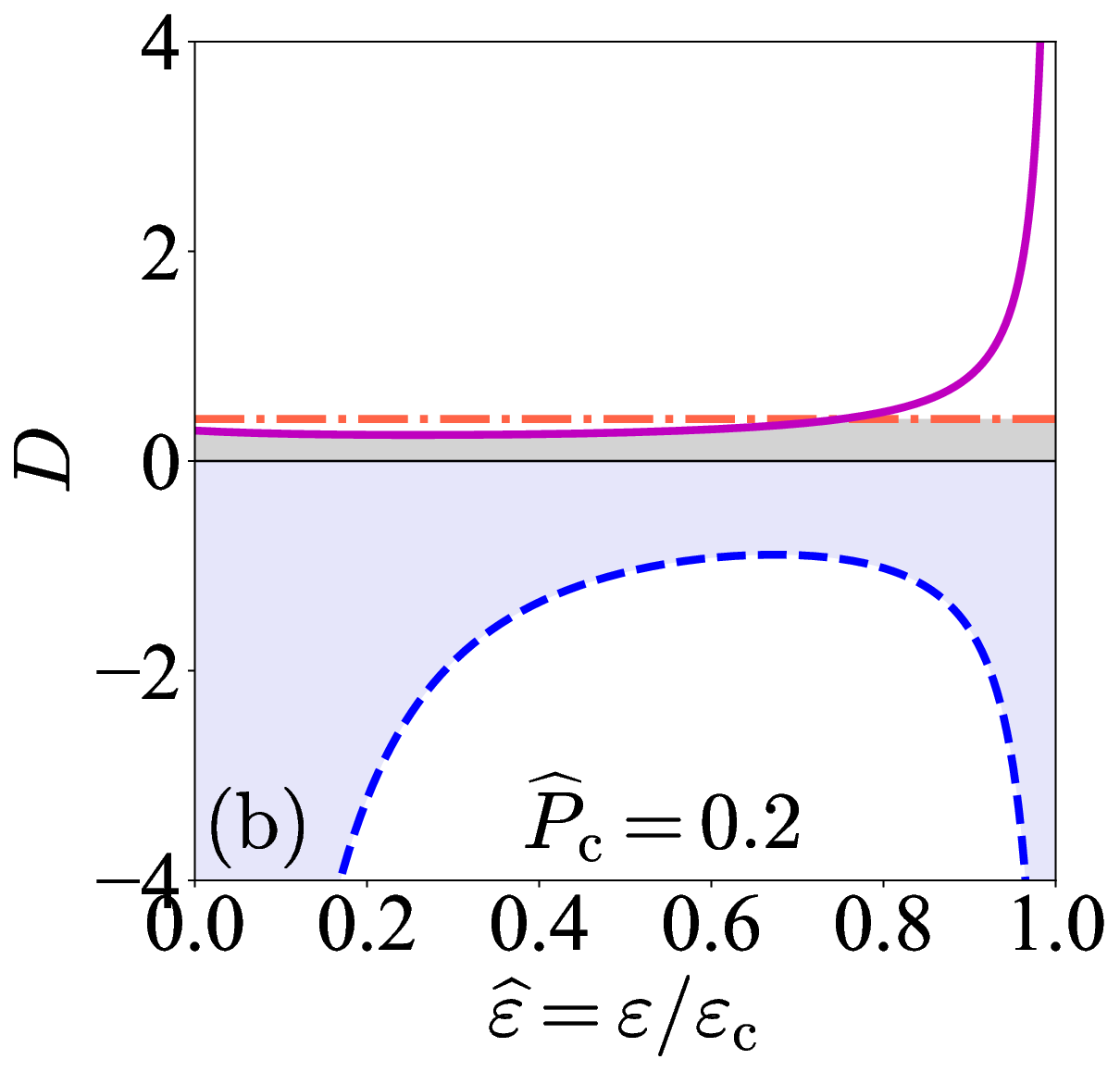}\\
\includegraphics[width=4.2cm]{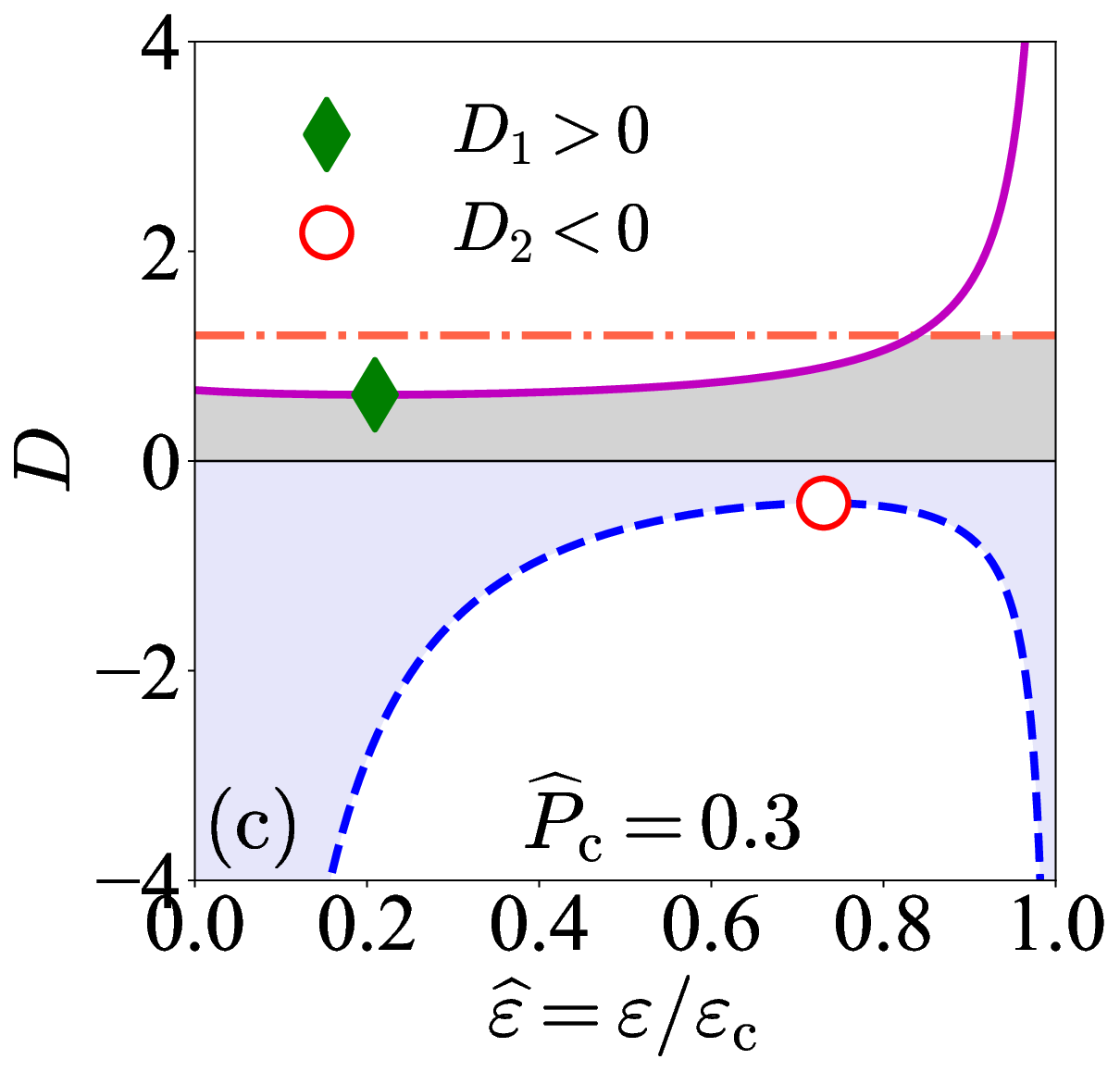}
\includegraphics[width=4.2cm]{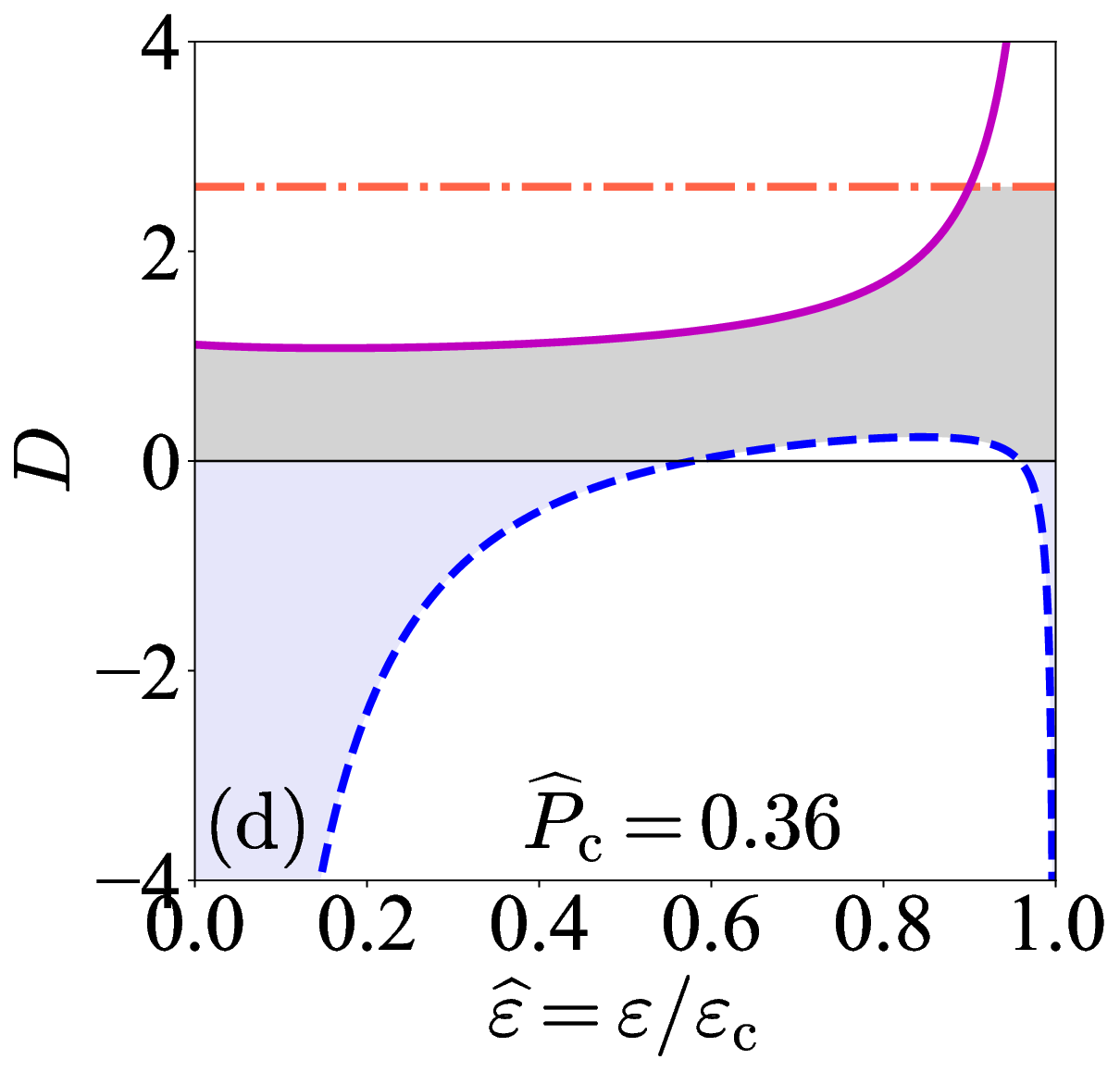}
\caption{Explanation on why $D$ tends to be negative (positive) for small (large) $\widehat{P}_{\rm{c}}$, using $K=4$. See text for details.
}\label{fig_WhyD}
\end{figure}

\begin{figure}[h!]
\centering
\includegraphics[width=4.2cm]{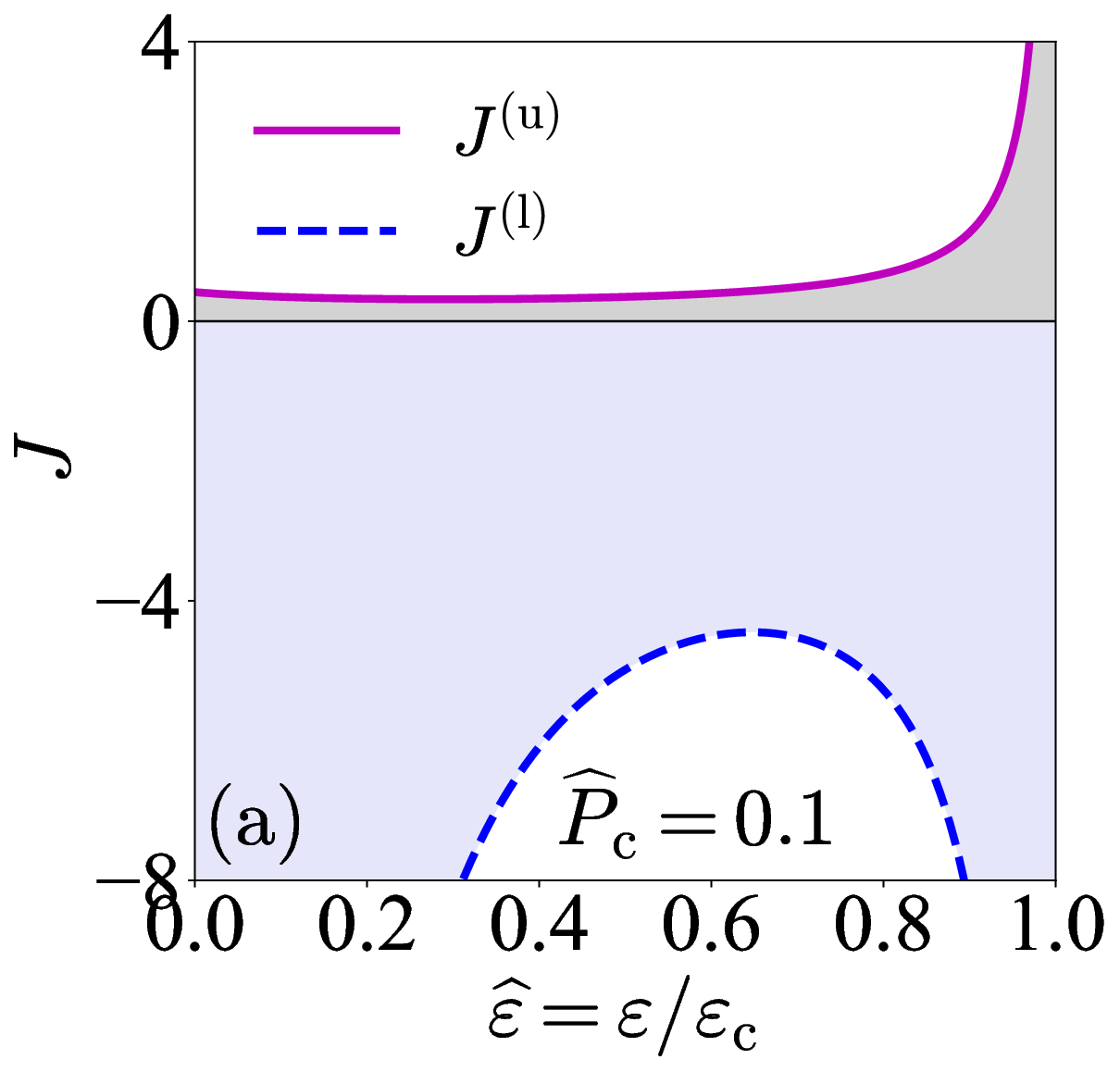}
\includegraphics[width=4.2cm]{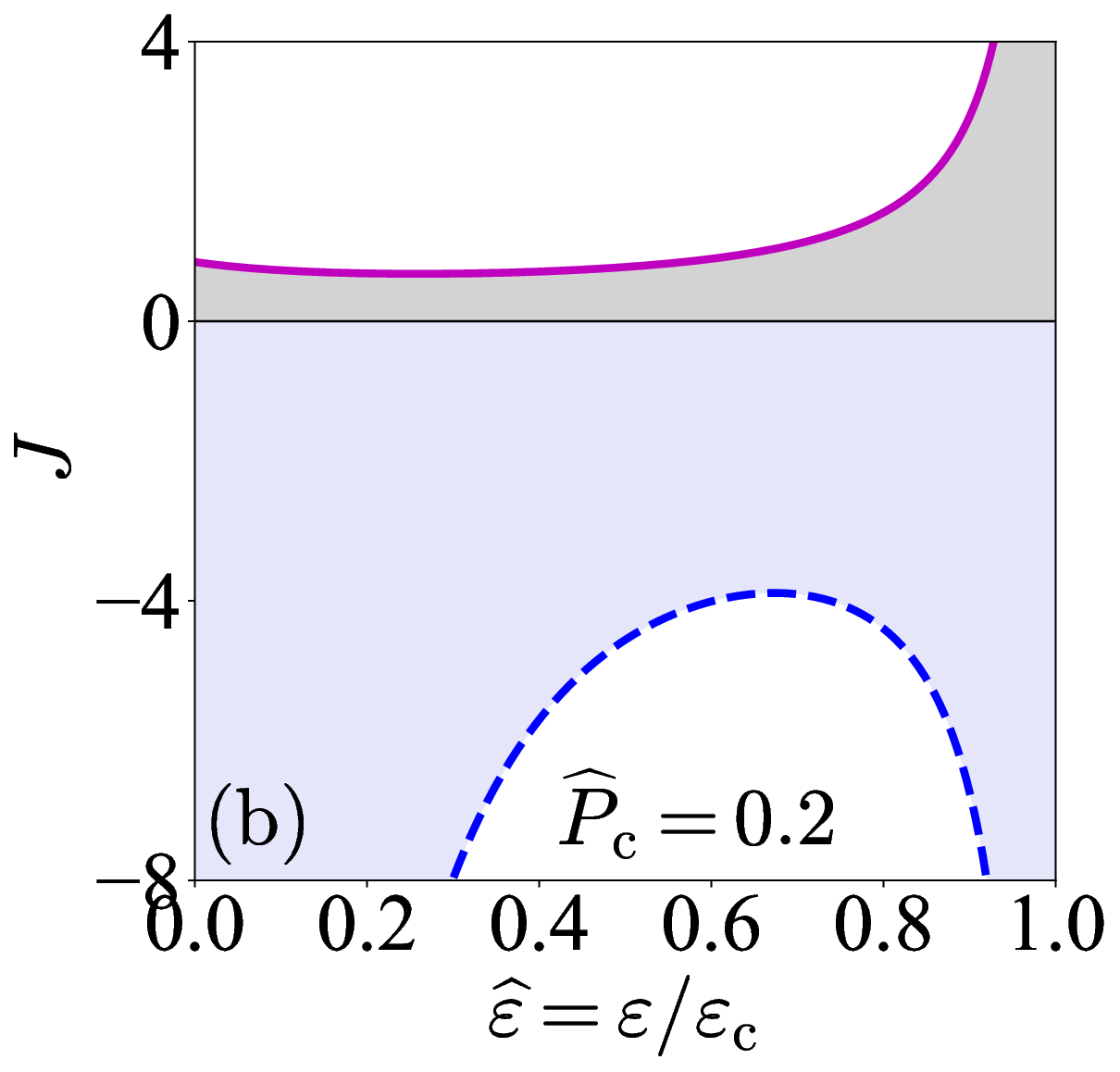}\\
\includegraphics[width=4.2cm]{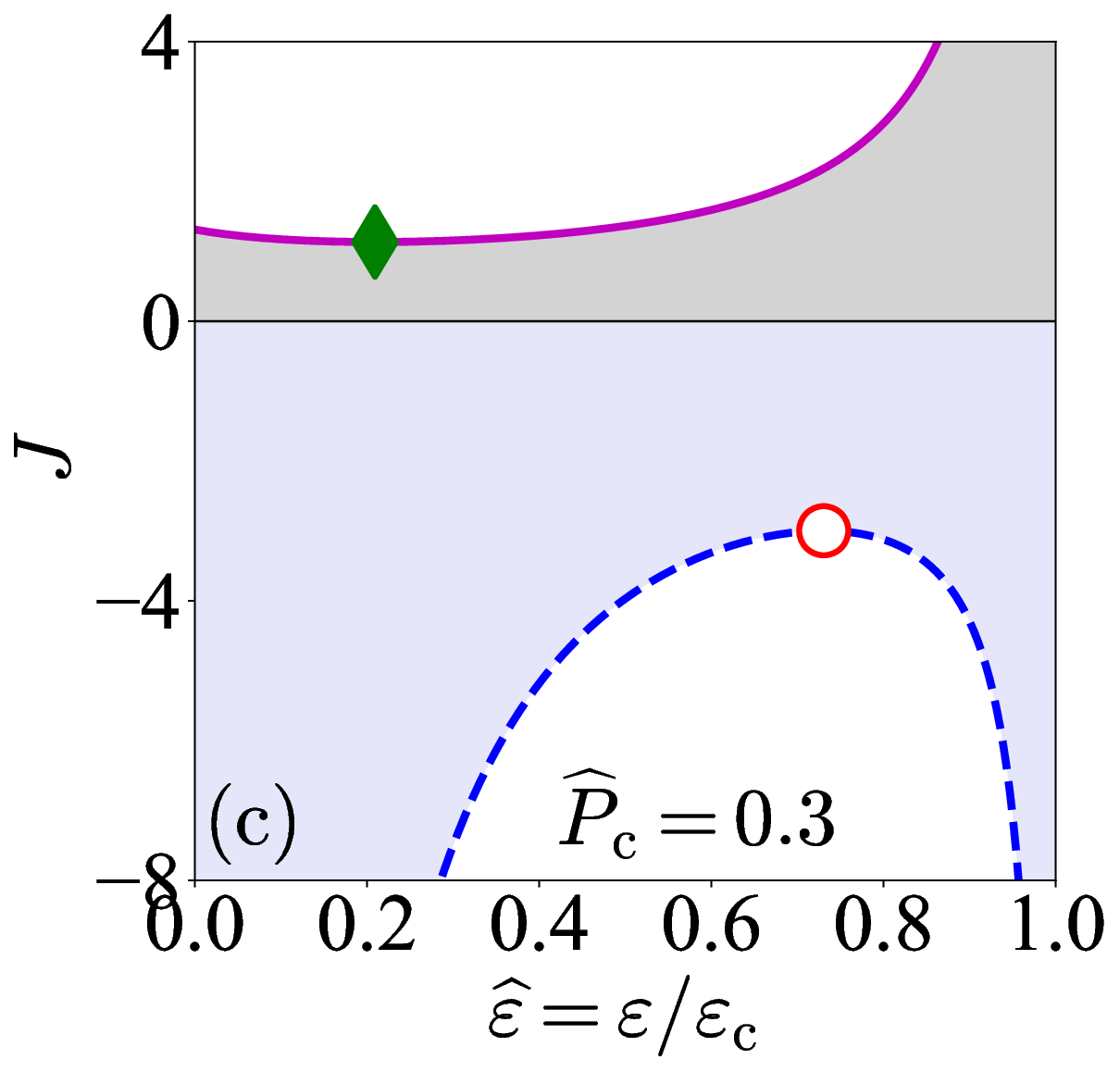}
\includegraphics[width=4.2cm]{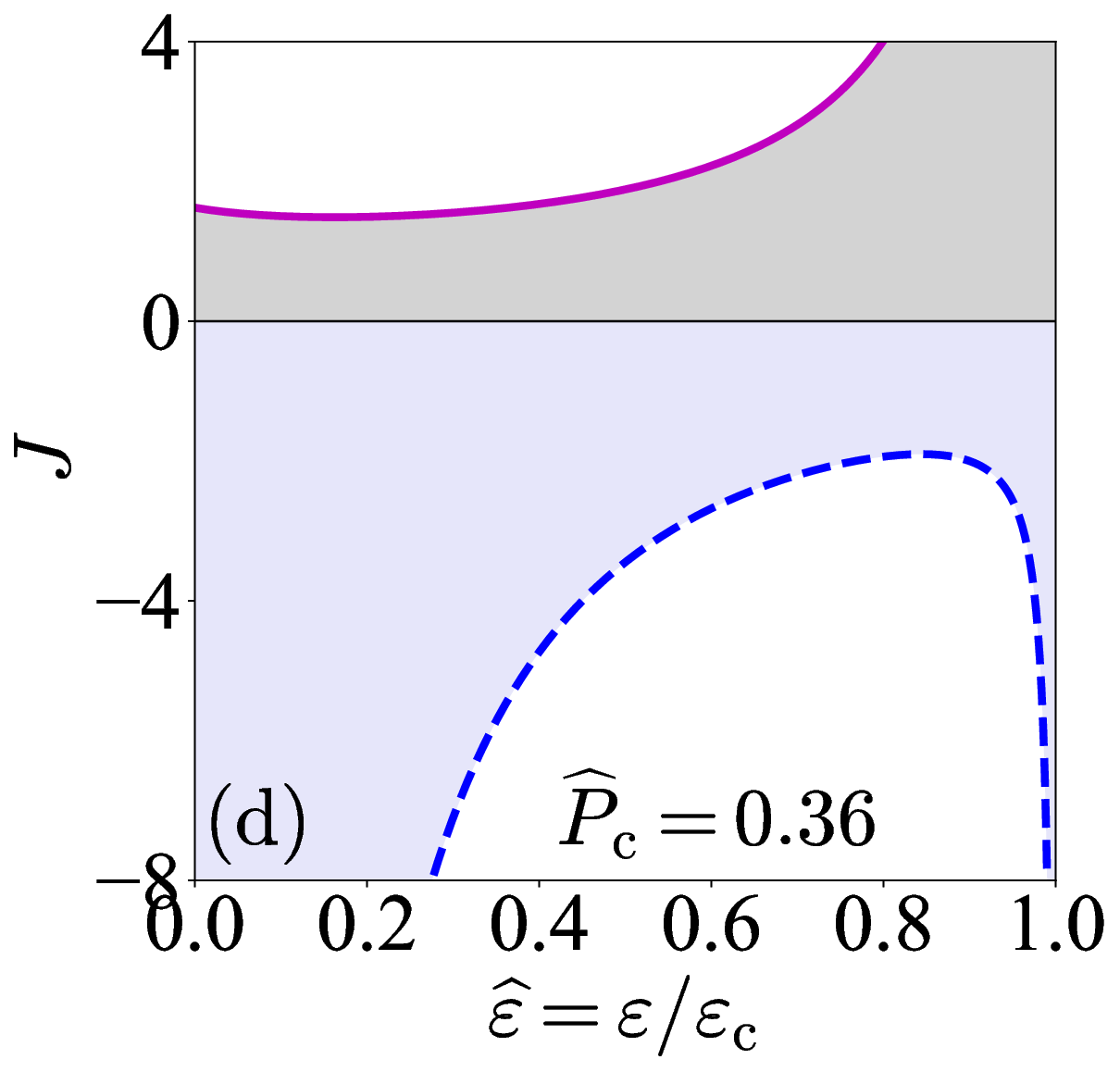}
\caption{The same as FIG.\,\ref{fig_WhyD} but for the coefficient $J$ of Eq.\,(\ref{ref-J}).
}\label{fig_WhyJ}
\end{figure}

The results are summarized in FIG.\,\ref{fig_WhyD}.
Here in each panel, the magenta (blue) line is for $D^{(\rm{u})}$ ($D^{\rm{(l)}}$), and the orange line marks the general boundary of $2D<\sigma_{\rm{c}}^2s_{\rm{c}}^2$ (see the estimate algorithm of (\ref{io-6})).
The positive and negative regions for $D$ are shown by the grey and lavender bands, respectively.
Moreover, $D\leq D^{(\rm{u})}$ indicates that $D$ should be smaller than the lowest point $D_1$ on the magenta line (solid green diamond in panel (c)).
Similarly, $D\geq D^{\rm{(l)}}$ implies that $D$ should be larger than the highest point $D_2$ on the blue line (shallow red circle in panel (c)).
When $\widehat{P}_{\rm{c}}$ is small (such as panels (a) and (b)), the magnitude of red circle is larger than that of the greed diamond, indicating that $D$ is more probable to be negative than be positive.
As $\widehat{P}_{\rm{c}}$ increases one eventually encounters $D_1\approx -D_2$ (as panel (c) shows), $D_2\approx0$ and $D_2\gtrsim 0$ (panel (d)).
The situation $D_1\approx -D_2$ roughly marks the value of $\widehat{P}_{\rm{c}}$ corresponding to equal probability of $D>0$ and $D<0$, this is about $\widehat{P}_{\rm{c}}\approx0.3$ (as verified by simulation given in the main text, see FIG.\,\ref{fig_NEG}).
When $\widehat{P}_{\rm{c}}$ increases even further, the probability of $D<0$ (of $D>0$) quickly decreases (increases).
The coefficient $J$ could be similarly analyzed, i.e., $J=3d_4+(s_{\rm{c}}^2-\widehat{P}_{\rm{c}})/3$ where $d_4^{\rm{(l)}}\leq d_4\leq d_4^{\rm{(u)}}$ (see (\ref{ref-J}) and (\ref{def_d4ul})).
Therefore $J^{\rm{(l)}}\leq J\leq J^{\rm{(u)}}$, and we probably have $J<0$ though the allowed region of $J<0$ (of $J>0$) eventually shrinks (expands) as $\widehat{P}_{\rm{c}}$ increases, see FIG.\,\ref{fig_WhyJ}.
For other values of $K$, the general features are similar as FIG.\,\ref{fig_WhyD} and FIG.\,\ref{fig_WhyJ} and are consistent with the results given in the main text (see FIG.\,\ref{fig_NEG} and FIG.\,\ref{fig_AD}).

\begin{figure}[h!]
\centering
\includegraphics[width=4.2cm]{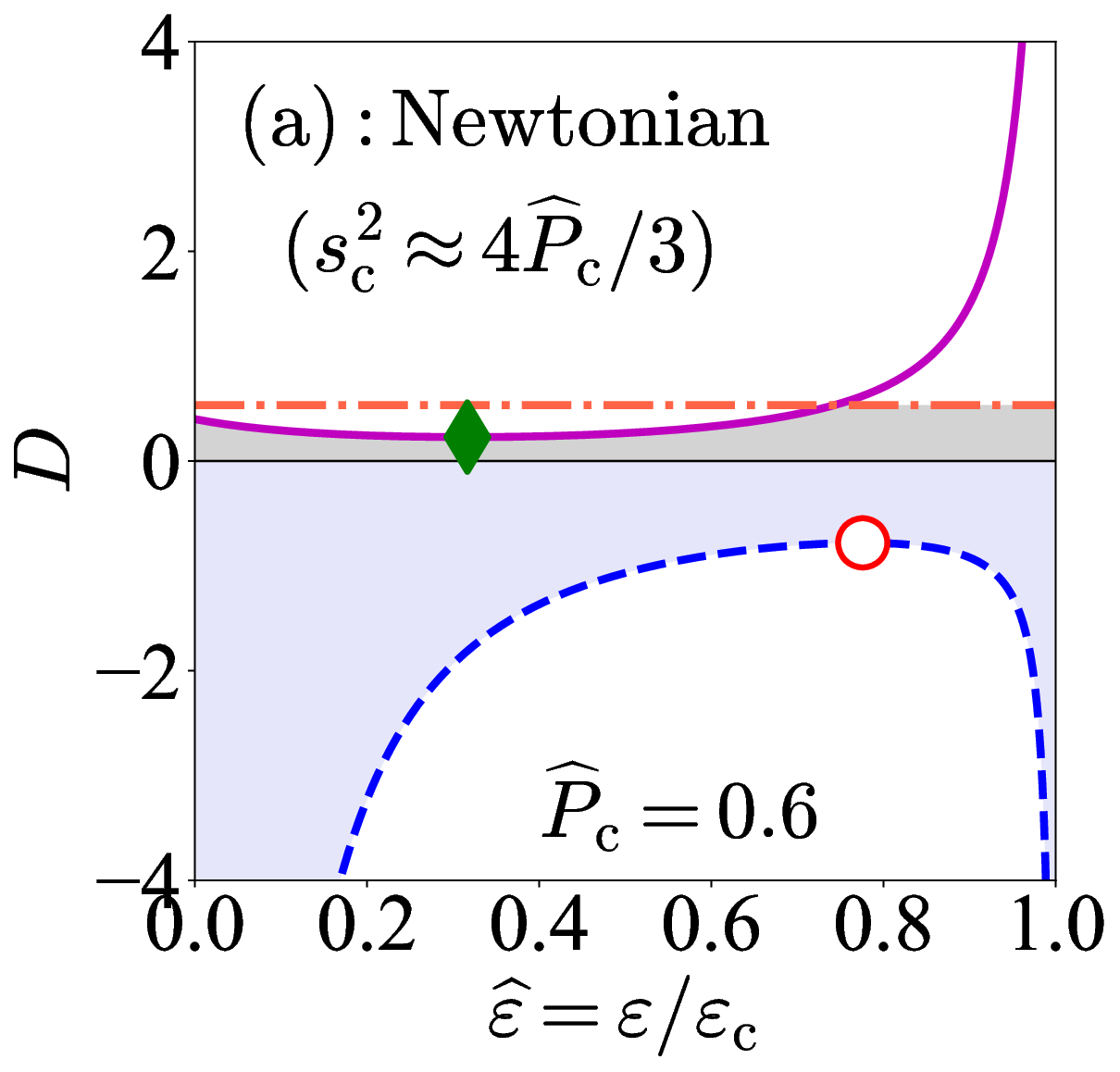}
\includegraphics[width=4.2cm]{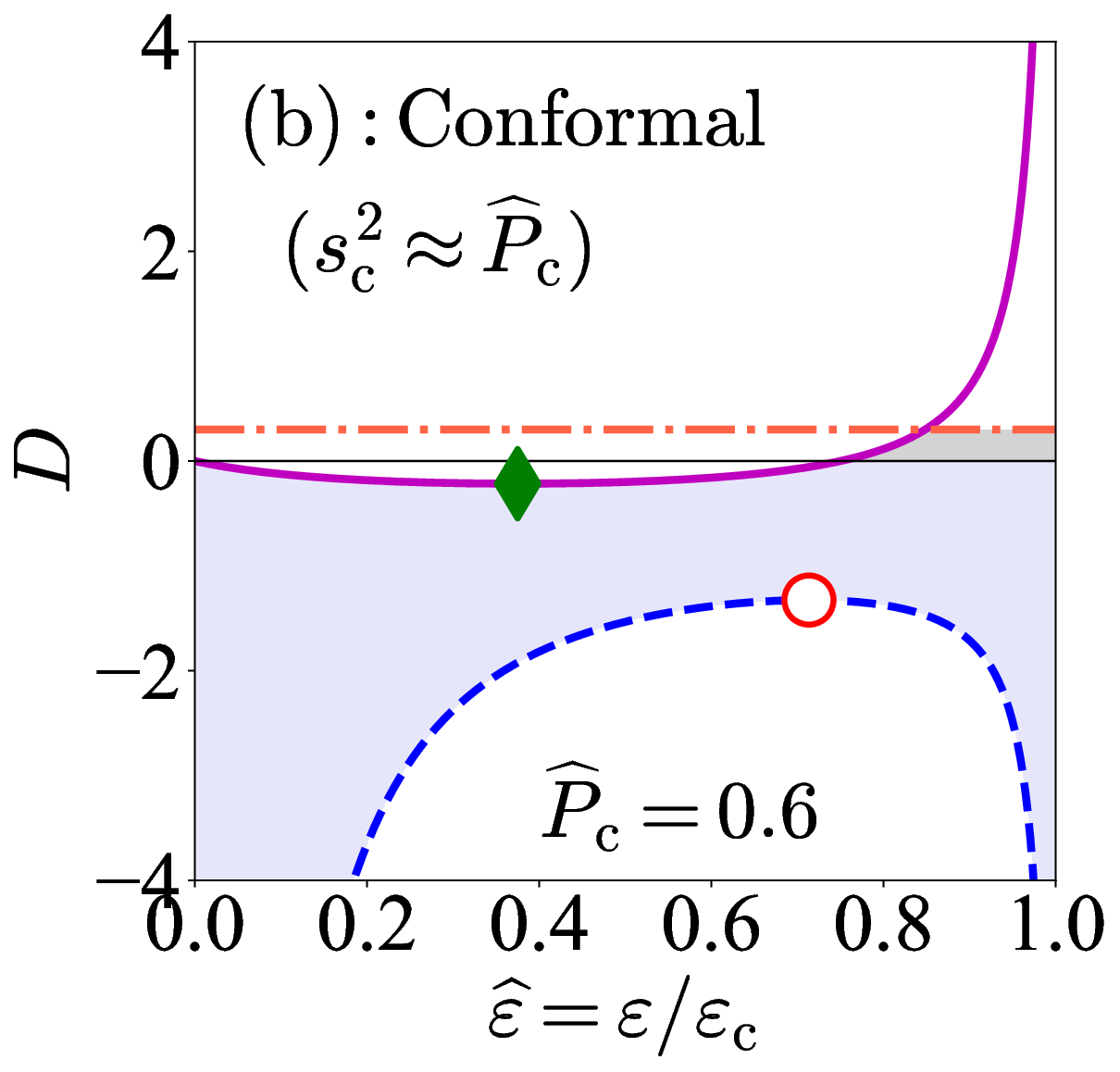}
\caption{The same as FIG.\,\ref{fig_WhyD} but adopting the Newtonian limit $s_{\rm{c}}^2\approx 4\widehat{P}_{\rm{c}}/3$ (left panel) and conformal limit $s_{\rm{c}}^2\approx\widehat{P}_{\rm{c}}$ (right panel). The principle of causality requires then $\widehat{P}_{\rm{c}}\lesssim3/4$ and $\widehat{P}_{\rm{c}}\lesssim1$ for these two situations, respectively.
Symbols and lines have the same meaning of FIG.\,\ref{fig_WhyD}.
}\label{fig_WhyDNL}
\end{figure}

Moreover, if one artificially adopts the Newton approximation for $s_{\rm{c}}^2$ as $s_{\rm{c}}^2\approx 4\widehat{P}_{\rm{c}}/3$ (panel (a) of FIG.\,\ref{fig_WhyDNL}) or the conformal limit $\gamma_{\rm{c}}=s_{\rm{c}}^2/\widehat{P}_{\rm{c}}\approx1\leftrightarrow s_{\rm{c}}^2\approx \widehat{P}_{\rm{c}}$ (panel (b) of FIG.\,\ref{fig_WhyDNL}), then even a large $\widehat{P}_{\rm{c}}\approx0.6$ may still tend to induce a negative $D$ (compared with FIG.\,\ref{fig_WhyD}).
In particular, the $D$ in conformal limit is very likely negative.
In addition, if $s_{\rm{c}}^2$ is nearly a constant (independent of $\widehat{P}_{\rm{c}}$), then the condition $D<\sigma_{\rm{c}}^2s_{\rm{c}}^2/2$ alone tells that $D<0$ since $\sigma_{\rm{c}}^2=\d s_{\rm{c}}^2/\d\widehat{P}_{\rm{c}}=0$ (see the inequality (\ref{io-7})).
In this case if $s_{\rm{c}}^2=1$, then its nearby $s^2(\widehat{\varepsilon})$ keeps 1, otherwise $s^2(\widehat{\varepsilon})>s_{\rm{c}}^2$.
These results clearly show that the nonlinear dependence of $s_{\rm{c}}^2$ on $\widehat{P}_{\rm{c}}$ (of Eq.\,(\ref{sc2})) is fundamental to account for the eventual change on the sign of $D$, and therefore it may influence the possible continuous crossover occurring in cores of NSs.

\section{Perturbative Corrections to $s_{\textmd{c}}^2$}\label{appA}

In this appendix, we estimate the first a few perturbative corrections to $s_{\rm{c}}^2$ from the $b_4$-term.
The scheme is given as follows: 
We first notice that\,\cite{CLZ23},
 \begin{equation}\label{Ap-1}
b_4=-\frac{1}{2}b_2\left(\widehat{P}_{\rm{c}}
+\frac{4+9\widehat{P}_{\rm{c}}}{15s_{\rm{c}}^2}\right),
\end{equation}
where $b_2=-6^{-1}(1+3\widehat{P}_{\rm{c}}^2+4\widehat{P}_{\rm{c}})$\,\cite{CLZ23}.
In (\ref{Ap-1}), the $s_{\rm{c}}^2$ should be modified (compared with Eq.\,(\ref{sc2})) as,
\begin{equation}\label{Ap-2}
s_{\rm{c}}^2\approx\widehat{P}_{\rm{c}}\left(1+\frac{1}{3}\frac{1+3\widehat{P}_{\rm{c}}^2+4\widehat{P}_{\rm{c}}}{1-3\widehat{P}_{\rm{c}}^2}\right)\left(1+\ell_1\widehat{P}_{\rm{c}}+\ell_2\widehat{P}_{\rm{c}}^2\right),
\end{equation}
where $\ell_1$ and $\ell_2$ are two coefficients to be determined.
When writing out (\ref{Ap-2}), the applicable region of the perturbative expansion should be set as $|\ell_2\widehat{P}_{\rm{c}}^2|\lesssim|\ell_1\widehat{P}_{\rm{c}}|$ and $|\ell_1\widehat{P}_{\rm{c}}|\lesssim1$, or equivalently $\widehat{P}_{\rm{c}}\lesssim\min\{|\ell_1\ell_2^{-1}|,|\ell_1^{-1}|\}$.
The reduced radius $\widehat{R}$ is estimated from $\widehat{P}_{\rm{c}}+b_2\widehat{R}^2+b_4\widehat{R}^4=0$, see treatments given in the Appendix of Ref.\,\cite{CLZ23}.

Similarly\,\cite{CLZ23}, the NS mass is given as $M_{\rm{NS}}\sim\widehat{R}^3/\sqrt{\varepsilon_{\rm{c}}}$.
Taking the derivative of $M_{\rm{NS}}$ with respect to $\varepsilon_{\rm{c}}$ and making it be zero gives the $s_{\rm{c}}^2$,  i.e.,
\begin{align}
s_{\rm{c}}^2\approx&\frac{4}{3}\widehat{P}_{\rm{c}}+\frac{4\ell_1+923}{708}\widehat{P}_{\rm{c}}^2+
\frac{37\ell_1-4\ell_1^2+4\ell_2+583}{354}\widehat{P}_{\rm{c}}^3.
\end{align}
By comparing it with the corresponding perturbative expansion of (\ref{Ap-2}) to the same order, namely,
\begin{equation}
s_{\rm{c}}^2\approx\frac{4}{3}\widehat{P}_{\rm{c}}+\left(\frac{4}{3}+\frac{4}{3}\ell_1\right)\widehat{P}_{\rm{c}}^2
+\left(2+\frac{4}{3}\ell_1+\frac{4}{3}\ell_2\right)\widehat{P}_{\rm{c}}^3,
\end{equation}
one obtains
$\ell_1=-21/940\approx-0.02$ and $\ell_2\approx-0.25$.
The effective applicable region is estimated as $\widehat{P}_{\rm{c}}\lesssim\widehat{P}_{\rm{c}}^{\rm{eff}}\approx\min\{|\ell_1\ell_2^{-1}|,|\ell_1^{-1}|\}=|\ell_1\ell_2^{-1}|\approx0.09$.

Although $\widehat{P}_{\rm{c}}\lesssim0.09$ is required,  one can find that for $\widehat{P}_{\rm{c}}\approx0.24$ the central SSS $s_{\rm{c}}^2\approx0.438$ which is very close to about 0.446 predicted by Eq.\,(\ref{sc2}).
Furthermore, setting $s_{\rm{c}}^2\leq1$ from the causality condition now leads to $\widehat{P}_{\rm{c}}\lesssim0.381$ (it should be noticed, however, this  upper bound far exceeds the effective region of $\widehat{P}_{\rm{c}}\lesssim\widehat{P}_{\rm{c}}^{\rm{eff}}\approx0.09$).

\section{Order of Magnitude Estimate for $a_4$}\label{appB}

In this appendix, we give an order of magnitude estimate for the coefficient $a_4$ appearing in the expansion of $\widehat{\varepsilon}\approx1+a_2\widehat{r}^2+a_4\widehat{r}^4$.
The starting point is that both $\widehat{\varepsilon}$ and $\widehat{\rho}=\rho/\rho_{\rm{c}}$ are decreasing functions of radial distance from the center.
We may use the following elementary result: If $y(x)=1+Ax+Bx^2$ ($A<0$) is a decreasing function defined for positive $x$, then the first-order derivative $y'(x)$ should be negative, and consequently,
\begin{equation}\label{eee}
B\leq-{A}/{2x_{\max}},
\end{equation}
with $x_{\max}$ the maximum value of $x$ the $y$ could take.

Applying the criterion (\ref{eee}) to $\widehat{\varepsilon}\approx1+a_2\widehat{r}^2+a_4\widehat{r}^4$ and expression (\ref{pk-4}) in the main text and treating them as functions of $\widehat{r}^2$ gives,
\begin{align}
a_4\leq&-\frac{b_2}{2s_{\rm{c}}^2}\frac{1}{\widehat{R}^2},~~\rm{and,}~~
a_4\leq\frac{b_2}{2s_{\rm{c}}^2}\left(\frac{b_2}{1+\widehat{P}_{\rm{c}}}-\frac{1}{\widehat{R}^2}\right),\label{est-1}
\end{align}
respectively, where $\widehat{R}=R/Q$ is the reduced NS radius and $a_2=b_2/s_{\rm{c}}^2$ is used here.
In order to estimate the order of magnitude of $a_4$ conservatively, we take both $\widehat{R}$ and $\widehat{P}_{\rm{c}}$ as small as possible (determined by the structure of (\ref{est-1})).
For instance, by taking $R\approx10\,\rm{km}$ and $\varepsilon_{\rm{c}}\approx0.5\,\rm{GeV}/\rm{fm}^3$, we obtain $\widehat{R}\approx0.91$ as $Q=(4\pi G\varepsilon_{\rm{c}})^{-1/2}\approx11\,\rm{km}$.
Similarly,  we adopt the reduced central pressure as $\widehat{P}_{\rm{c}}\approx0.16$ (for massive NSs), so the first inequality of (\ref{est-1}) leads to $a_4\lesssim0.67$ while the second gives $a_4\lesssim0.80$.
Therefore,  $a_4\sim\mathcal{O}(1)$.
On the other hand,  both $\widehat{\varepsilon}$ and $\widehat{\rho}$ are obviously decreasing functions of $\widehat{r}$ if $a_4<0$.

\end{document}